\documentclass[article]{aastex631}
\usepackage{graphicx}	
\usepackage{amsmath}	
\usepackage{amssymb}	
\usepackage{xcolor}
\usepackage{soul}
\usepackage{booktabs}
\usepackage[caption=false]{subfig}
\usepackage{float}

\usepackage{newtxtext,newtxmath}
\usepackage[utf8]{inputenc}
\usepackage[polish,english]{babel}

\shorttitle{A bias-free cosmological analysis with Quasars} 

\begin{document}


\title{A bias-free cosmological analysis with quasars alleviating $H_0$ tension}

\author{Aleksander \L{}ukasz Lenart}
\altaffiliation{First and second authors share the same contribution}

\affiliation{Astronomical Observatory, Jagiellonian University, ul. Orla 171, 31-501 Kraków, Poland}
\author{Giada Bargiacchi}
\altaffiliation{Corresponding author, giada.bargiacchi@unina.it}
\affiliation{Scuola Superiore Meridionale,
                Largo S. Marcellino 10,
                80138, Napoli, Italy}
\affiliation{Istituto Nazionale di Fisica Nucleare (INFN), Sez. di Napoli,
                Complesso Univ. Monte S. Angelo, Via Cinthia 9,
                80126, Napoli, Italy}
\author{Maria Giovanna Dainotti}
\affiliation{National Astronomical Observatory of Japan, 2 Chome-21-1 Osawa, Mitaka, Tokyo 181-8588, Japan}
\affiliation{The Graduate University for Advanced Studies, SOKENDAI, Shonankokusaimura, Hayama, Miura District, Kanagawa 240-0193, Japan}
\affiliation{Space Science Institute, 4765 Walnut St, Suite B, 80301 Boulder, CO, USA}
\author{Shigehiro Nagataki}
\affiliation{Interdisciplinary Theoretical \& Mathematical Science Program, RIKEN (iTHEMS), 2-1 Hirosawa, Wako, Saitama, Japan 351-0198}
\affiliation{RIKEN Cluster for Pioneering Research, Astrophysical Big Bang Laboratory (ABBL), 2-1 Hirosawa, Wako, Saitama, Japan 351-0198}
\affiliation{Astrophysical Big Bang Group (ABBG), Okinawa Institute of Science and Technology Graduate University (OIST),
1919-1 Tancha, Onna-son, Kunigami-gun, Okinawa, Japan 904-0495}
\author{Salvatore Capozziello}
\affiliation{Scuola Superiore Meridionale, 
                Largo S. Marcellino 10,
                80138, Napoli, Italy}
\affiliation{Istituto Nazionale di Fisica Nucleare (INFN), Sez. di Napoli,
                Complesso Univ. Monte S. Angelo, Via Cinthia 9,
                80126, Napoli, Italy}
\affiliation{Dipartimento di Fisica "E. Pancini" , Universit\'a degli Studi di  Napoli "Federico II"\\
                Complesso Univ. Monte S. Angelo, Via Cinthia 9
                80126, Napoli, Italy}



\begin{abstract}
Cosmological models and their parameters are widely debated because of theoretical and observational mismatches of the standard cosmological model, especially the current discrepancy between the value of the Hubble constant, $H_{0}$, obtained by Type Ia supernovae (SNe Ia), and the Cosmic Microwave Background Radiation (CMB). Thus, considering high-redshift probes like quasars (QSOs), having intermediate redshifts between SNe Ia and CMB, is a necessary step. In this work, we use SNe Ia and the most updated QSO sample, reaching redshifts up to $z\sim7.5$, applying the Risaliti-Lusso QSO relation based on a non-linear relation between ultraviolet and X-ray luminosities. We consider this relation both in its original form and corrected for selection biases and evolution in redshift through a reliable statistical method also accounting for the circularity problem. We also explore two approaches: with and without calibration on SNe Ia. We then investigate flat and non-flat standard cosmological models and a flat $w$CDM model, with a constant dark energy equation of state parameter $w$. Remarkably, when correcting for the evolution as a function of cosmology, we obtain closed constraints on $\Omega_M$ using only non-calibrated QSOs. We find that considering non-calibrated QSOs combined with SNe Ia and accounting for the same correction, our results are compatible with a flat $\Lambda$CDM model with $\Omega_M = 0.3$ and $H_0 = 70 \, \mathrm{km\,s^{-1}\,Mpc^{-1}}$. Intriguingly, the $H_0$ values obtained place halfway between the one from SNe Ia and CMB, paving the way for new insights into the $H_0$ tension.
\end{abstract}



\section{Introduction}

Quasars (QSOs) are extraordinarily luminous active galactic nuclei (AGNs) currently observed up to redshift $z = 7.642$ \citep{2021ApJ...907L...1W}. These features make them potentially the next rung of the cosmic distance ladder beyond Supernovae Type Ia (SNe
Ia), that have been observed only up to $z=2.26$ \citep{Rodney}.
Using QSOs as cosmological tools requires a full understanding of their physical mechanisms that are still being debated by the scientific community.

A non-linear relation between the ultraviolet (UV) and X-ray luminosities in QSOs was ﬁrst discovered by the ﬁrst X-ray surveys
\citep{1979ApJ...234L...9T,1981ApJ...245..357Z,1986ApJ...305...83A}, and has been conﬁrmed using various samples of QSOs observed with the main X-ray observatories over a wide redshift range and wide ranges of UV luminosity that span over 5 orders of magnitudes \citep[e.g.][]{steffen06,just07,2010A&A...512A..34L,lr16,2021A&A...655A.109B}. 
One possible physical explanation of this relation is the following one: QSO accretion disk on a central supermassive black hole emits photons in the UV band, which are then processed through the inverse Compton effect by an external plasma of relativistic electrons, giving rise to the X-ray emission. This physical explanation,
while plausible, lacks accounting for the stability of the X-ray emission, as the external electrons should cool down due to the inverse Compton effect and fall onto the central region. Ultimately, one needs an efficient energy transfer between the accretion disk and the external region to describe such a stable emission. The nature of this link between the two regions is not known yet. However, some models have been proposed \citep[see e.g.][]{2017A&A...602A..79L}.

This X-UV relation has been recently used to provide independent measurements of QSO distances \citep[see e.g.][for details]{rl15,rl19,2020A&A...642A.150L} making them standardized cosmological tools. In its cosmological application, this relation is referred to as the Risaliti-Lusso (RL) relation. To use this relation in cosmology, we need to properly select the QSO sample addressing as many observational issues as possible, such as dust reddening, X-ray absorption, galaxy contamination, and Eddington bias, as specified in \citet{2020A&A...642A.150L} and \citet{DainottiQSO}. Indeed, in its first applications, the X-UV relation showed a very large intrinsic dispersion $sv \sim 0.35/0.40$ dex in logarithmic units \citep[][]{2010A&A...512A..34L}. Only recently, it has been discovered that this dispersion is not completely intrinsic, but has mainly an observational origin \citep{lr16}, thus an accurate selection of the sources is crucial. This finding has allowed reducing the intrinsic scatter to $sv \sim 0.2$ dex \citep{2020A&A...642A.150L} and has rendered this relation suitable for cosmological analyses, turning QSOs into more reliable cosmological tools. In addition, in \citet{DainottiQSO} it has been proved through well-established statistical tests that this relation is not induced or distorted by selection biases and/or redshift evolution, but it is intrinsic to the physics of QSOs. 
The methodology used to standardize QSOs is complementary to the one  traditionally applied for SNe Ia to estimate the cosmological parameters, yet it extends the Hubble-Lema\^itre diagram (i.e. the distance modulus-redshift diagram) to a redshift range currently inaccessible to SNe Ia ($z=2.26-7.54$). Indeed, extending the cosmological computations with high-redshift data is crucial to distinguish between different cosmological models that are degenerate at low redshifts, to allow better constraints on the dark energy (DE) behaviour, and to explore possible extensions of the standard cosmological model. 

As a matter of fact, the most widely adopted parameterization of the observed Universe is based on the so-called $\Lambda$ cold dark matter ($\Lambda$CDM) model \citep{peebles1984}, which relies on the existence of cold dark matter (CDM) and DE ($\Lambda$) associated with a cosmological constant \citep{2001LRR.....4....1C} in a spatially flat geometry.
Predictions from this model have been found to agree with most of the observational probes such as the cosmic microwave background \citep[CMB; e.g.][]{planck2018}, the baryon acoustic oscillations \citep[BAOs; e.g.][]{eboss2021}, and the present accelerated expansion of the Hubble flow, based on the Hubble-Lema\^itre diagram of SNe Ia \citep[e.g.][]{riess1998,perlmutter1999}, where the dominant dynamical contribution of DE related to the cosmological constant should drive such an acceleration. However, the fundamental physical origin and the properties of DE are still unknown, as the interpretation of $\Lambda$ is hampered by a severe  fine-tuning issue to obtain the right amount of DE observed today. Moreover, the data sets mentioned above do not fully exclude a spatially non-flat Universe \citep{Park:2017xbl,2020NatAs...4..196D,Handley:2019tkm,DiValentino:2020hov,2021JCAP...10..008Y}. 
Relevant deviations from the spatially flat $\Lambda$CDM model have already been found using high-redshift probes; \citet{Dainotti2008,Dainotti11b,Dainotti2013a,Dainotti2020a,Dainotti2020b,2022MNRAS.514.1828D} have worked extensively on the standardization of Gamma-Ray Bursts (GRBs) as cosmological candles, while QSOs combined with SNe Ia have been studied by \citet{rl19,lusso2019,2020A&A...642A.150L,2021A&A...649A..65B,2022MNRAS.515.1795B}. 

In addition, the search for high-redshift standard candles has also been enhanced by the Hubble tension, a $4-6 \, \sigma$ discrepancy between the direct measurements of the local Hubble constant $H_0$ and the one inferred from cosmological models, remarkably the value from the Planck data based on the CMB within the $\Lambda$CDM model. The restricted number of cosmological probes at intermediate redshifts and selection biases are the major shortcomings that prevent a solution to this problem. Thus, additional high-$z$ standardized probes beyond SNe Ia, such as GRBs and QSOs, could be instrumental in shedding light on this problem \citep{cardone09,cardone10,postnikov14,Benetti,Spallicci,Rocco,2021A&A...649A..65B,2021ApJ...912..150D,2022PASJDainotti,2022MNRAS.515.1795B,2022arXiv220109848D, Moresco:2022phi}. Several groups worldwide are investigating if this tension is due to selection biases, or due to new physics. The $H_0$ tension is mainly discussed within the $\Lambda$CDM model, characterized by a constant equation of state of DE ($w = -1$), but it is vital to investigate the extensions of this model, such as the $w$CDM model, in which $w \neq -1$, or non-flat models. 
Relaxing the standard $\Lambda$CDM model, rigorously flat in spatial curvature and with non-evolving $\Lambda$, is conceptually useful both for solving the cosmological constant problem and as a possible arena for new physics asking for the evolution of DE (\textbf{\citealt{2014Galax...2...22P}}, \citealt{Luongo}, \textbf{\citealt{2019IJMPD..2842001P}, \citealt{2021PhRvD.104l3511P}, \citealt{2022Univ....8..502P}}).

Therefore, we here conduct a detailed analysis of both these extensions of the flat $\Lambda$CDM model using QSOs, through the RL relation, and SNe Ia as cosmological probes. For the first time in the literature, the so-called circularity problem can be solved for QSOs corrected for evolutionary effects. The circularity means that the computations of cosmological parameters depend on the underlying the cosmological model assumed, and hence this implies the assumption of given cosmological parameters. In general the parameters of correlations depend on values of cosmological parameters if they involve the luminosities or energies, thus fixed parameters of the correlations obtained under the assumption of some cosmological model, would bias cosmological results computed with such fixed parameters. This work shows important points of originality compared to previous analyses based on QSOs \citep[e.g.][]{rl15,rl19,2020A&A...642A.150L,2020MNRAS.492.4456K,2020MNRAS.497..263K,2021A&A...649A..65B,2022MNRAS.515.1795B,2022arXiv220310558C}: 1) we apply the RL relation not only in its original form, the one commonly used for cosmological studies, but we also take into account selection biases and the evolution in redshift of the luminosities through reliable statistical methods overcoming the so-called circularity problem; 2) we investigate how cosmological results change upon the choice of calibrating or non-calibrating QSOs with SNe Ia; 3) we discuss how the $H_0$ tension will be impacted by the inclusion of QSOs in the analyses and to what extent non-flat cosmological models can be a viable explanation for describing the current results. Our point of view is empirical and conservative and shows that improving data samples and distance indicators, first of all in the redshift range between SNe Ia and CMB, could alleviate the $H_0$ tension.

Recent studies have investigated the reliability of the application of the X-UV relation in cosmology.  In \citet{2022ApJ...935L..19P}, the authors argued that the RL procedure is circular. In this work, we completely overcome, for the first time in the literature on QSOs, the circularity problem while also accounting for the correction for evolution in redshift of luminosities. Indeed, we apply this correction contemporaneously to the variation of cosmological parameters, as detailed in Section \ref{evolution}. Moreover, the authors of the above-mentioned paper base their criticism on a binned analysis of the parameters of the RL relation, which is not an appropriate approach due to the relative large dispersion of the correlation. Although we agree that without correcting the relation there exists a clustering of data points in relation to the redshift, as shown in the left panel of Figure \ref{figRL}, we overcome this issue by applying the correction for redshift evolution and selection biases, as visible in the right panel of the same figure. Thus, we agree, that one cannot apply calibration methodology in a simple straightforward way without any correction. Our results presented in this manuscript show, that calibration methodology becomes unnecessary in many cases, when our innovative approach is applied, but we still show results of the calibration methodology without the correction, for the sake of comparison on how our correction changes the results. Indeed, in this manuscript, we show that QSOs have a high potential to be standalone cosmological probes.
From another point of view, \citet{2021MNRAS.502.6140K,2022MNRAS.510.2753K} doubted the applicability of the QSO sample in its entirety for constraining cosmological parameters stating that the parameters of the RL relation depend on the cosmological model assumed and the redshift. \citet{2021MNRAS.502.6140K,2022MNRAS.510.2753K} state that due to those issues some QSO sub-samples cannot be standardised. Thus, their effort is in the direction of pinpointing a sub-sample with a higher potential for standardizing QSOs. Our approach is different and we focus on the correction of the whole sample in order to use its full potential of extending the Hubble diagram. We agree that the non-corrected RL relation is changing with redshift and we solve this issue further in the manuscript. Conversely, we here find compatibility in $1 \, \sigma$ (except for one case with a $2 \, \sigma$ discrepancy) between the values of all the parameters of the relation when using non-calibrated QSOs combined with SNe Ia, independently of the model studied. \citet{2021MNRAS.502.6140K,2022MNRAS.510.2753K} found incompatibility in more than 3 $\sigma$ only in very exotic models. Moreover, our correction for selection bias and redshift evolution, treated as a function of evolution leads to the compatibility of all parameters in less than $0.2 \sigma$. These results are shown and discussed in Sections \ref{flatLCDM}, \ref{nonflatLCDM}, and \ref{flatwCDM}. Thus, these discussions on the RL relation and its applicability are not a concern given our innovative approach and results.

The manuscript is organized as follows. In Section \ref{data} we present the data sets used for the cosmological analyses and their selection and in Section \ref{methodology} the methodology applied for the fits and the treatment of selection biases, redshift evolution of luminosities, circularity problem, and calibration of QSOs. In Section \ref{models} we describe the cosmological models studied in this work and our assumptions on their corresponding parameters, whilst in Section \ref{results} we present the main results for all data sets and different approaches comparing them and discussing the implications. Finally, in Section \ref{conclusions}, we summarize our findings.

\section{The Data}
\label{data}
Here, we work by selecting and combining samples of QSO and SNe Ia measurements to investigate the late-time Universe. In this section, we describe each data set used for the cosmological analyses. These are carried out using both QSOs alone and the two probes combined. 

For SNe Ia, we consider the collection of 1048 sources from the \textit{Pantheon} sample \citep{scolnic2018}. These are collected by different surveys and span from $z=0.01$ up to $z=2.26$.
The sample of QSOs used is the one described in \citet{2020A&A...642A.150L}. This is composed of 2421 sources that cover the redshift range up to $z =7.54$ \citep{banados2018}. These sources have been carefully selected for cosmological studies and we refer for a detailed description to \citet{rl15}, \citet{lr16}, \citet{rl19}, \citet{salvestrini2019} and \citet{2020A&A...642A.150L}. Here and in Section \ref{fitting method}, we summarize the crucial points required by the present work.

This QSO sample is properly selected to remove as many observational biases as possible. First, a selection is made by removing the sources for which the signal-to-noise ratio does not guarantee well-sampled photometry (i.e. signal-to-noise ratio S/N $< 1$). After this first screening, all QSOs with a spectral energy distribution (SED) that shows UV reddening and near-infrared host-galaxy contamination, are discarded, requiring only sources with extinction E(B$-$V)$\leq 0.1$. This corresponds to selecting only the sources that satisfy $\sqrt{(\Gamma_{1,\mathrm{UV}}-0.82)^2 + (\Gamma_{2,\mathrm{UV}}-0.40)^2} \leq 1.1$, where $\Gamma_{1, \mathrm{UV}}$ and $\Gamma_{2, \mathrm{UV}}$ are the slopes of a log($\nu$)-log($\nu \, L_\nu$) power-law in the rest frame 0.3-1$\mu$m and 1450-3000 \AA \, ranges, respectively, and $\nu$ and $L_\nu$ are the frequency and the luminosity per unit of frequency. The values  $\Gamma_{1, UV}=0.82$ and $\Gamma_{2, UV}=0.4$ refer to a SED with zero extinction \citep[see][]{2006AJ....131.2766R}. In addition, X-ray observations where photon indices ($\Gamma_X$) are peculiar or indicative of absorption are excluded by requiring $\Gamma_X + \Delta \Gamma_X \geq 1.7$ and $\Gamma_X \leq 2.8 $ if $z < 4$ and $\Gamma_X \geq 1.7$ if $z \geq 4$, where $\Delta \Gamma_X$ is the uncertainty on the photon index. Finally, the remaining observations are corrected for the Eddington bias. 
Indeed, the cleaned sample consists only of sources fulfilling $\text{log}F_{X,exp} - \text{log}F_{min} \geq {\cal F}$, where ${\cal F}$ is a threshold value and $F_{X,exp}$ is the X-ray flux computed from the observed UV flux assuming the RL relation with fixed parameters within the flat $\Lambda${CDM} model with $\Omega_{M}=0.3$ and $H_{0} = 70\, \mathrm{km\,s^{-1}\,Mpc^{-1}}$. $F_{lim}$ is the flux limit of the specific observation estimated from the catalogue. More precisely, as detailed in \citet{lr16}, for each object the minimum detectable flux is computed from the total exposure time of the charge-coupled device (CCD) where the source is detected using the functions plotted in Figure 3 by \citet{2001A&A...365L..51W}. The value of ${\cal F}$ required in this filter is ${\cal F} = 0.9$ for the Sloan Digital Sky Survey Data Release 14 SDSS DR14– 4XMM Newton \citep{2018A&A...613A..51P,2020A&A...641A.136W} and XXL \citep{2016yCat..74570110M} sub-samples and  ${\cal F} = 0.5$ for the SDSS-\textit{Chandra} \citep{2010ApJS..189...37E}. 
All the surviving multiple X-ray observations are ultimately averaged to reduce the effects of the X-ray variability. The sample we use is the result of all these selection filters.

Here, we also remark that, as opposed to what has been done by other authors \citep[e.g.][]{2020A&A...642A.150L,2021A&A...649A..65B,2022MNRAS.515.1795B}, we consider the full sample of QSOs, without the filter of $z>0.7$ proposed in \citet{2020A&A...642A.150L}, thus highlighting that the results shown here are not biased by any cut in redshift in the sample nor suffer from artificial truncation of the current data sample. In addition, this QSO sample is the most suitable one for cosmological studies, compared to others previously used in the literature \cite[see e.g.][]{rl15,2016ApJ...831...60S,rl19,lusso2019}, for several reasons: 1) it is obtained by matching recent UV and X-ray surveys (e.g. SDSS DR14, 4XXM Newton), 2) it is carefully analyzed and selected against observational biases \citep{2020A&A...642A.150L}, 3) it presents observations of 29 luminous QSOs in the high-redshift range of $z= 3.0-3.3$ which have been obtained from an XMM–Newton campaign (cycle 16, proposal ID: 080395, PI: Risaliti), and had been specifically selected as suitable for cosmological analyses because they represented the most luminous quasar population with homogeneous optical/UV properties, \citep{2019A&A...632A.109N}, 4) it includes two samples of $z > 4$ QSOs published by \citet{salvestrini2019} and \citet{2019A&A...630A.118V} and a local sample in the redshift range $0.009 < z < 0.1$. These factors guarantee a high quality of measurements, a huge number of sources, high-redshift points to extend the Hubble-Lema\^itre diagram far beyond the one of SNe Ia, and coverage at very low-$z$ that allows a better calibration with SNe Ia. 

\section{Methodology}
\label{methodology}

\subsection{Fitting methodology}
\label{fitting method}
All the analyses presented in this work are obtained using our own codes in Mathematica 12.2 \citep{Mathematica} and Jupyter notebooks \citep{jupyter}, in which we computed the investigated parameters using a Bayesian technique, the D'Agostini method \citep{2005physics..11182D}. This technique makes use of Markov Chain Monte Carlo (MCMC) approach. The D'Agostini method has the advantage of accounting for error bars on both variables considered and also for an intrinsic dispersion $sv$ of the relation fitted. Our applied algorithm starts from a given uniform prior of the parameters. Then the algorithms searches in a loop for solutions of values of the parameters within the priors that maximize the likelihood. With every iteration of the loop the prior are updated remembering previous results based on the application of the Bayes theorem. With this technique we can obtain a distribution of probability of all parameters at once without fixing any of them. This allows to overcome the circularity problem without the aid of any calibrator.

The likelihood function ($LF$) used in this method to fit the SNe Ia sample is defined as\footnote{For the sake of simplicity we always use $\text{ln}$ instead of $\text{log}_{\mathrm{e}}$ and $\text{log}$ instead of $\text{log}_{10}$.}
\begin{equation} \label{lfsne}
\text{ln}(LF)_{\mathrm{SNe}} = -\frac{1}{2} \Bigg[\left(\mathbf{y}-\boldsymbol{\mu}\right)^{T} \, \textit{C}^{-1} \, \left(\mathbf{y}-\boldsymbol{\mu}\right)\Bigg]
\end{equation}
where $\mathbf{y}$ is the distance modulus measured, $\textit{C}$ is the associated 1048x1048 covariance matrix that includes both statistical and systematic uncertainties, and $\boldsymbol{\mu}$ is the distance modulus predicted by the cosmological model assumed, yet depending both on the free parameters of the model and the redshift.

As anticipated, the strategy to compute QSO distances makes use of the non-linear relation between their UV and X-ray luminosity \citep{steffen06,just07,2010A&A...512A..34L,lr16,2021A&A...655A.109B}, namely 
\begin{equation}\label{RL}
\mathrm{log} L_{X} = g \, \mathrm{log} L_{UV} + b
\end{equation}
where $L_X$ and $L_{UV}$ are the luminosities (in $\mathrm{erg \, s^{-1} \, Hz^{-1}}$) at 2 keV and 2500 \AA, respectively. In \cite{DainottiQSO} this relation has been corrected for selection biases and redshift evolution using the Efron \& Petrosian method (\citealt{1992ApJ...399..345E}, hereafter EP) and it has been proved that it is intrinsic to QSO's properties so that it can be reliably used to standardize QSOs as cosmological tools.
To make use of Equation \eqref{RL}, we compute luminosities from measured flux densities $F$ according to $L_{X,UV}= 4 \pi d_{l}^{2}\ F_{X, UV}$, where $d_l$ is the luminosity distance. $d_l$ is computed under the assumption of a cosmological model using the corresponding parameters as free parameters of the fit. Usually, luminosities of the sources have to be corrected for the K-correction, defined as $1/(1+z)^{1-\alpha}$, where $\alpha$ is the spectral index of the sources, but for QSOs it is assumed to be $\alpha = 1$, leading to a $K=1$. So the K-correction has been omitted following \citet{2020A&A...642A.150L}.
The $LF$ function used for QSOs is \citep[see also][]{2021MNRAS.502.6140K,2022MNRAS.510.2753K,2022MNRAS.515.1795B,2022arXiv220310558C}:
\begin{equation} \label{lfqso}
\text{ln}(LF)_{\text{QSO}} = -\frac{1}{2} \sum_{i=1}^{N} \left[ \frac{(y_{i}-\phi_{i})^{2}}{s^{2}_{i}} + \text{ln}(s^{2}_{i})\right].
\end{equation}
In this case, the data $y_{i}$ correspond to $\text{log}L_{X}$, while $\phi_{i}$ to the logarithmic X-ray luminosity predicted by the X-UV relation. Moreover, $s^{2}_{i} = \sigma ^{2}_{y_{i}} + g^{2} \sigma ^{2}_{x_{i}} + sv^{2}$ and it takes into account the statistical uncertainties on $\text{log}L_{X}$ ($y$) and $\text{log}L_{UV}$ ($x$), but also the intrinsic dispersion $sv$ of the X-UV relation, which is another free parameter of the fit. Practically, $LF_{\text{QSO}}$ is just the same $LF$ function used for SNe Ia in Equation \eqref{lfsne}, but modified to include the contribution of the intrinsic dispersion of the X-UV relation.
In all models studied in this work, we fit also QSOs combined with SNe Ia, thus the joint $LF$ function used for the combined sample of QSOs and SNe Ia is given by $\text{ln}(LF)_{\text{QSO+SNe}} = \text{ln}(LF)_{\text{SNe}}+\text{ln}(LF)_{\text{QSO}}$.
In this work, we explore different methods to apply the QSO relation to cosmology. Specifically, we use 1) Equation \eqref{RL} in its form, 2) Equation \eqref{RL} corrected for a fixed evolution in redshift of luminosities, 3) Equation \eqref{RL} corrected for a redshift evolution of luminosities that varies together with cosmological parameters. For each of these approaches, we also consider two different cases: with and without calibration with SNe Ia. Details of all these different approaches are given in the following subsections.

\subsection{Selection biases and redshift evolution with a circularity-free treatment }
\label{evolution}

\begin{figure}
\centering
    \includegraphics[width=0.49\textwidth]{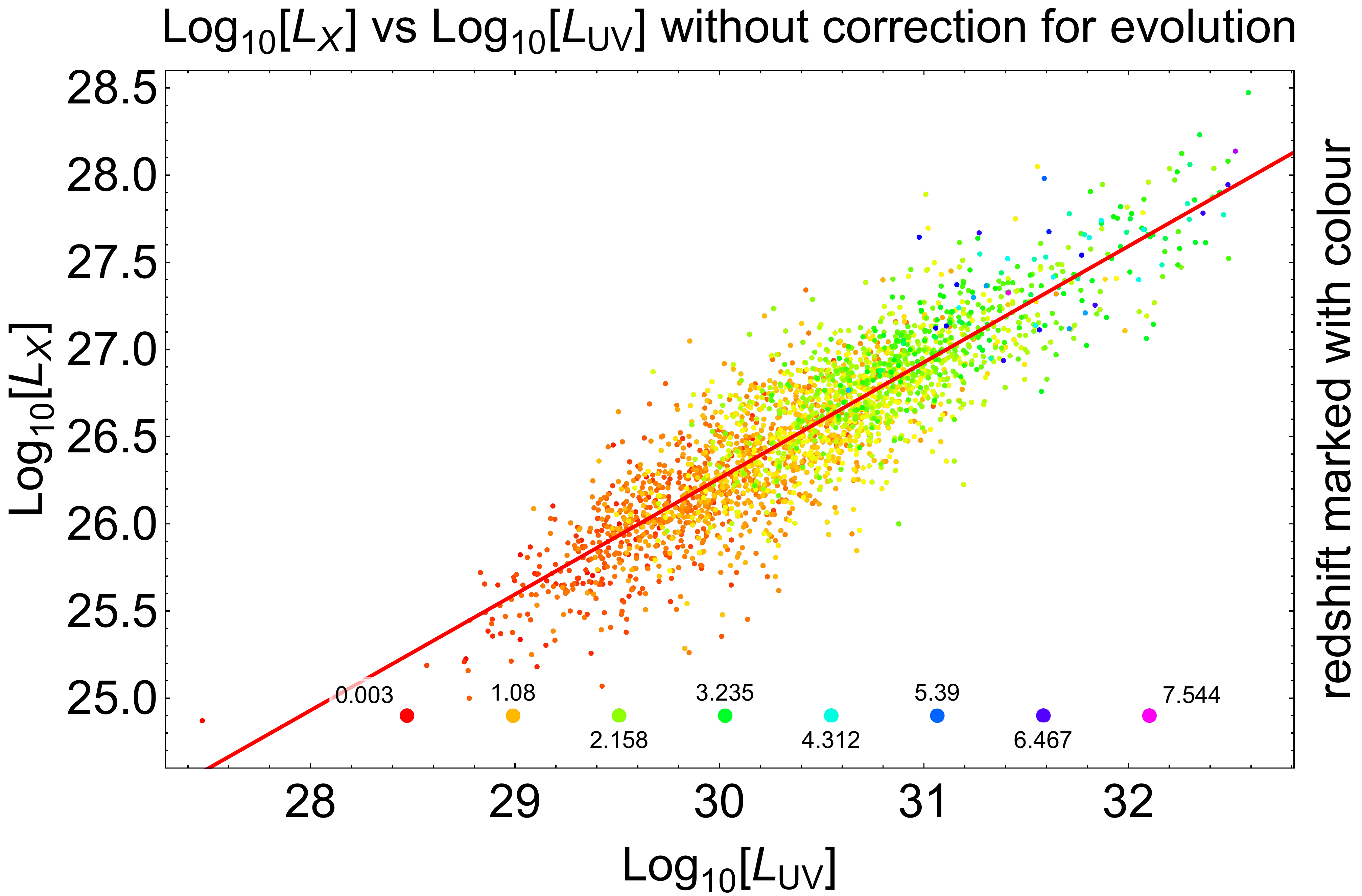}
    \includegraphics[width=0.49\textwidth]{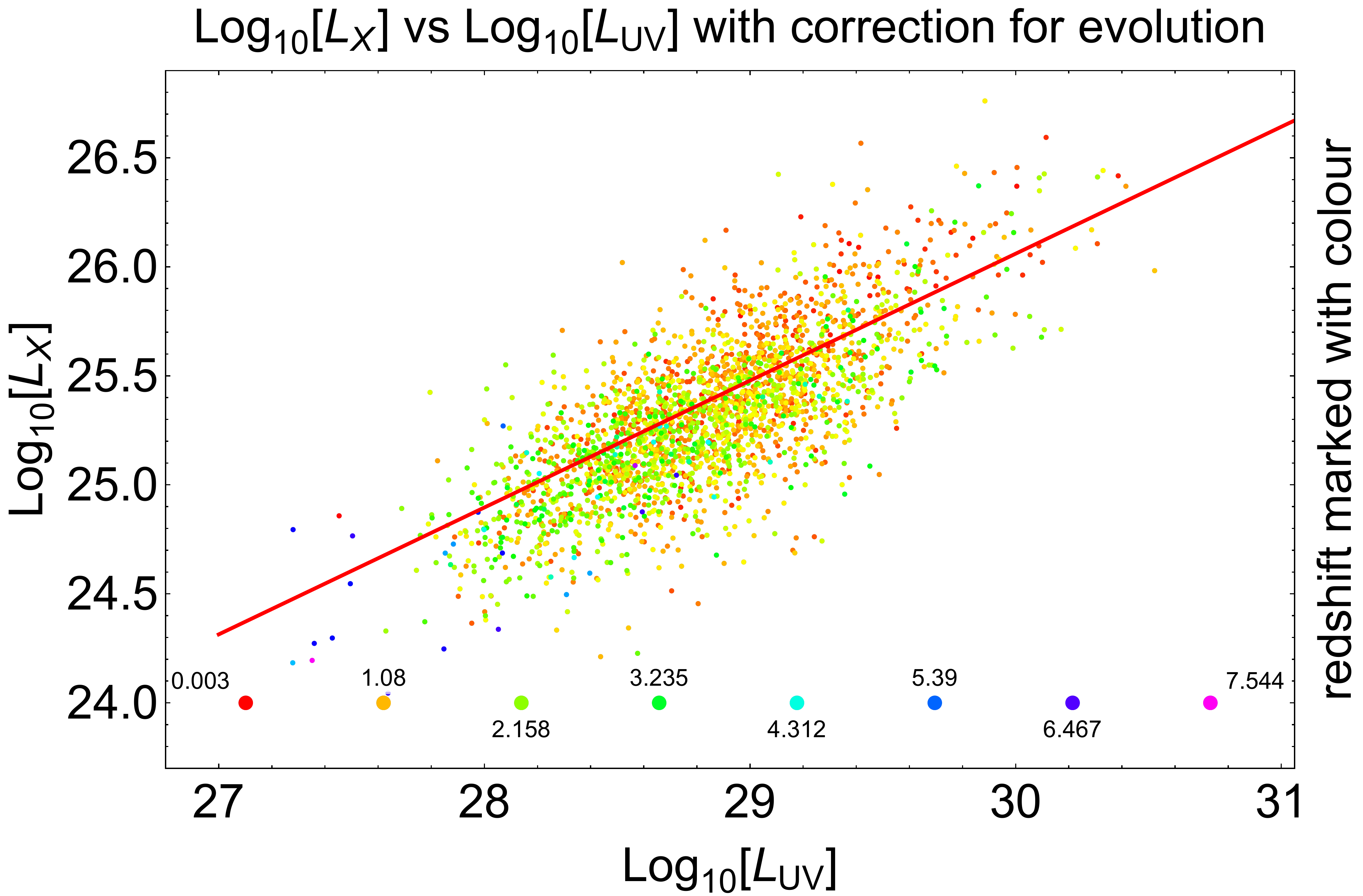}
    \caption{Comparison of the RL relation with and without the correction for evolution in redshift of luminosities, as shown in the right and left panels, respectively. Points are marked with different colours according to their redshift, as depicted in the pictures, and the red line represents the best-fit linear RL relation. }
    \label{figRL}
\end{figure}

As already stressed, QSOs are observed at very high redshifts, which is their huge advantage for cosmological applications, but also the reason for significant selection biases. Indeed, as it has been pinpointed in astrophysics, such selection biases can deform a correlation between the physical parameters of a source (see Figure \ref{figRL}), distort the cosmological parameters \citep{Dainotti2013a}, or even induce an artificial correlation. Thus, it is vital to test correlations against this type of effects.

\begin{figure}
\centering
\includegraphics[width=.5\columnwidth]{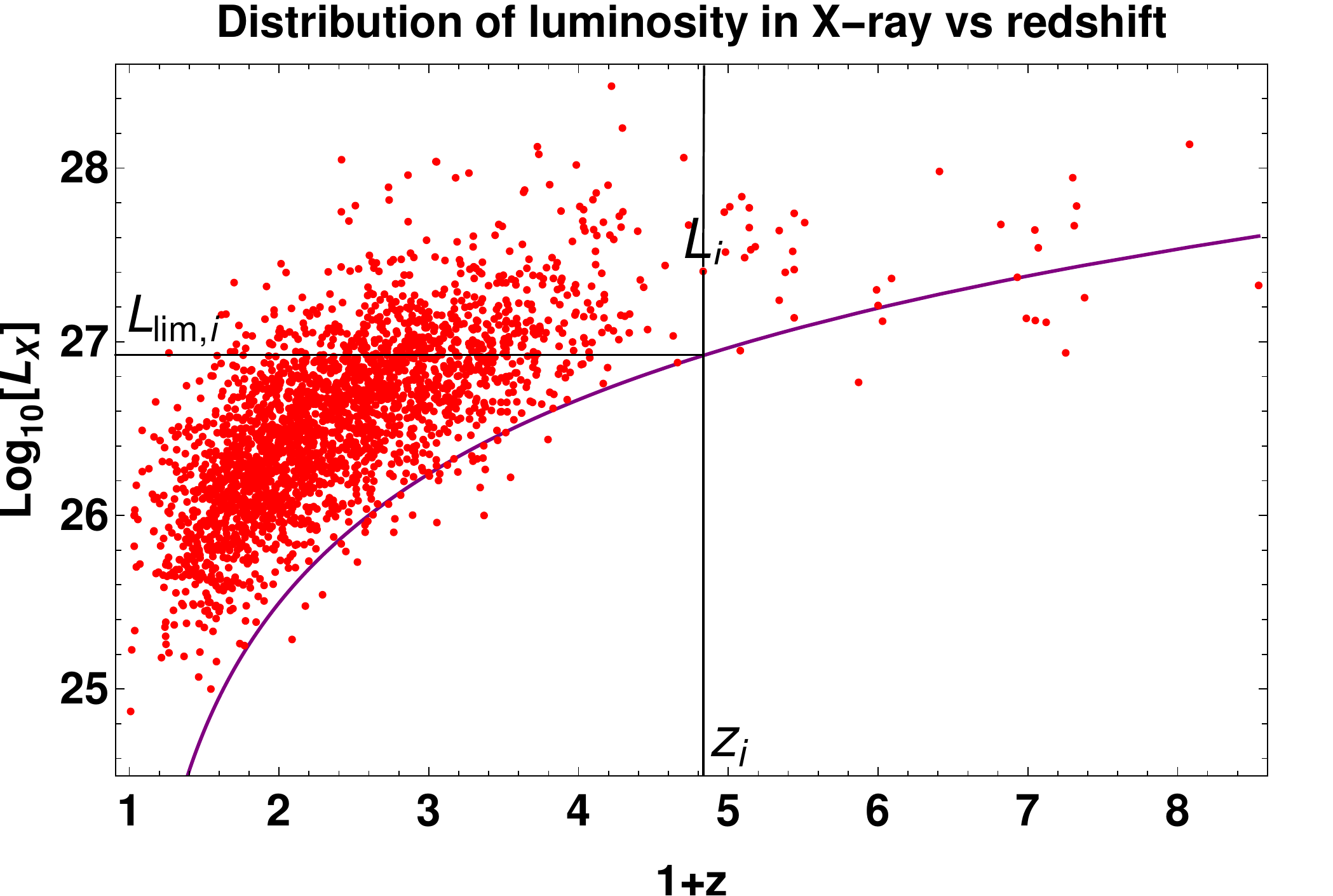}
\caption{X-ray luminosity (in logarithmic units) vs. $(1+z)$. Solid purple curve shows the truncation due to the flux limit, which is here assumed to be $F_{lim} = 6\times 10^{-33}  \mathrm{erg \, s^{-1} \, cm^{-2} \, Hz^{-1}}$ \citep{DainottiQSO}. The rank $\mathcal{R}_i$ of the point of luminosity $L_{i}$ at redshift $z_{i}$ is the number of points in the rectangle within black lines, with points under the purple curve being discarded. This figure is computed under the assumption of a flat $\Lambda$CDM model with $\Omega_{M} = 0.3$.}
\label{fig:EP}
\end{figure}

The most commonly used method to correct data for selection biases and redshift evolution is the EP method, whose reliability has already been demonstrated for GRBs, also via Monte Carlo simulations \citep[see e.g.][]{Dainotti2013a,Dainotti2013b,Dainotti2015b,2017A&A...600A..98D,2021Galax...9...95D}. This method assumes that the corrected, de-evolved, physical quantity (in our case the luminosity) $L'$ is equal to the observed one $L$ divided by some function of the redshift $\xi (z)$ of an assumed form, $L' = \frac{L}{\xi (z)}$. The simplest function that mimics the evolution is the power-law $\xi (z) = (1+z)^{k}$ \citep[see e.g.][]{Dainotti2013a,2017A&A...600A..98D}. As addressed in \cite{Dainotti2015b}, the functional form for the evolution can be a power-law of this form or a more complex function \citep[see also][]{2011ApJ...743..104S}, but both these functions result in computed parameters that are compatible in 1 $\sigma$ for the case of QSOs \citep[see Table 1 of][]{DainottiQSO}. To determine the $k$ parameter, one has to compute a grid of values for $k$ and find the one that corresponds to the absence of correlation between the corrected quantity and the redshift. Therefore, the Kendall's $\tau$ coefficient is perfect for this task. Namely, we are looking for such a $k$ for which $\tau = 0$, where, following the EP method, we define $\tau$ as:

\begin{equation}
    \tau =\frac{\sum_{i}{(\mathcal{R}_i-\mathcal{E}_i)}}{\sqrt{\sum_i{\mathcal{V}_i}}}.
\end{equation}
In this formula, for each redshift $z_{i}$ in our sample, we compute the number of data points in a rectangle built intersecting the limiting luminosity $L_{lim, i}$ (i.e. the lowest possible luminosity observed at a given redshift computed with the assumed limiting flux from $L_{lim, i} = F_{lim} \times 4 \pi d^2_l(z_{i})$) and the redshift $z_{i}$ itself. The associated set for $z_i$ contains all QSOs verifying $L_{z_j} \geq  L_{min,i}$ and $z_j \leq z_i$, where $j$ and $i$ refer to objects of the associated set and the complete QSO sample, respectively. The rank $\mathcal{R}_i$ of the data point $y_i$ with luminosity $L_i$ at redshift $z_i$ is computed as the number of these data points in the corresponding associated set. Then, we subtract from the rank of each data point its expectation value corresponding to a distribution with no correlation: $\mathcal{E}_i = \frac{1}{2}(i+1)$. After summing the obtained differences, the correlation is removed when the sum is 0 (i.e. $\tau = 0$). To make this concept clearer, we present the visualization of the computation of the rank $\mathcal{R}_i$ in the above-described rectangle in Figure \ref{fig:EP}, where the case of the X-ray luminosity is shown as an example. In Figure \ref{fig:EP} the assumed limiting flux is $F_{lim} = 6\times 10^{-33}  \mathrm{erg \, s^{-1} \, cm^{-2} \, Hz^{-1}}$ \citep[see][]{DainottiQSO} and in our computation, we discard the data points which are below the value of the corresponding limiting luminosity (purple curve). This figure was obtained for demonstration purposes under the assumption of a flat $\Lambda$CDM model with $\Omega_{M} = 0.3$.
For completeness, we have to consider the normalization for the variance $\mathcal{V}_i = \frac{1}{12}(i^{2}+1)$. To this end, we compute $\tau$ by dividing the above-described sum by the sum of variances for each data point.
To find the $1 \, \sigma$ uncertainty on the investigated $k$ parameter, we compute the $k$'s values corresponding to $\tau = 1$ and $\tau = -1$.

This method has already been successfully applied in the literature to many probes, including QSOs. In particular, using the same QSO sample as the one used in this work, \cite{DainottiQSO} tested two different functional forms of $\xi (z)$ obtaining numerically compatible results. Assuming $\xi (z) = (1+z)^{k}$, the authors obtained $L'_{X} = L_{X}/(1+z)^{3.36\pm0.07}$ and $L'_{UV} = L_{UV}/(1+z)^{4.36\pm0.08}$, with $F_{lim} = 4.5\times \, 10^{-29} \,\mathrm{erg \, s^{-1} \, cm^{-2} \, Hz^{-1}}$ for UV and $F_{lim} = 6\times \, 10^{-33}\, \mathrm{erg \, s^{-1} \, cm^{-2} \, Hz^{-1}}$ for X-rays. These choices are legitimated by the authors showing that the evolutionary coefficient $k$ depends only weakly on these assumptions: even spanning over one order of magnitude in $F_{lim}$ both in UV and X-rays, the evolutionary coefficient's results remain compatible within 2 $\sigma$. This was already proved in \citet{2021ApJ...914L..40D} for several sample sizes of GRBs. 

Looking at Figure \ref{figRL}, it is visible that the application of the above-defined correction removes the dependence of the RL relation on redshift and that, once corrected, the data points are evenly dispersed in redshift. Indeed, in the case without correction (left panel), the redshift increases with higher luminosities, while, once the relation is corrected for evolution (right panel), this trend completely disappears and redshifts are blended over the whole range of luminosities. 

\begin{figure}
\centering
\gridline{
\fig{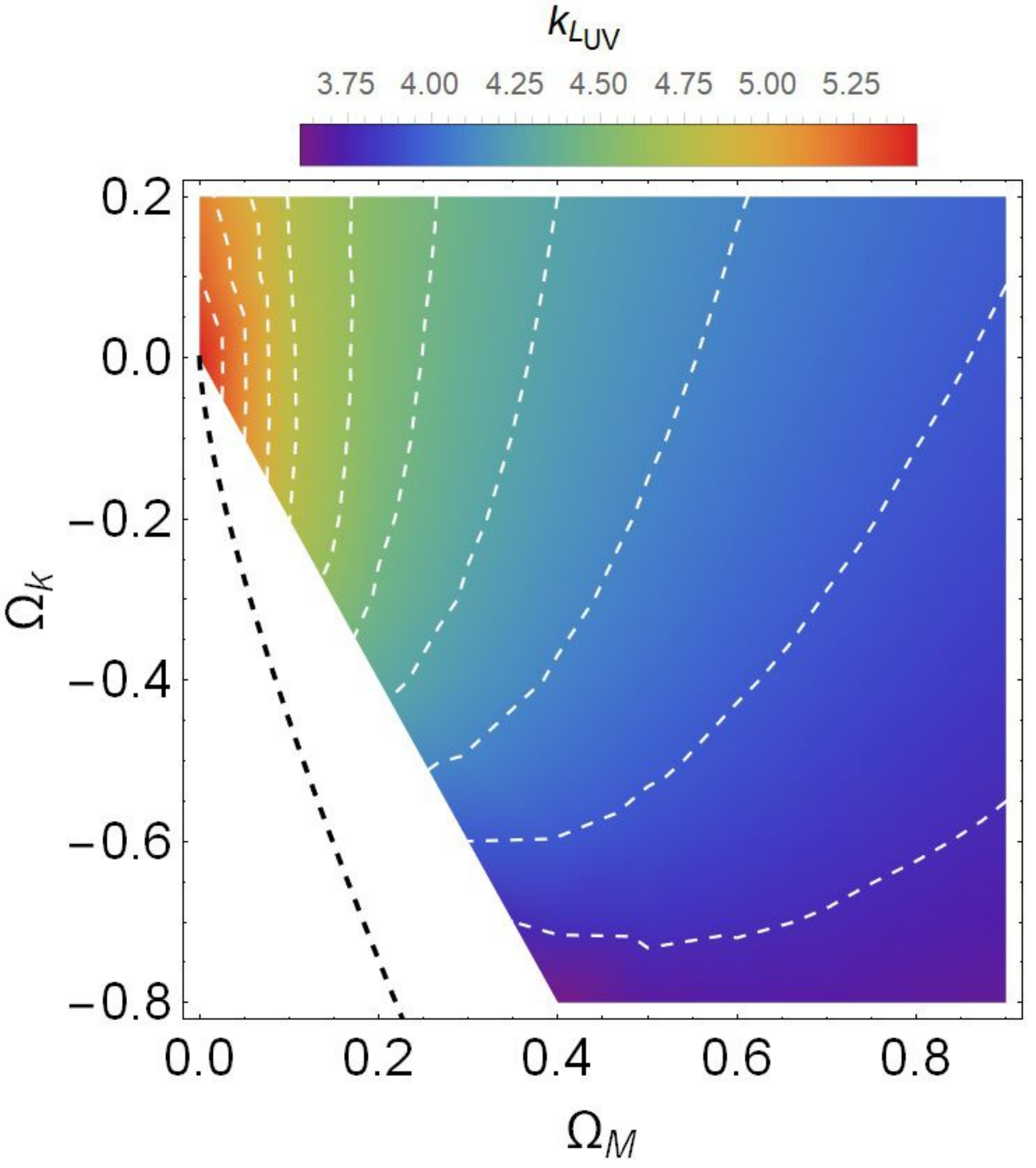}{0.45\textwidth}{(a) Behaviour of the $k$ parameter in the parameter space $\{\Omega_{M}, \, \Omega_{k}\}$ for UV. The dashed black line shows the no Big-Bang constraint (see Section \ref{assumptions} and Equation \ref{nobigbang}).}
\fig{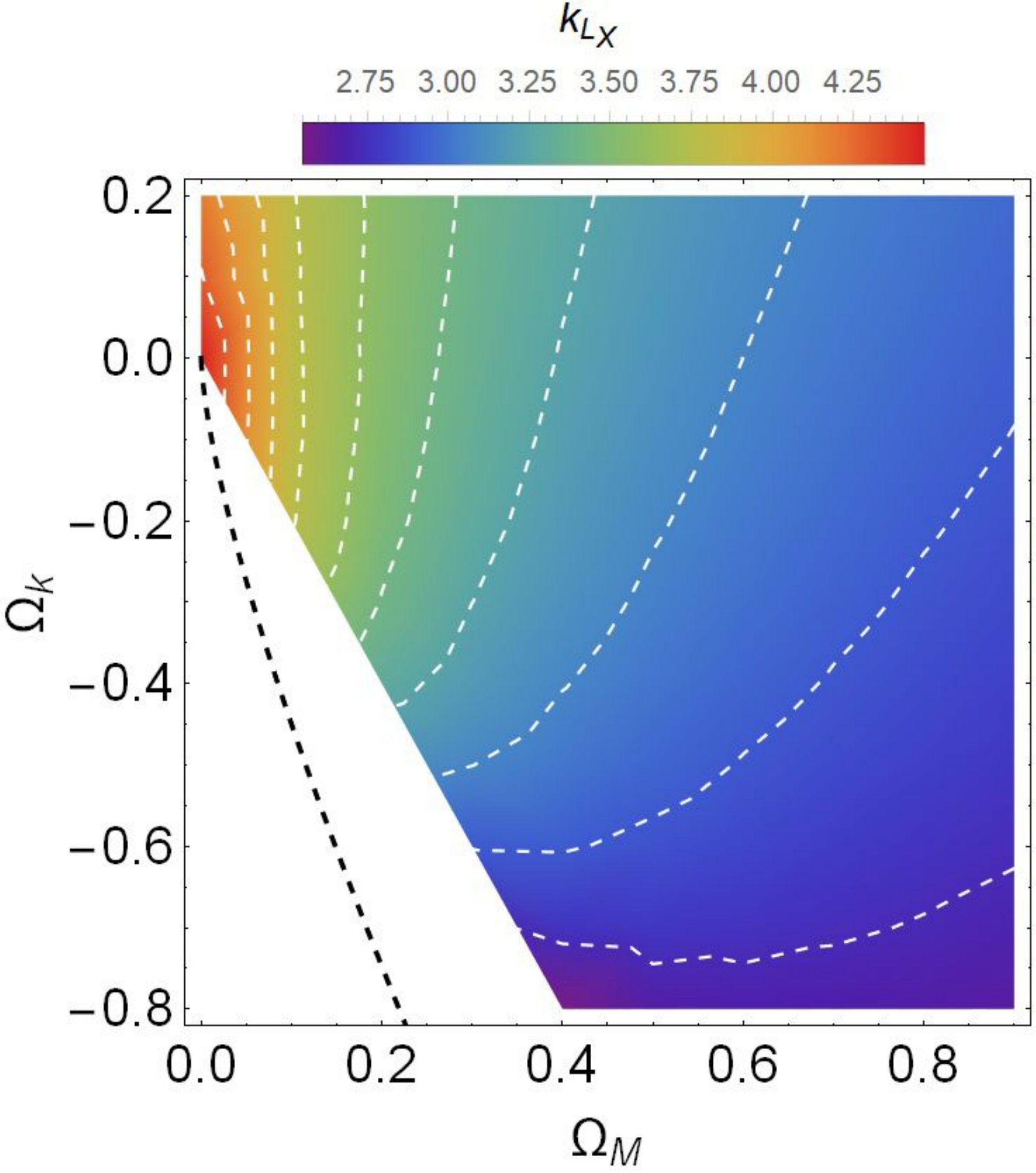}{0.45\textwidth}{(b) Behaviour of the $k$ parameter in the parameter space $\{\Omega_{M}, \, \Omega_{k}\}$ for X-rays. The dashed black line shows the no Big-Bang constraint (see Section \ref{assumptions} and Equation \ref{nobigbang}).}}
\gridline{
\fig{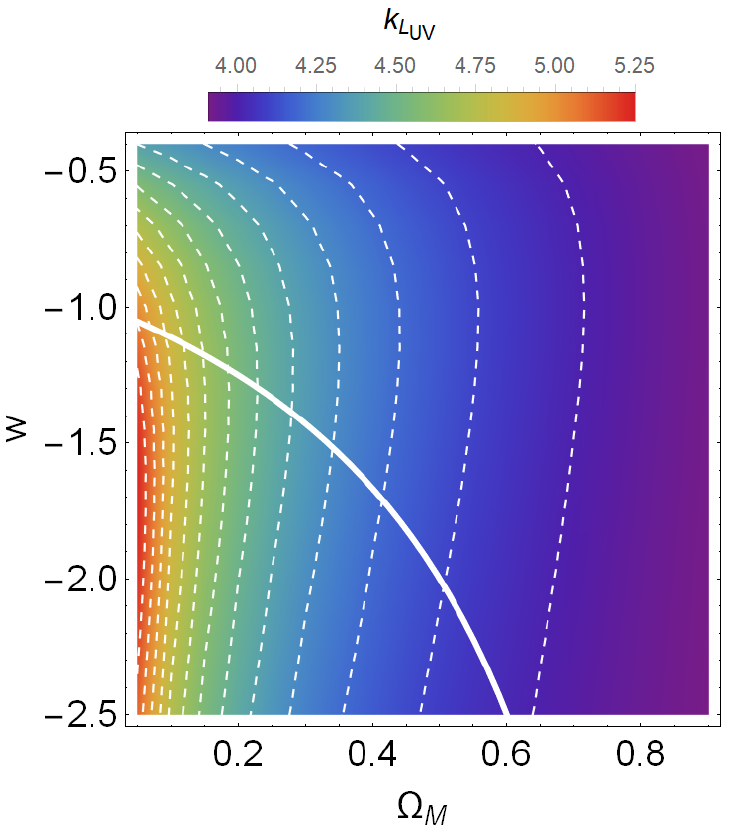}{0.45\textwidth}{(c) The behaviour of the $k$ parameter in the parameter space $\{\Omega_{M}, \, $w$\}$ for UV. The region under the thick white line corresponds to the scenario of $H'(z)|_{z=0} \leqslant 0$, while the region above the following line to $H'(z)|_{z=0} > 0$.}
\fig{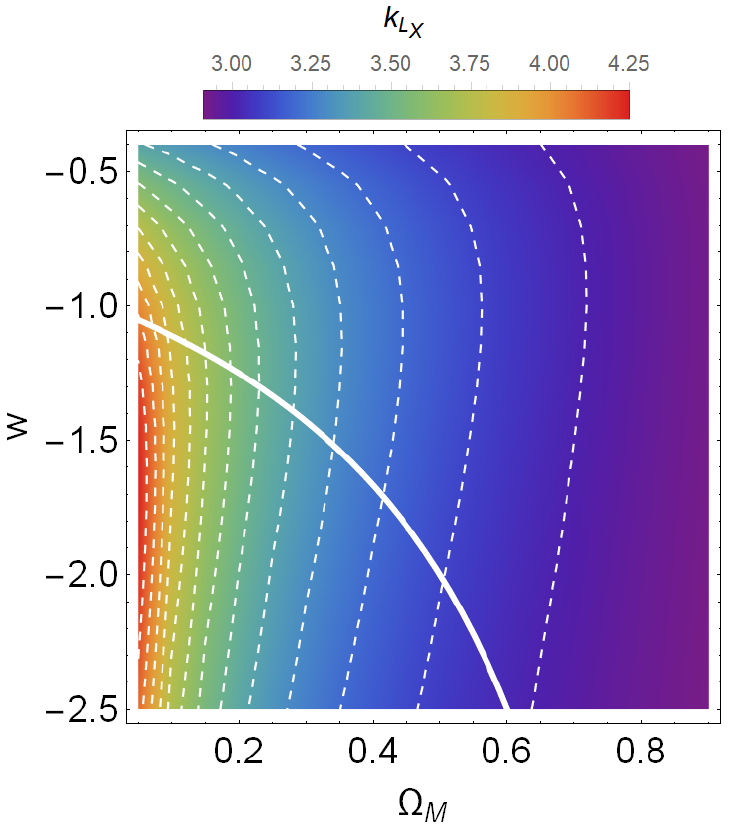}{0.45\textwidth}{(d) The behaviour of the $k$ parameter in the parameter space $\{\Omega_{M}, \, $w$\}$ for X-rays. The region under the thick white line corresponds to the scenario of $H'(z)|_{z=0} \leqslant 0$, while the region above the following line to $H'(z)|_{z=0} > 0$.}}
    \caption{Behaviour of the $k$ parameter in the parameter spaces $\{\Omega_{M}, \, \Omega_{k}\}$ (panels (a) and (b)) and $\{\Omega_{M}, \, w\}$ (panels (c) and (d))  for both UV (left panels) and X-rays (right panels).}
    \label{fig:EPvsOmOk}
\end{figure}

It has to be pinpointed here that the execution of such a method is possible only with an assumed cosmology, required by the computation of luminosities, thus this technique cannot be straightforwardly applied in cosmological computations. Such a calculation of cosmological parameters would be indeed affected by the so-called circularity problem. To overcome this issue, we repeat the same procedure for a grid of cosmological parameters that fall in reasonable physical ranges of values and we study how the $k$ parameter behaves with cosmology. Thus, when applying this correction in the MCMC method for obtaining the values of cosmological parameters such as $\Omega_{M}$, $\Omega_{k}$, $H_0$, and $w$, we consider $k=k(\Omega_{M})$, $k=k(\Omega_{k})$, $k=k(\Omega_{M}, \,\Omega_{k})$, $k=k(\Omega_{M}, \, H_0)$, $k=k(w)$, or $k=k(\Omega_{M}, \,w)$ according to the free parameters of the model considered. Indeed, we here remind that $k$ does not depend on $H_{0}$, since this parameter is responsible only for a scaling of the luminosity's value and does not change the correlation, as shown in \cite{DainottiQSO}, so we do not have to consider also $k=k(H_0)$. In our computations, the numerical functions $k=k(\Omega_{M})$, $k=k(\Omega_{k})$, and $k=k(w)$ are always created with a cubic-spline method, while $k=k(\Omega_{M}, \,\Omega_{k})$ and $k=k(\Omega_{M},w)$ are created with a 'quintic' type of interpolation. Since cosmological parameters usually depend on each other and have to be fitted  simultaneously in the MCMC method, we compute two maps that show how $k$ changes in the parameter spaces $\{\Omega_{M}, \, \Omega_{k}\}$ and $\{\Omega_{M}, \, w\}$, which covers all cases studied in our cosmological computations (see Figure \ref{fig:EPvsOmOk}). We comment on the results of these investigations in Section \ref{kvscosmology}. 

\subsection{Calibration methodology}
\label{calibration}

Many attempts have been done in the literature to improve cosmological constraints by applying the so-called calibration method. This approach is based on anchoring a probe to another one that presents an overlapping redshift range and has measured distances. This procedure is commonly applied, for example, to calibrate SNe Ia with the distances of Cepheids measured through their pulsation period-luminosity relation \citep[see e.g.][]{1996ApJ...460L..15S,2009ApJS..183..109R}. This has lead to the so-called cosmic distance ladder to measure distances in the Universe.
At redshifts higher than the ones explored by SNe Ia, the most commonly used technique assumes that high-$z$ cosmological probes follow the same cosmology as SNe Ia in their common redshift range. Thus, the sub-sample composed only of sources observed up to the maximum redshift of SNe Ia is fitted under the assumption of a cosmological model with parameter values based on the computations with SNe Ia. The parameters of the correlation are then fixed to the values obtained assuming the cosmology obtained by SNe Ia in that particular overlapping region. This approach is the one we follow in this work to calibrate QSOs. As the maximum redshift of our SNe Ia data set is $z=2.26$, we create our calibrating sub-sample composed of 2066 QSOs (out of the initial 2421 sources) and fit the RL relation on this sub-sample combined with SNe Ia. The results of the MCMC fitting without correcting for evolution in redshift of luminosities are as follows: $g = 0.648 \pm 0.009 $, $b = 6.817 \pm 0.265 $, $sv = 0.234 \pm 0.004 $. While once the correction for evolution is applied, we obtain: $g' = 0.591 \pm 0.013 $, $b' = 8.278 \pm 0.0362 $, $sv' = 0.231 \pm 0.004 $, where the ' stands for the same parameters as before but once the correction is applied.

\section{The cosmological models}
\label{models}

In this work, we investigate three cosmological models, which are the most commonly studied in recent analyses. In both cases, the considered cosmological components are: dark energy (indicated by the subscript $_{\Lambda}$), non-relativistic matter ( $_M$ ), including both baryons ( $_b$ ) and cold dark matter ( $_{CDM}$ ), and the relativistic component ( $_r$ ), composed of radiation ( $_{\gamma}$ ) and neutrinos ( $_{\nu}$ ). The last one makes a negligible contribution to the late Universe (in which $\Omega_{r}= \Omega_{\gamma}+\Omega_{\nu} =9 \, \mathrm{x}\, 10^{-5}$), thus we set the current relativistic density parameter $\Omega_{r}$ equal to 0 in our computation. 

In the first case, we consider a Universe with a potential non-zero curvature of space-time, in which all parameters are connected via the relation $1 = \Omega_{r} + \Omega_{M} + \Omega_{\Lambda} + \Omega_{k} $. Within this model, the luminosity distance $d_l$ is given by:

\begin{equation}
    d_{l} = (1+z) \frac{c}{H_{0}} 
    \begin{cases}
  \frac{1}{\sqrt{|\Omega_{k}|}} sin(\sqrt{|\Omega_{k}|} \int^{z}_{0} \frac{d \zeta}{E(\zeta)} ) & \text{ if } \Omega_{k} < 0,\\ 
 \int^{z}_{0} \frac{d \zeta}{E(\zeta)} & \text{ if } \Omega_{k} = 0, \\ 
 \frac{1}{\sqrt{\Omega_{k}}} sinh(\sqrt{\Omega_{k}} \int^{z}_{0} \frac{d \zeta}{E(\zeta)} ) & \text{ if } \Omega_{k} > 0 ,
\end{cases}
\label{generalmod}
\end{equation}
where $E(\zeta)$ stands for the dimensionless Hubble parameter defined as

\begin{equation}
    E(\zeta) = \frac{H(\zeta)}{H_{0}} = \sqrt{\Omega_{r} (1+\zeta)^{4} + \Omega_{M} (1+\zeta)^{3} + \Omega_{k} (1+\zeta)^{2} + \Omega_{\Lambda}}. 
\end{equation}
$\Omega_{k} = 0$ corresponds to a flat space-time and we refer to the model with this fixed value as flat $\Lambda$CDM model, while the model with any possible value of $\Omega_{k}$ is referred to as non-flat $\Lambda$CDM model.

The other model investigated in this work is the most natural extension of the $\Lambda$CDM scenario. Indeed, we consider an equation of state of DE  $w= P_{\Lambda}/\rho_{\Lambda}$, with $P_{\Lambda}$ and $\rho_{\Lambda}$ the pressure and energy density of DE, respectively, that can assume any constant value. The case with $w=-1$ corresponds to the $\Lambda$CDM model. In this model, Equation \eqref{generalmod} changes only in the form of the $E(\zeta)$ function, which becomes
\begin{small}
\begin{equation}
    E_{w}(\zeta) = \sqrt{\Omega_{r} (1+\zeta)^{4} + \Omega_{M} (1+\zeta)^{3} + \Omega_{k} (1+\zeta)^{2} + \Omega_{\Lambda} (1+\zeta)^{3(1+w)}}.
    \label{equ:Ew}
\end{equation}
\end{small}In this case, we assume a flat Universe fixing $\Omega_{k} = 0$ and, as a consequence, $\Omega_{\Lambda} = 1-\Omega_{M}-\Omega_{r}$. This assumption is consistent with the most recent cosmological observations on the CMB \citep{planck2018} and other recent studies (e.g. \citealt{eboss2021,2021JCAP...11..060G}), where non-flat universes are consistent with zero curvature.

\subsection{Assumptions on cosmological parameters}
\label{assumptions}

It is worth mentioning some of the assumptions we use in our cosmological analyses. First of all, in all models, we use flat uniform priors on the free parameters as follows: $ 0 \leq \Omega_{M} \leq 1$, $60 \leq H_0 \leq 80 $, $-2.5 \leq w \leq -0.34$, and $-0.9 \leq \Omega_k \leq 0.6$. The limits on $\Omega_k$ have been increased to obtain convergence in all cases studied.
The upper limit on $w$ is imposed following the second Friedmann's equation, according to which we need $ w(z)<-1/3$ to explain the present accelerated expansion of the Universe \citep{riess1998,perlmutter1999} as a DE dominant effect.
In addition to these priors, we do not look for solutions to the values of cosmological parameters in the whole parameter space, as there are regions that lead to non-physical or non-reasonable solutions. Thus, in our analyses, we discard the part of the $\{\Omega_{M}, \Omega_{k}\}$ parameter space which leads to the no Big-Bang solutions, that do not admit an initial singularity. Following \citet{1992ARA&A..30..499C}, the region that admits physical solutions with an initial singularity in a $\Lambda$CDM model corresponds to:

\begin{equation}
\label{nobigbang}
\Omega_{k} \geq
\left\{\begin{matrix}
 1 - \Omega_{M} - 4 \, \Omega_{M}\, \mathrm{cosh}^{3} \Bigg[\frac{1}{3} \, \mathrm{arccosh} \left( \frac{1- \Omega_{M}}{\Omega_{M}}\right)\Bigg] & \, \Omega_{M}\leq\frac{1}{2}, \\ \\
 1 - \Omega_{M} - 4 \, \Omega_{M}\, \mathrm{cos}^{3} \Bigg[\frac{1}{3} \, \mathrm{arccos} \left( \frac{1- \Omega_{M}}{\Omega_{M}}\right)\Bigg] & \, \Omega_{M}>\frac{1}{2}.
\end{matrix}\right.
\end{equation}
Part of the discarded region can be seen in both panels of the upper row of Figure \ref{fig:EPvsOmOk} under the dashed black line. 

Furthermore one can consider only the region of {$\Omega_{M}$, $w$} space which does not lead to present values of the Hubble constant higher than its value in the past. This would allow only for the case of $\frac{d\, H(z)}{d\, z}|_{z=0} \geq 0$, what satisfies the null energy condition \citep{doi:10.1142/9789812792129_0014}. Taking the derivative of equation \ref{equ:Ew} and substituting $z=0$ one can obtain the following criteria:

\begin{equation}
    w \geq \frac{1}{\Omega_{M}-1}.
    \label{wcond}
\end{equation}

The border set up by this condition is shown in both panels of the bottom row in Figure \ref{fig:EPvsOmOk} with a thick white line. The depicted region can be a test of violation of the null energy condition regarding the present discussion about its viability.

\begin{figure}
\centering

\gridline{
\fig{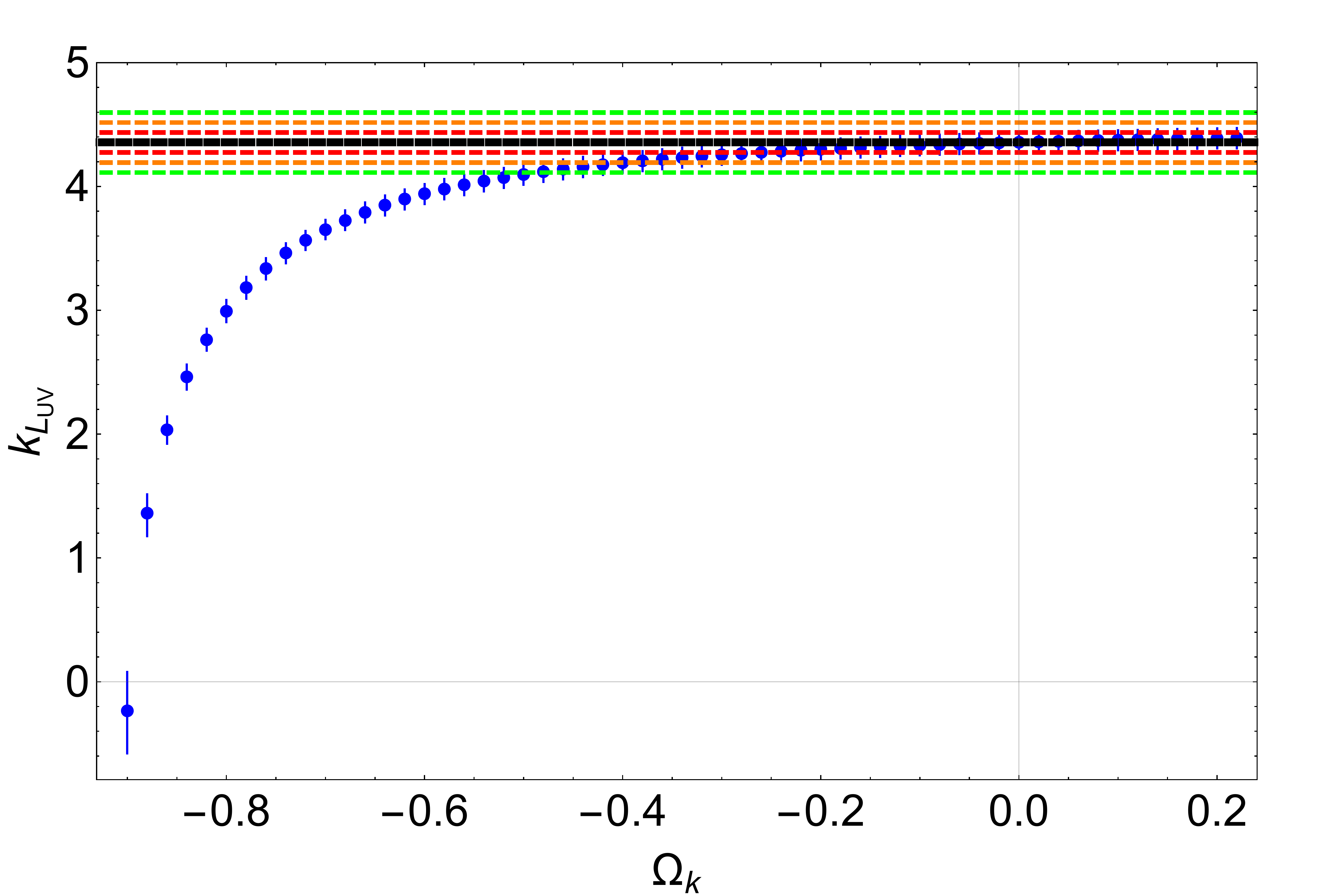}{0.45\textwidth}{(a) Behaviour of $k$ parameter for the correction of $L_{UV}$ in relation to $\Omega_{k}$ for a non-flat $\Lambda$CDM model with $\Omega_M=0.3$.}
\fig{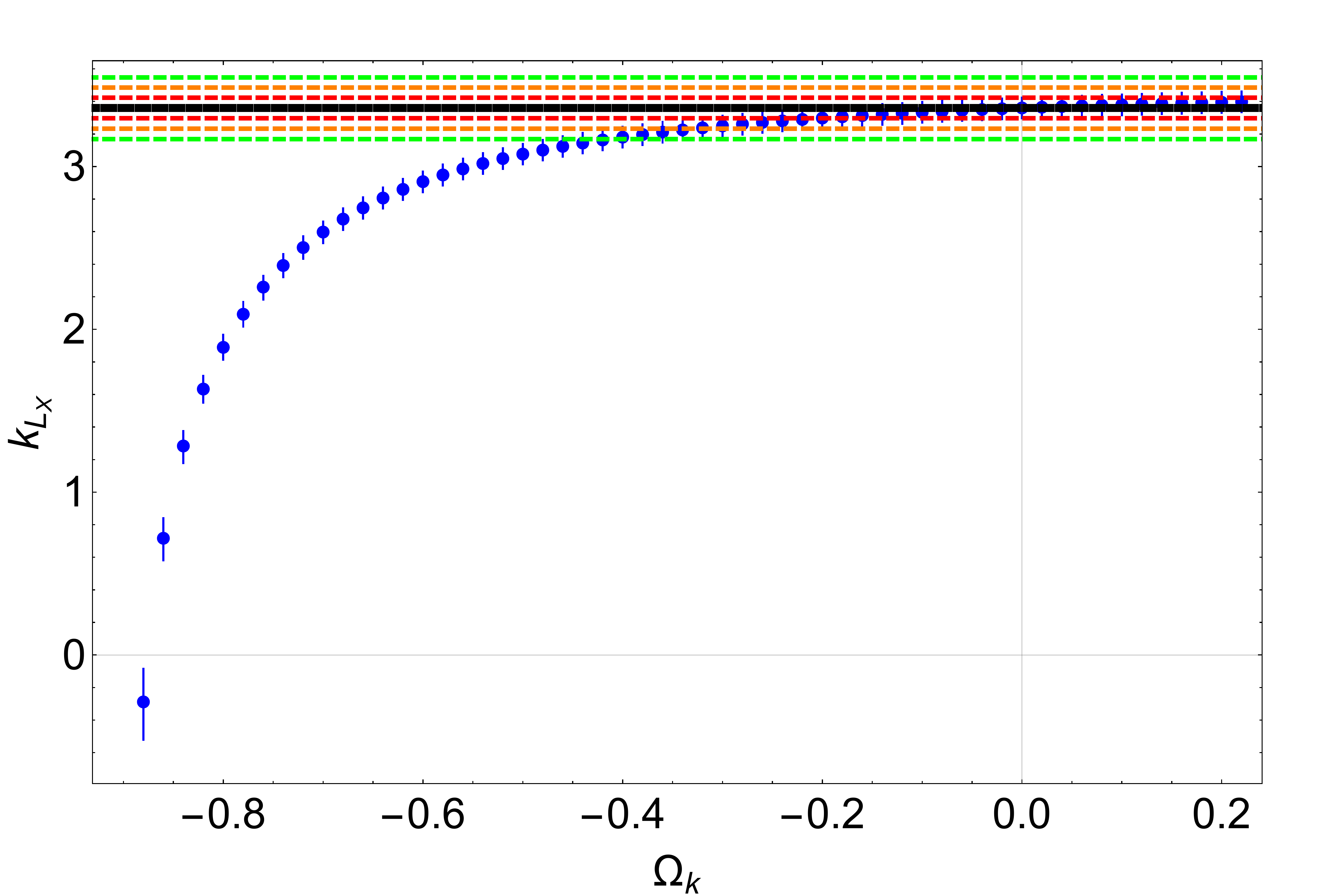}{0.45\textwidth}{(b) Behaviour of $k$ parameter for the correction of $L_{X}$ in relation to $\Omega_{k}$ for a non-flat $\Lambda$CDM model with $\Omega_M=0.3$.}}
\gridline{
\fig{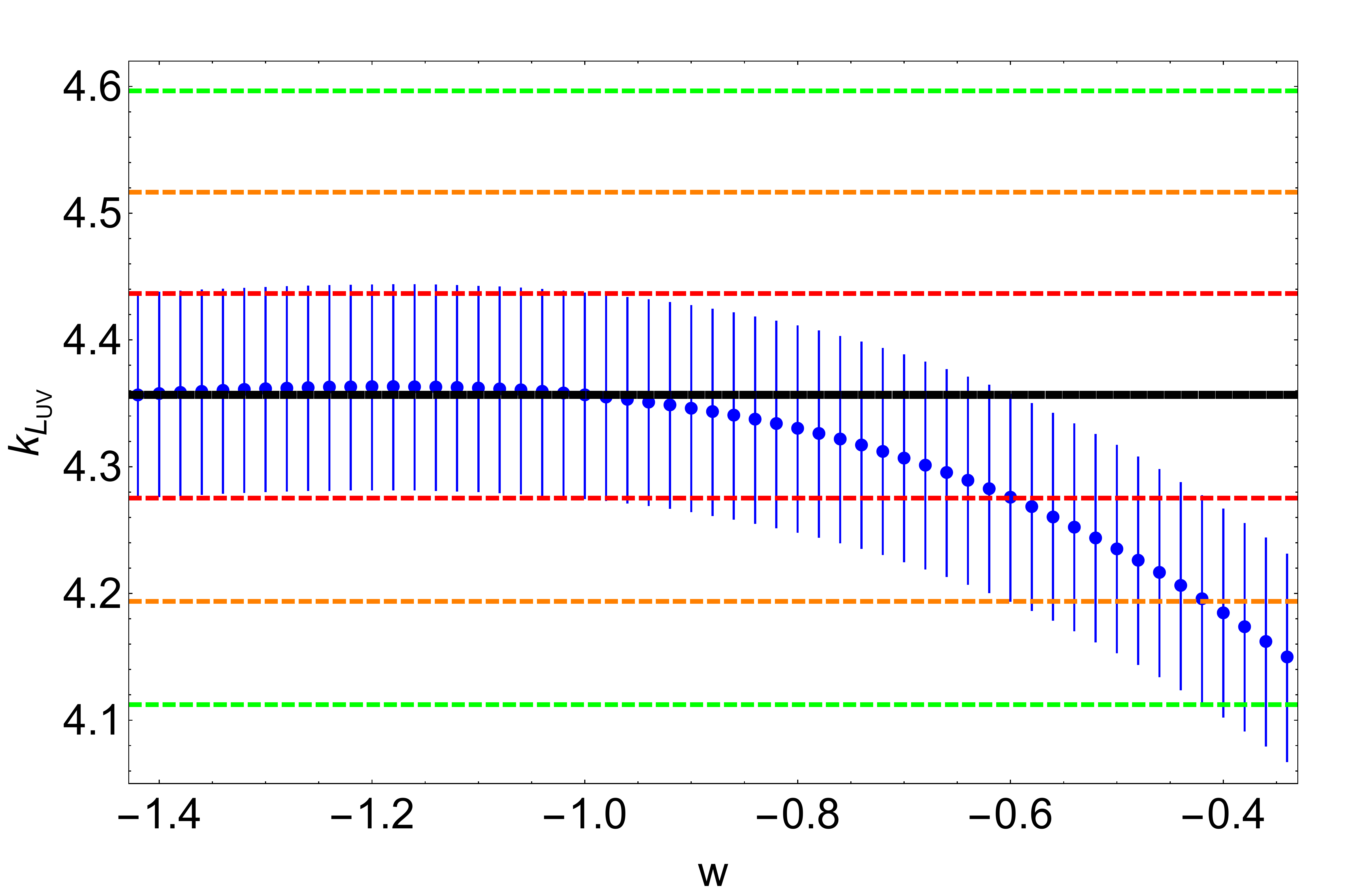}{0.45\textwidth}{(c) Behaviour of $k$ parameter for $L_{UV}$ in relation to $w$ for a flat $w$CDM model with $\Omega_M=0.3$.}
\fig{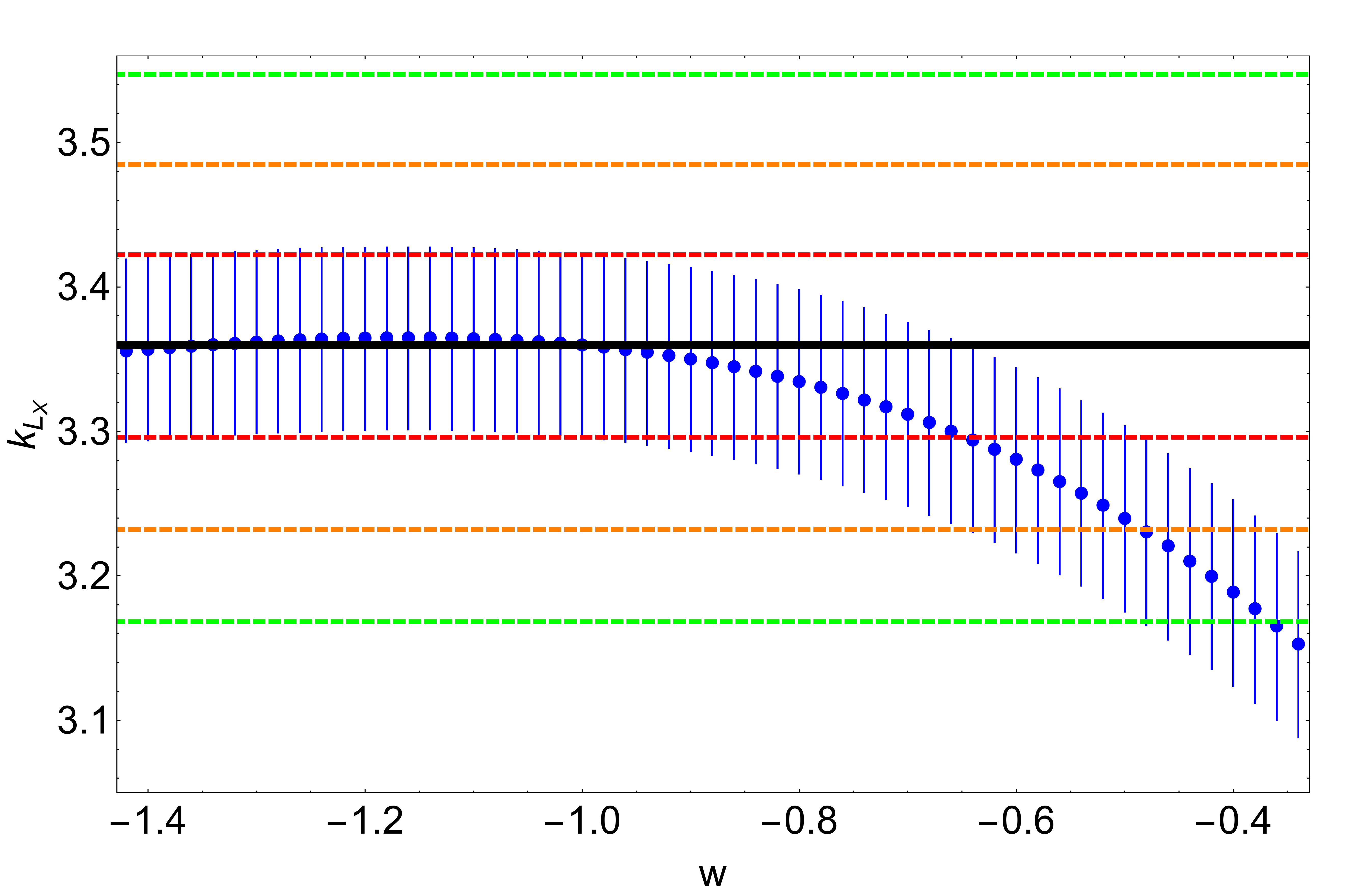}{0.45\textwidth}{(d) Behaviour of $k$ parameter for $L_{X}$ in relation to $w$ for a flat $w$CDM model with $\Omega_M=0.3$.}}
    \caption{ We mark with a thick, black line the values of $k_{L_{X}}$ and $k_{L_{UV}}$ computed for the $\Omega_{M}=0.3$, $\Omega_{k}=0$ and $w=-1$ together with 1, 2, and 3 $\sigma$ error bars marked with red, orange, and green dashed lines, respectively.}
    \label{fig:EPOk}
\end{figure}


\section{Results}
\label{results}

In this section, we present and comment on the results of our analyses. Section \ref{kvscosmology} is focused on how the correction for evolution in redshift of luminosities is impacted by cosmological parameters (see Figures \ref{fig:EPvsOmOk} and \ref{fig:EPOk}). Sections \ref{flatLCDM}, \ref{nonflatLCDM}, and \ref{flatwCDM} show the cosmological constraints obtained in all the three models studied in this work considering all the data sets and approaches used.
Tables \ref{tab:1}, and \ref{tab:3} report for each case studied the mean values of free cosmological parameters with their corresponding $1 \, \sigma$ error. Figures \ref{cornerplot1}, \ref{cornerplot1.1}, \ref{cornerplot2}, and \ref{cornerplot3} show the corresponding corner plots obtained in the cosmological computations. 

\subsection{Impact of cosmology on correction for evolution}
\label{kvscosmology}

All the cases considered in this work present a smooth dependence of $k$ on cosmological parameters (i.e. there are no bumps or discontinuities in computations) for both $k_{L_{X}}$ and $k_{L_{UV}}$. The upper row of Figure \ref{fig:EPOk} shows how $k$ depends on the parameter determining the curvature of the Universe $\Omega_{k}$ with fixed $\Omega_{M}=0.3$ for both UV (left panel) and X-ray (right panel) cases. In both panels, the computed dependence has a similar trend, with values of $k$ that do not vary much for values of $\Omega_{k}$ close to $\Omega_k=0$ (the ones expected from the most recent observations and studies).
For a matter of a pure more theoretical discussion on how close these solutions are to the non-physical regions, also showing peculiar behaviour in terms of the evolution, we can observe the following features:
at the negative values far from $\Omega_k=0$ a rapid decrease of $k$'s values is present and $k$ becomes incompatible within $3 \, \sigma$ with the value computed for $\Omega_{k} = 0$ at $\Omega_{k} \sim -0.7$ for both quantities considered.
In all the figures we mark with a thick, black line the value of $k_{L_{X}}$ and $k_{L_{UV}}$ computed for $\Omega_{M}=0.3$, $\Omega_{k}=0$, $w=-1$ together with 1, 2, and 3 $\sigma$ error bars marked with red, orange, and green dashed lines, respectively. We computed also values of the $k$ parameters for values of $\Omega_{k}$ close to $1$ and we observe that these values do not change significantly from the highest value shown on the plot ($\Omega_{k}=0.2$), thus since we do not exceed this region in our computations we show on the plot only a range of $\Omega_{k}$ between $-0.9$ and $0.2$. The upper row of Figure \ref{fig:EPvsOmOk} shows instead the results of a more general approach, in which we compute the value of $k$ over a grid of values of both $\Omega_{M}$ and $\Omega_{k}$, as described in Section \ref{evolution}. The variation of both $k_{L_{X}}$ and $k_{L_{UV}}$ with these cosmological parameters is very similar. Unsurprisingly, the most significant evolution of the luminosities appears in the region close to the point $\{\Omega_M=0,\Omega_k=0\}$, because in such a Universe $E(\zeta)$ would become $E(\zeta)=1$, and it would result in a quadratic function of the distance luminosity with redshift, which would lead to still quite dispersed but a power of four relation of the luminosity with the redshift. Values of $k$ close to the restricted region marked with the black dashed line rapidly decrease with the distance from the origin of axes. We can conclude that, in reasonable regions of this parameter space (i.e. near $\{\Omega_M=0.3, \Omega_k=0\}$), the values of $k$ are compatible with the ones obtained for $\Omega_{k}=0$, $\Omega_{M}=0.3$. As visible from the plots, we discard a part of the $\{\Omega_{M},\, \Omega_{k}\}$ space bigger than the region defined by Equation \ref{nobigbang}. Indeed, the rapid decrease of $k$'s values close to this region affects the precision of the interpolation method that we use to compute the function $k (\Omega_M,\Omega_k)$ that we then use for our cosmological computations. Nevertheless, it has also to be pointed out that this discarded region does not impact our analysis, as our results do not fall in this cut-off region in any of the cases.

Results of the investigation of the dependence of $k$ on $w$ parameter fixing $\Omega_{M}=0.3$ are shown in the bottom row of Figure \ref{fig:EPOk} for both UV (left panel) and X-ray cases (right panel). Within the whole range of $w$ values explored $k$'s values are compatible with the value obtained for $w=-1$ within less than $2\, \sigma$. Thus, we do not expect a significant difference in cosmological results between the computation of $w$ with fixed correction for evolution and the one with $k=k(w)$. If we look at the more general investigation in the parameter space of both $\Omega_{M}$ and $w$ presented in the bottom row of Figure \ref{fig:EPvsOmOk}, we see that, as in the previous case, the variation of both $k_{L_{X}}$ and $k_{L_{UV}}$ with the cosmological parameters presents very similar features, and in reasonable regions of this parameter space $\{\Omega_M, w\}$ (i.e. near $\{\Omega_M=0.3, w = -1\}$), the values of $k$ are compatible with the ones obtained for $\Omega_{M}=0.3$ and $w=-1$, while $k$'s values vary more significantly in exotic regions of this space. Due to high dispersion of the RL relation our analyses still cover a large range of the cosmological parameters. Thus, the change of evolutionary parameter, $k$, with cosmology could still be significant in some computations. For example, when using $k=k(\Omega_{M},\,w)$ for the joint sample of SNe Ia and QSOs, the $3 \, \sigma$ contours in the  $\{\Omega_M, w \}$ space extend from $\Omega_{M}=0.2$ and $w=-0.8$ to $\Omega_{M}=0.5$ and $w=-1.6$ (see plot (i) in Figure \ref{cornerplot3}). In this case we find that the values of $k_{L_{UV}}$ vary from $4.13 \pm 0.08$ (for $\Omega_{M}=0.5$ and $w=-1.6$) up to $4.47 \pm 0.08$ (for $\Omega_{M}=0.2$ and $w=-0.8$). Thus, the values of $k$ result to be compatible with each other within $3\, \sigma$, while $k_{L_{X}}$ varies from $3.13 \pm 0.06$ (for $\Omega_{M}=0.5$ and $w=-1.6$) up to $3.48 \pm 0.06$ (for $\Omega_{M}=0.2$ and $w=-0.8$), being compatible with each other within $4\, \sigma$. The explored range of $k$'s values in the MCMC fitting becomes even wider when we consider the case of QSOs alone, both calibrated and non-calibrated. Thus, the variation of $k$ with cosmology is often not negligible.

In the approach presented in this manuscript, we also compute in all considered cases the error on $k$'s value as a function of cosmology. In reasonable ranges, we always observe very small variations of the error with cosmology.
As visible in the upper row of Figure \ref{fig:EPOk}, the error on $k$ for $\Omega_{k}<-0.8$ is about twice the size of the one for $\Omega_{k}=0$, but for values of $\Omega_{k}>-0.55$ the errors are less than $8\%$ bigger than the one for $\Omega_{k}=0$ for both $k_{L_{X}}(\Omega_{k})$ and $k_{L_{UV}}(\Omega_{k})$. Concerning the error $\Delta_{k}(w)$, we do not observe variations of the errors
bigger than $4\%$ in whole considered region for both $k_{L_{X}}(w)$ and $k_{L_{UV}}(w)$ (see bottom row of Figure \ref{fig:EPOk}). For $\Delta_{k}(\Omega_{M}, \Omega_{k})$, we always observe variations smaller than $7\%$ for every computed point for both $k_{L_{X}}(\Omega_{M}\,\Omega_{k})$ and $k_{L_{UV}}(\Omega_{M},\Omega_{k})$. The same result is obtained for the errors $\Delta_{k}(\Omega_{M}, w)$, for both $k_{L_{X}}(\Omega_{M},w)$ and $k_{L_{UV}}(\Omega_{M},w)$. Thus, in future analyses, the evolution of the errors on the $k$ parameter with cosmology could be possibly neglected.

\subsection{Results on flat $\Lambda$CDM model}
\label{flatLCDM}

\subsubsection{Calibrated QSOs alone}
Assuming a flat $\Lambda$CDM model, the case with QSOs alone calibrated with SNe Ia and without any correction for evolution favours high values of $\Omega_M$, even if compatible in $3 \, \sigma$ with $\Omega_{M} = 0.3$, when $H_0$ is fixed, and high values of $H_0$ when $\Omega_M$ is fixed, while, when both parameters are free, we do not obtain convergence on the $\Omega_M$ parameter which tends to have very high values even outside our uniform prior bound. Including a fixed correction for the evolution instead, $\Omega_M$ always goes to lower values, both for fixed and free $H_0$ ($\Omega_M= 0.251 \pm 0.054$ and $\Omega_M=0.167 \pm 0.062$, respectively), while $H_0$ ranges from 67.55 to 72.09 $ \mathrm{km} \, \mathrm{s^{-1}} \,\mathrm{Mpc^{-1}}$ in $1 \, \sigma$ when $\Omega_M$ is fixed, corresponding to intermediate values between the one measured from SNe Ia ($H_0 = 74.03 \, \pm 1.42  \mathrm{km} \, \mathrm{s^{-1}} \,\mathrm{Mpc^{-1}}$) and the one obtained from the Planck data on the CMB under the assumption of the same cosmological model ($H_0 = 67.4 \pm 0.5 \, \mathrm{km} \, \mathrm{s^{-1}} \,\mathrm{Mpc^{-1}}$). The case with both parameters free does not converge for any of the two parameters leading to too low $\Omega_M$ values ($ \Omega_M = 0.167 \pm 0.062$), compatible within $3 \, \sigma$ with the theoretical De Sitter Universe in which $\Omega_M=0$, and too high $H_0$ values ($H_0 = 76.40 \pm 3.33  \, \mathrm{km} \, \mathrm{s^{-1}} \,\mathrm{Mpc^{-1}}$).
Taking into account the variation of the correction for evolution together with the cosmological parameter $\Omega_M$, $\Omega_M$ becomes consistent with $\Omega_M = 0.3$, when $H_0$ is fixed, also due to large uncertainties, but $\Omega_M$ does not converge when also $H_0$ is free to vary, going to the upper limit of our uniform prior. In this case, $H_0$ is compatible in $1 \, \sigma$ with the value from the CMB.

\begin{table}[b!]
    \caption{Cosmological results obtained for all considered cases.}
    \label{tab:1}
    \centering
    
    \begin{tabular}{c|c|c|c||c|c|c|c}
        \toprule[1.2pt]
        \multicolumn{4}{|c|}{Only QSOs calibrated on SNe Ia.}&\multicolumn{4}{|c|}{Combination of SNe Ia with QSOs without calibration.}\\
        \toprule[1.2pt]
        \multicolumn{8}{|c|}{Results without correction for evolution}\\
        \toprule[1.2pt]
        $\Omega_{M}$ & $H_{0}$ & $w$ & $\Omega_{k}$ & $\Omega_{M}$ & $H_{0}$ & $w$ & $\Omega_{k}$\\\hline 
        $ 0.443 \pm 0.054 $ & \bf{70} & \bf{-1} & \bf{0}& $ 0.305 \pm 0.008 $ & \bf{70} & \bf{-1} & \bf{0}\\\hline
        \bf{0.3} & $ 73.76 \pm 2.18 $ & \bf{-1} & \bf{0}& \bf{0.3} & $ 69.97 \pm 0.14 $ & \bf{-1} & \bf{0}\\\hline
        $ 0.663 \pm 0.108 $ & $ 62.77 \pm 2.52 $ & \bf{-1} & \bf{0}& $ 0.338 \pm 0.022 $ & $ 69.44 \pm 0.32 $ & \bf{-1} & \bf{0}\\\hline
        \bf{0.3} & \bf{70} & $ -0.696 \pm 0.132 $ & \bf{0}& \bf{0.3} & \bf{70} & $ -1.003 \pm 0.019 $ & \bf{0}\\\hline
        \bf{0.3} & \bf{70} & \bf{-1} & $ -0.531 \pm 0.275 $& \bf{0.3} & \bf{70} & \bf{-1} & $ -0.01 \pm 0.018 $\\\hline
        $ 0.522 \pm 0.052 $ & \bf{70} & $ -1.966 \pm 0.395 $ & \bf{0}& $ 0.462 \pm 0.034 $ & \bf{70} & $ -1.552 \pm 0.166 $ & \bf{0}\\\hline
        $ 0.517 \pm 0.06 $ & \bf{70} & \bf{-1} & $ -0.740 \pm 0.106 $& $ 0.656 \pm 0.048 $ & \bf{70} & \bf{-1} & $ -0.762 \pm 0.101 $\\\hline
        \multicolumn{8}{|c|}{Results with fixed correction for evolution}\\
        \toprule[1.2pt]
        $\Omega_{M}$ & $H_{0}$ & $w$ & $\Omega_{k}$ & $\Omega_{M}$ & $H_{0}$ & $w$ & $\Omega_{k}$ \\\hline 
        $ 0.251 \pm 0.040 $ & \bf{70} & \bf{-1} & \bf{0}& $ 0.291 \pm 0.008 $ & \bf{70} & \bf{-1} & \bf{0}\\\hline
        \bf{0.3} & $ 69.82 \pm 2.27 $ & \bf{-1} & \bf{0}& \bf{0.3} & $ 69.99 \pm 0.14 $ & \bf{-1} & \bf{0}\\\hline
        $ 0.167 \pm 0.062 $ & $ 76.40 \pm 3.33 $ & \bf{-1} & \bf{0}& $ 0.251 \pm 0.02 $ & $ 70.72 \pm 0.34 $ & \bf{-1} & \bf{0}\\\hline
        \bf{0.3} & \bf{70} & $ -1.113 \pm 0.182 $ & \bf{0}& \bf{0.3} & \bf{70} & $ -1.006 \pm 0.018 $ & \bf{0}\\\hline
        \bf{0.3} & \bf{70} & \bf{-1} & $ 0.01 \pm 0.122 $& \bf{0.3} & \bf{70} & \bf{-1} & $ 0. \pm 0.018 $\\\hline
        $ 0.151 \pm 0.099 $ & \bf{70} & $ -0.765 \pm 0.267 $ & \bf{0}& $ 0.059 \pm 0.038 $ & \bf{70} & $ -0.662 \pm 0.041 $ & \bf{0}\\\hline
        $ 0.24 \pm 0.05 $ & \bf{70} & \bf{-1} & $ 0.076 \pm 0.104 $& $ 0.118 \pm 0.032 $ & \bf{70} & \bf{-1} & $ 0.377 \pm 0.069 $ \\\hline
        \multicolumn{8}{|c|}{Results with correction for evolution as function of cosmology}\\
        \toprule[1.2pt]        
        $\Omega_{M}$ & $H_{0}$ & $w$ & $\Omega_{k}$ & $\Omega_{M}$ & $H_{0}$ & $w$ & $\Omega_{k}$ \\\hline 
        $0.382 \pm 0.178$ & \bf{70} & \bf{-1} & \bf{0}& $ 0.299 \pm 0.008 $ & \bf{70} & \bf{-1} & \bf{0}\\\hline
        $0.543 \pm 0.251$  & $67.50 \pm 2.80$ & \bf{-1} & \bf{0}& $ 0.3 \pm 0.022 $ & $ 69.99 \pm 0.35 $ & \bf{-1} & \bf{0}\\\hline
        \bf{0.3} & \bf{70} & $ -1.074 \pm 0.205 $ & \bf{0}& \bf{0.3} & \bf{70} & $ -1.005 \pm 0.019 $ & \bf{0}\\\hline
        \bf{0.3} & \bf{70} & \bf{-1} & $ -0.041 \pm 0.102 $& \bf{0.3} & \bf{70} & \bf{-1} & $ -0.006 \pm 0.018 $\\\hline
        $ 0.546 \pm 0.172 $ & \bf{70} & $ -1.591 \pm 0.500 $ & \bf{0}& $0.358 \pm 0.054$ & \bf{70} & $ -1.168 \pm 0.156 $ & \bf{0}\\\hline
        $ 0.591 \pm 0.217 $ & \bf{70} & \bf{-1} & $ -0.235 \pm 0.191$& $0.415 \pm 0.063$ & \bf{70} & \bf{-1} & $ -0.250 \pm 0.134 $\\\hline
    \end{tabular}
    \tablecomments{In the upper part of the table, we show the results without the correction for evolution, in the middle part the results obtained with fixed evolution, and at the bottom the results obtained with correction for evolution as a function of cosmology. Errors reported in the table correspond to $1 \, \sigma$ uncertainties. Bold values represent cosmological parameters that are fixed in the analysis.}
\end{table}

\subsubsection{Non-calibrated QSOs together with SNe Ia}

When combining non-calibrated QSOs with SNe Ia we do not deal with convergence issues in any of the cases. Specifically, without accounting for the correction for evolution, we find $\Omega_M$ close to 0.3 (in $1 \, \sigma$ and $2 \, \sigma$ when $H_0$ is fixed and free, respectively) and $H_0$ close to 70 (in $1 \, \sigma$ and $2 \, \sigma$ when $\Omega_M$ is fixed and free, respectively). When accounting for a fixed correction for evolution, we still have  $\Omega_M$ close to 0.3 even if in $2 \, \sigma$ and $3 \, \sigma$ when $H_0$ is fixed and free, respectively, and $H_0$ close to 70, in $1 \, \sigma$ and $3 \, \sigma$ when $\Omega_M$ is fixed and free, respectively. Letting the evolution vary together with the cosmological parameter (i.e. $k(\Omega_m)$), we obtain $\Omega_M$ and $H_0$ always compatible within $1 \, \sigma$  with 0.3 and 70, respectively. Indeed, we here remind that $k$ does not depend on $H_0$, as already stressed, so we do not have to study again the case in which only $H_0$ is free. 

Since in these cases we are not calibrating QSOs with SNe Ia, we also fit $g$, $b$, and $sv$ contemporaneously with the cosmological parameters. Without evolution, we always obtain $g=0.66 \pm 0.01$ independently of the number of free parameters, $b$ always compatible within $1 \, \sigma$ with a  central value of $b = 6.33$, and $sv=0.230 \pm 0.004$ in all cases. When we account for the correction for evolution, both fixed or as a function of cosmological parameters, we obtain $g=0.59 \pm 0.01$, $b$ always compatible within $1 \, \sigma$ with a central value of $b = 8.24$, and $sv= 0.225 \pm 0.003$ in all cases. We can note that the correction for evolution reduces the intrinsic dispersion by 2.2 \% making the RL relation tighter. 

\subsubsection{Non-calibrated QSOs alone}

Only in this case of a flat $\Lambda$CDM model, we also consider the data set composed of QSOs alone non calibrated on SNe Ia to investigate how this sample can impact the determination of the $\Omega_M$ parameter in the cosmological analysis. These results are shown in Table \ref{tab:3} and Figure \ref{cornerplot1.1}, where $H_0$ is fixed to $H_{0}=70\, \mathrm{km} \, \mathrm{s^{-1}} \,\mathrm{Mpc^{-1}}$. If we do not consider the correction for the evolution, $\Omega_M$ is not constrained and hits the upper limit $\Omega_M=1$ in the range of uniform priors. Once accounting for a fixed correction for the evolution instead, we obtain closed contours on this parameter, but with values shifted to very small values $\Omega_M = 0.067 \pm 0.017$, although not compatible with $\Omega_M=0$ within 3$\, \sigma$. Unsurprisingly, the case with the correction $k=k(\Omega_M)$ allows us to overcome the convergence issue and the non-physical behaviour with $\Omega_M$ close to 0, leading to $\Omega_M = 0.500 \pm 0.201$, which is compatible within $1 \, \sigma$ with $\Omega_M = 0.3$. Concerning the values of the parameters of the RL relation, we obtain consistency with the ones from non calibrated QSOs combined with SNe Ia, as well as a similar reduction in the intrinsic dispersion when accounting for the correction for the evolution. 
Concerning this, we have also tried to compute other cosmological parameters such as $H_{0}$, $\Omega_{k}$, and $w$ for the sample composed of QSOs alone without calibration, but we have not obtained closed contours in any of these cases, thus we do not show these results here. This can be ascribed to the fact that the RL relation, even after the correction for evolution, still retains an intrinsic scatter, thus non-calibrated QSOs alone are too weak to constrain parameters of more complex models. Our results show a path, that further investigations of the properties of this probe could lead to tighter, more reliable results.

\subsection{Results on non-flat $\Lambda$CDM model}
\label{nonflatLCDM}

\subsubsection{Calibrated QSOs alone}

When allowing for $\Omega_k \neq 0$, QSOs alone calibrated with SNe Ia point toward negative values of $\Omega_k$, if we do not correct for luminosity evolution in redshift. More precisely, $\Omega_k = -0.531 \pm 0.275$ when $\Omega_M$ is fixed, while when both are free parameters of the fit $\Omega_M$ tends to 0.52, but $\Omega_k$ does not converge due to too low values in our range of uniform prior. 
Very interestingly, when we correct for a fixed evolution the convergence is reached for both $\Omega_k$ and $\Omega_M$ free to vary, however with large uncertainties, and $\Omega_M$ shifts toward 0.3 and $\Omega_k$ toward 0. Instead, when $\Omega_k$ is the only free parameter of the fit it is not constrained, as happens also accounting for a correction that varies with $\Omega_k$. This can be ascribed to the fact that QSOs are more sensitive to the value of $\Omega_M$, which is better determined by probes at intermediate redshifts, rather than the value of $\Omega_k$, which is instead constrained at very high redshifts as the one of CMB; thus when also $\Omega_M$ is free QSOs can constrain $\Omega_M$ and, as a consequence of their degeneracy, also $\Omega_k$.
Varying both parameters and taking into account a correction for evolution that varies together with these parameters, does not constrain $\Omega_M$ that tends to have values close to $\Omega_M=1$.

\begin{table}[t!]
\caption{Cosmological results obtained with only the QSO sample without calibration.}
    \centering
    \begin{tabular}{c|c}
    \toprule[1.2pt]
        Results obtained with QSOs alone without calibration & $\Omega_{M}$ \\
        \toprule[1.2pt]
        Results without correction for evolution & $0.934 \pm 0.059$ \\\hline
        Results with fixed correction for evolution & $0.067 \pm 0.017$ \\\hline
        Results with correction for evolution as a function of $\Omega_{M}$ & $0.500 \pm 0.210$ \\\hline
    \end{tabular}
    \label{tab:3}
    \tablecomments{We show the results without the correction for evolution, with fixed correction for evolution, and with correction for evolution as a function of $\Omega_{M}$. Errors reported in the table correspond to $1 \, \sigma$ uncertainties. In this computation we fix the parameters as: $H_{0}=70\, \mathrm{km} \, \mathrm{s^{-1}} \,\mathrm{Mpc^{-1}}$, $\Omega_{k}=0$, $w=-1$.}
\end{table}

\subsubsection{Non-calibrated QSOs together with SNe Ia}

In the case of non-calibrated QSOs combined with SNe Ia, $\Omega_k$ is always compatible with $\Omega_k = 0$ within $1 \, \sigma$ when $\Omega_M$ is fixed to 0.3 (as expected for a flat Universe), both with and without the correction for the evolution, while, when also $\Omega_M$ is free, it tends to negative values (i.e. $\Omega_k = -0.762 \pm 0.101$) if we do not correct for evolution, and to positive values (i.e. $\Omega_k = 0.377 \pm 0.069$) when considering a fixed correction. In these two cases, $\Omega_M$ is shifted to high and low values, respectively. When accounting for the correction with $k(\Omega_{M},\,\Omega_{k})$, we observe an intermediate behaviour: $\Omega_M = 0.415 \pm 0.063$, within $2 \, \sigma$ from $\Omega_M=0.3$, and $\Omega_k=-0.250 \pm 0.134$, within $3 \, \sigma$ from a zero curvature. 

Concerning the free parameters of the RL relation, the results are the same as the ones obtained for the flat $\Lambda$CDM model, showing that the values of these parameters do not depend on the curvature of the Universe. In addition, comparing the values of $g$ and $b$ obtained when only $\Omega_k$ is the free parameter and when both $\Omega_M$ and $\Omega_k$ are free parameters, we note that there are slight discrepancies ($\sim 2 \, \sigma$) in the case without evolution, that are completely removed once the evolution is taken into account. This shows that the correction for the luminosity evolution makes the parameters of the RL relation compatible within $1 \, \sigma$ comparing the cases with different numbers of free parameters.
Many computations involving QSOs through the literature concerning the value of $\Omega_{k}$ show traces of its value being smaller than 0 \citep{2022MNRAS.510.2753K}. Our results could indicate that this behaviour is due to selection bias and redshift evolution since after this correction treated as a function of cosmology is applied, the obtained values of $\Omega_{k}$ become more compatible with 0. For the computations with non-calibrated QSOs together with SNe for the case without correction for evolution varying only $\Omega_{k}$, we obtain $\Omega_{k} = -0.010\pm 0.018$, while in the analogous case with applied correction for evolution as a function of cosmology $\Omega_{k} = -0.006\pm 0.018$. For the computations with non-calibrated QSOs together with SNe for the case without correction for evolution varying $\Omega_{k}$ together with $\Omega_{M}$, we obtain $\Omega_{k} = -0.010\pm 0.018$ and $\Omega_{M} = 0.656\pm 0.048$, while in the analogous case with applied correction for evolution as a function of cosmology $\Omega_{k} = -0.250\pm 0.134$ and $\Omega_{M} = 0.415\pm 0.063$. Very similar behaviour can be seen in the cases with calibrated QSOs. We see here that the results after the correction for evolution as a function of cosmology is applied, are much closer to the flat $\Lambda$CDM model parameters obtained with SNe alone than the ones without this correction. This could indicate that the previous results in the literature showing value of $\Omega_{k}$ being incompatible with 0 were possibly driven simply by the selection bias and redshift evolution. In order to drive clear conclusions on the flatness of the universe more work still needs to be done. This issue could be solved in future with a larger sample of QSOs and the addition of more high-redshift probes like GRBs.

\subsection{Results on flat $w$CDM model}
\label{flatwCDM}

Considering the flat $w$CDM model, we need to distinguish between two different DE regimes in the values of the equation of state parameter $w$. Indeed, the regime $w>-1$ is referred to as a ``quintessence'', while the one with $w<-1$ as ``phantom''. The phantom DE scenario predicts a final ``Big Rip'' for the Universe in which all the matter is ripped apart by the accelerated expansion. In relation to a region of the viability of the null energy condition given by equation \ref{wcond} none of our computations shows any trace of violation of this condition.

\subsubsection{Calibrated QSOs alone}

If we consider QSOs alone calibrated with SNe Ia and we do not take into account any correction for evolution, we obtain very different results concerning the DE scenario, depending on if we fixed or not the parameter $\Omega_M$. In particular, if $\Omega_M=0.3$ we obtain $w \sim -0.7$ corresponding to the quintessence behaviour (even if compatible within $3 \, \sigma$ with $w=-1$), while, if $\Omega_M$ is a free parameter of the fit, we run into the phantom DE region with $w \sim -2$ and $\Omega_M \sim 0.5$, even if in this case $w$ does not converge going to too negative values compared to our uniform prior on this parameter. On the other hand, we observe the opposite trend when accounting for a fixed correction for evolution. In this case, indeed, the phantom regime is obtained with $\Omega_M$ fixed (but at only $1 \, \sigma$ from a quintessence case with $w=-0.931$) and the quintessence regime (at $1 \, \sigma$ from $w=-1$) with $\Omega_M$ as a free parameter, even if in this latter case $\Omega_M$ does not converge hitting the barrier at $\Omega_M=0$.
When considering $k=k(w)$ with $\Omega_M$ fixed, $w$ results compatible with both scenarios in $1 \, \sigma$.
When $\Omega_M$ is free to vary instead both parameters are unconstrained. 

\subsubsection{Non-calibrated QSOs together with SNe Ia}

When fitting jointly non-calibrated QSOs and SNe Ia, we obtain that $w$ is always consistent within $1 \, \sigma$ with $w=-1$ if $\Omega_M$ is fixed to 0.3 (as expected in a flat $\Lambda$CDM model), independently of the treatment of the correction for evolution. Instead, when varying also $\Omega_{M}$, we obtain $w \sim -1.6$ and $\Omega_{M} \sim 0.46$ without correction, $w \sim -0.7$ and $\Omega_{M} \sim 0.06$ with a fixed correction, even if in this last case $\Omega_M$ does not converge, and $w \sim -1.2$ and $\Omega_{M} \sim 0.36$ with a correction that varies with the cosmological parameters. This last case is compatible with $\Omega_{M} = 0.3$ and $w=-1$ within $2 \, \sigma$.
The results presented for the flat $\Lambda$CDM model on the values of $g$, $b$, and $sv$ are once again valid also in this case of a flat $w$CDM model. This strongly stresses the independence of the parameters of the RL relation on the assumed cosmological model.

\subsection{Impact on the $H_0$ tension}

We also would like to stress that if we consider non-calibrated QSOs + SNe both with $H_{0}$ and $\Omega_{M}$ changing together or only changing $H_{0}$, then we can safely state that the resulting $H_{0}$ values are all compatible with each other, and they are in the middle between the values of the $H_{0}$ determined by the Planck collaboration and the ones of the SNe Ia. This result is very intriguing since it is similar to the results obtained by \cite{Freedman2021}, see figures 10 and 11 in their paper. In \cite{Freedman2021}, the authors use a different methodology due to a local characteristic of the red giants. Their approach is following the traditional, cosmic-ladder treatment, which means that distant objects such as SNe Ia are calibrated by nearby standard candles such as Tip of the Red Giant Branch (TRGB). In this approach, $H_{0}$ is determined for nearby objects (i.e. redshift of the objects is much smaller than 1) for which the luminosity distance mainly depends on $H_{0}$ and z only and does not depend on other cosmological parameters such as $\Omega_{M}$, $\Omega_{\Lambda}$. In that sense, Freedman's approach is different from our new approach that overcomes the circularity problem where $H_{0}$ is obtained simultaneously with other cosmological parameters. Those results could imply that the Hubble constant tension could be due to observational biases, but one needs to have a larger sample to verify such a hypothesis \citep{Freedman2021}.
When we consider the QSOs alone but calibrated with SNe Ia we have mixed results about the $H_{0}$ values. If we consider that the evolutionary function for $\Omega_{M}$ provides the most reliable estimates, then we can conclude that values closer to the Planck results are the most favored. On the other hand, from Figure 11 we can still learn that when we vary only one cosmological parameter at time again the results are mixed. Indeed, if we consider only $H_{0}$ varying without evolution then we have compatibility for $H_{0}$ with SNe Ia, while when we fix the evolution, the $H_{0}$ value lies in the middle of the values obtained by CMB and SNe Ia. From this analysis, we still cannot conclude that evolution is the driving factor that pushes the $H_{0}$ closer to the Planck values. Indeed, for the case of both parameters free to vary the results with and without evolution lies on the opposite side of the Planck (no evolution) and SNe Ia values for $H_{0}$ (for fixed evolution). 
{\bf We also would like to stress that the $H_0$ trend toward the value of 70 $ \mathrm{km} \, \mathrm{s^{-1}} \,\mathrm{Mpc^{-1}}$ may be also induced by the fact that SNe Ia are uncalibrated (namely calibrated arbitrary for an absolute magnitude $M=-19.35$ and thus consequently at $H_0=70 \, \mathrm{km} \, \mathrm{s^{-1}} \,\mathrm{Mpc^{-1}}$, see also \citealt{2023arXiv230101024P}), thus the trend may not be induced by the underlying physics, but due to this calibration choice (private communication with Adam Riess). Notwithstanding the importance of this discussion, this topic is beyond the scope of the current paper, and we will address it more in detail in a forthcoming publication.}

\begin{figure}
    \centering
    
\gridline{
\fig{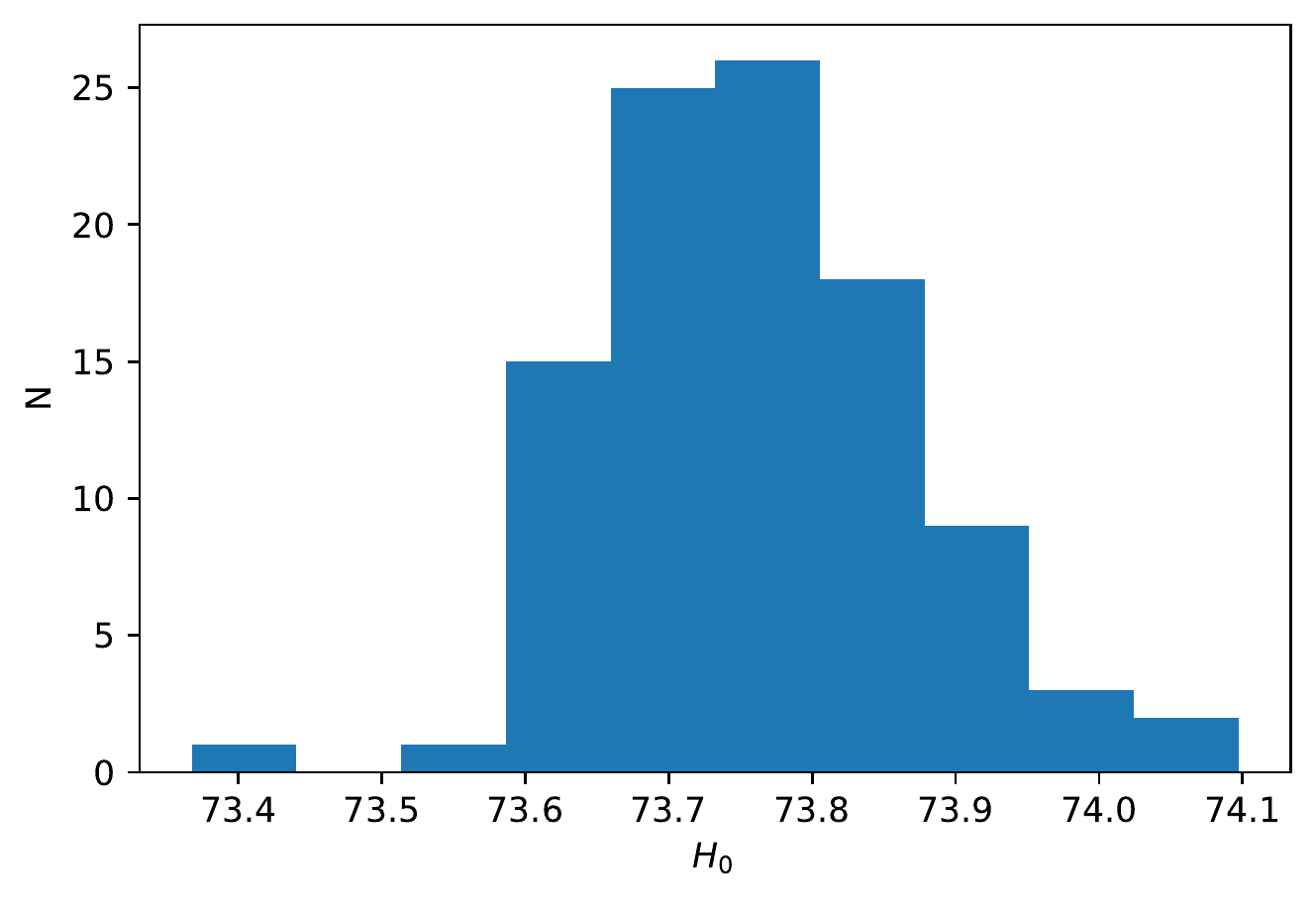}{0.45\textwidth}{(a) Only QSOs calibrated without evolution, varying only $H_{0}$}
\fig{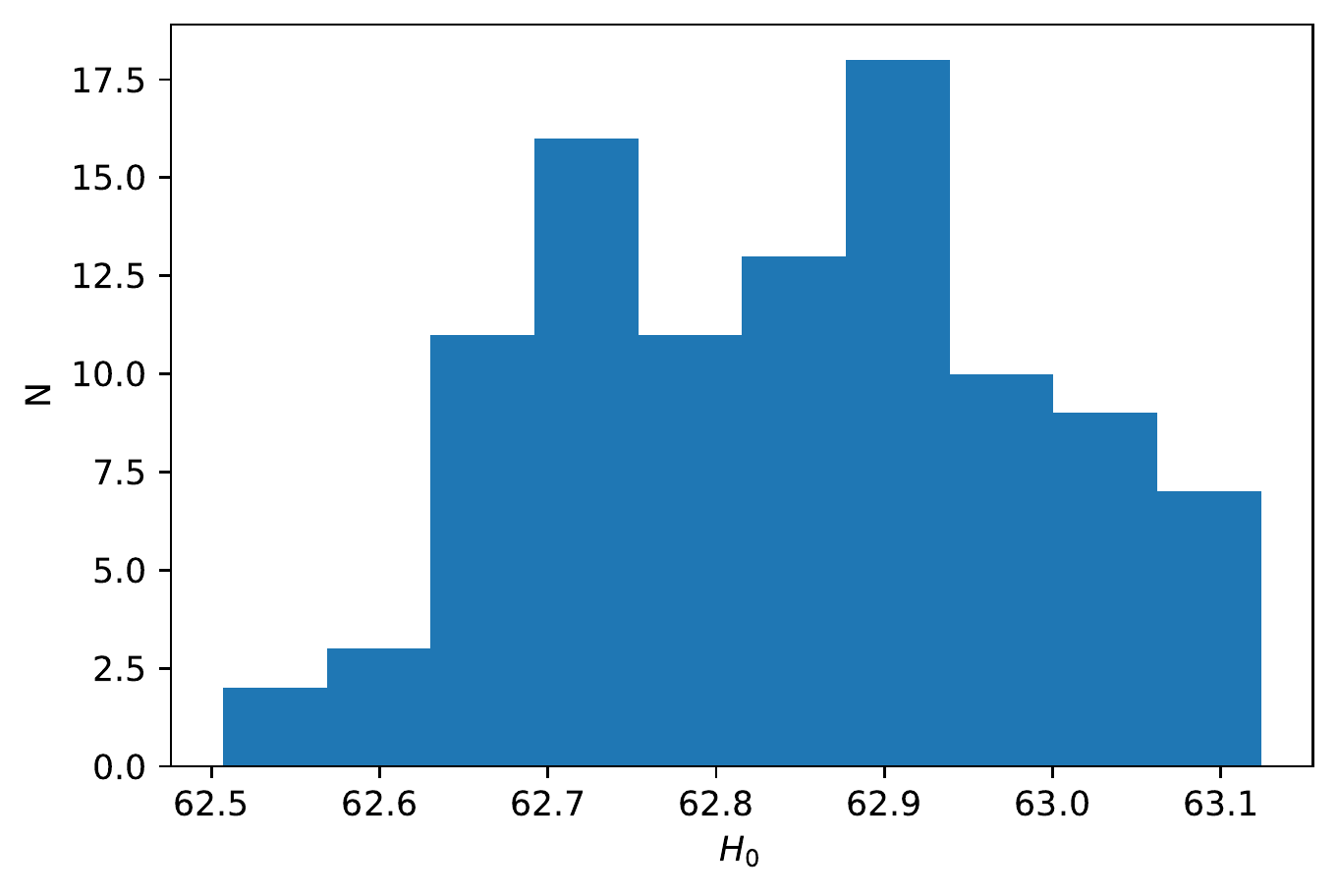}{0.45\textwidth}{(b) Only QSOs calibrated without evolution, varying $H_{0}$ together with $\Omega_{M}$}}
\gridline{
\fig{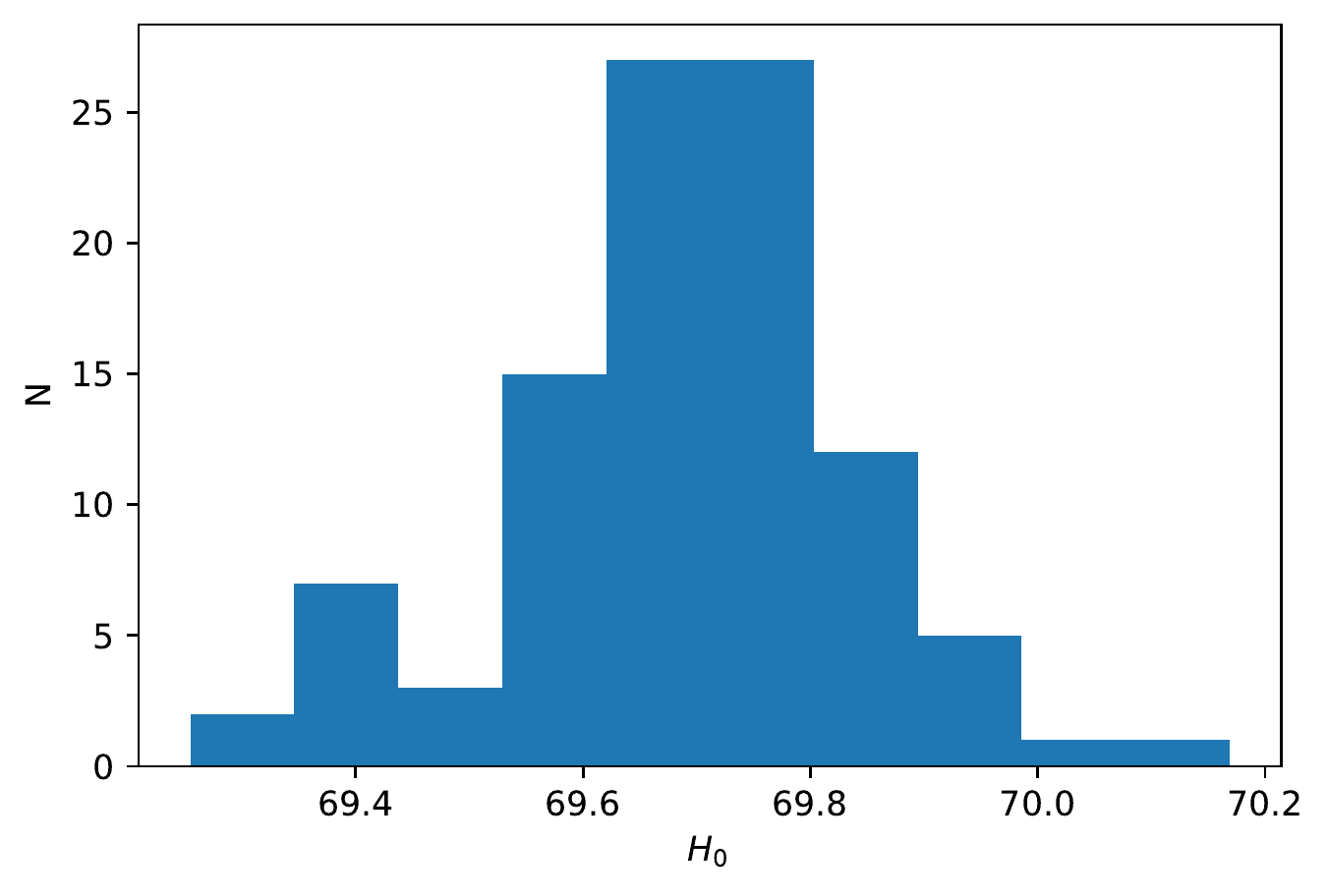}{0.45\textwidth}{(c) Only QSOs calibrated with fixed evolution, varying only $H_{0}$}
\fig{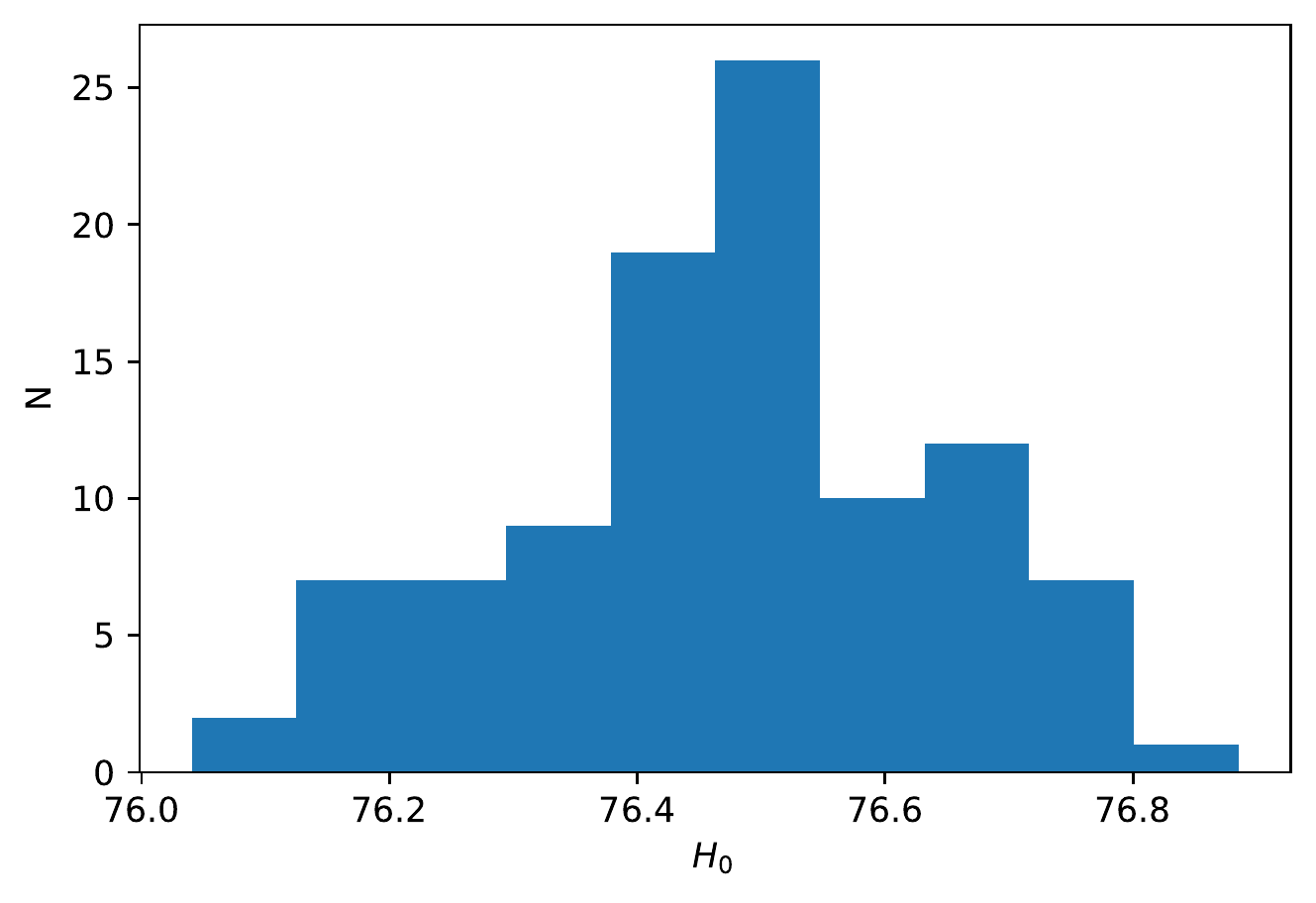}{0.45\textwidth}{(d) Only QSOs calibrated with fixed evolution, varying $H_{0}$ together with $\Omega_{M}$}}
\gridline{
\fig{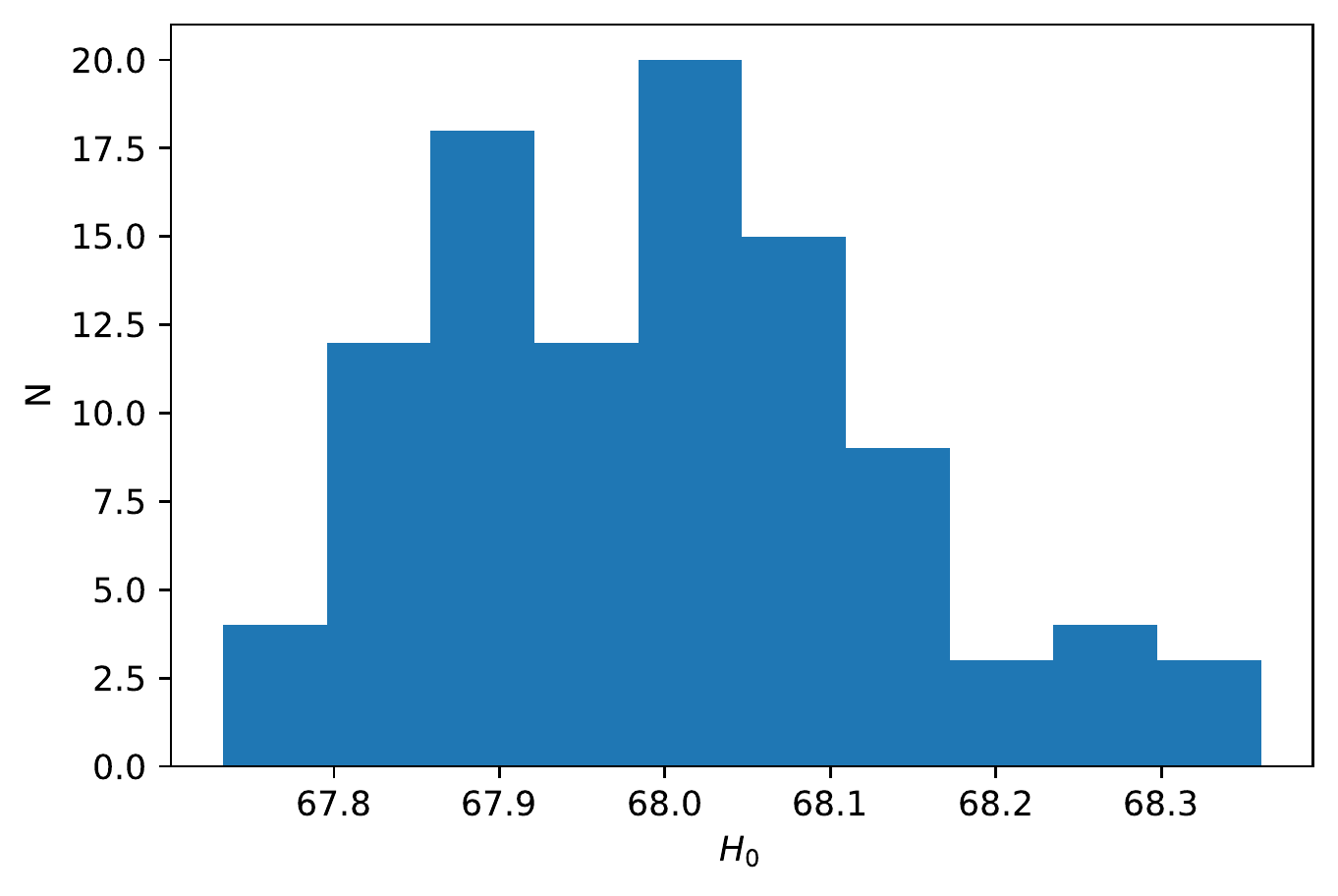}{0.45\textwidth}{(e) Only QSOs calibrated with evolution as a function of cosmology, varying $H_{0}$ together with $\Omega_{M}$}}
    \caption{The results of the 100 looped fits of the $H_{0}$ parameter for all cases with calibrated QSOs. $H_{0}$ is in units of $\mathrm{km} \, \mathrm{s}^{-1} \, \mathrm{Mpc}^{-1}$.}
    \label{fig:H0loop}
\end{figure}

In order to verify statistically these results, due to a large uncertainty of the parameters in cases with calibrated QSOs, we have run the Monte Carlo simulations 100 times to better evaluate the uncertainties and the true value of the cosmological parameters. The achieved distributions of the values of the $H_{0}$ obtained in looped computations are shown in Figure \ref{fig:H0loop} and the mean values are present in Table \ref{tab:zscores}. This approach has been already successfully used for GRBs in \citet{grbcosmology} and we here show the histograms of the distributions of $H_{0}$ for the five cases which are detailed in the lower part of Table \ref{tab:zscores}.

\begin{table}[]
    \centering
    \begin{tabular}{|c|c|C|}
    \hline
        Considered case & $H_{0}$ $\left[\frac{km}{Mpc\times s}\right]$ & $z$-score \\\hline\hline
        \multicolumn{3}{|c|}{Without calibration QSOs+SNe}\\\hline\hline
         $k=k(\Omega_{M})$, $H_0$ varied with $\Omega_{m}$ & $69.99 \pm 0.35$ & $4.24$ \\
         Fixed evolution, $H_0$ varied with $\Omega_{m}$ & $70.72 \pm 0.34$ & 5.49 \\
         Fixed evolution, $H_0$ varied alone & $69.99 \pm 0.14$ & 4.99 \\
         No evolution, $H_0$ varied with $\Omega_{m}$ & $69.44 \pm 0.32$ & 3.44 \\
         No evolution, $H_0$ varied alone & $69.97 \pm 0.14$ & 4.95 \\\hline\hline
         \multicolumn{3}{|c|}{Calibrated QSOs alone}\\\hline\hline
         $k=k(\Omega_{M})$, $H_0$ varied with $\Omega_{m}$ & $68.00 \pm 2.61$ & $0.23$ \\
         Fixed evolution, $H_0$ varied with $\Omega_{m}$ & $76.48 \pm 3.04$ & 2.95 \\
         Fixed evolution, $H_0$ varied alone & $69.68 \pm 2.19$ & 1.01 \\
         No evolution, $H_0$ varied with $\Omega_{m}$ & $62.84 \pm 2.41$ & 1.86 \\
         No evolution, $H_0$ varied alone & $73.76 \pm 2.18$ & 2.84 \\\hline
    \end{tabular}
    \caption{ Comparison of results for the $H_{0}$ parameter obtained in this work with different approaches. The $z$-score marking the compatibility of our results with the results obtained with Planck data is computed as $z_{i} = \frac{|H_{0,CMB}-H_{0,i}|}{\sqrt{\sigma_{CMB}^{2}+\sigma_{i}^{2}}}$, where $H_{0,CMB} \pm \sigma_{CMB} = 67.4\pm 0.5\, \frac{km}{Mpc\times s}$ is the measurement of the Hubble constant with the Planck data, and $H_{0,i} \pm \sigma_{i}$ is one of our measurements presented in above table.}
    \label{tab:zscores}
\end{table}

\begin{figure}
\gridline{\fig{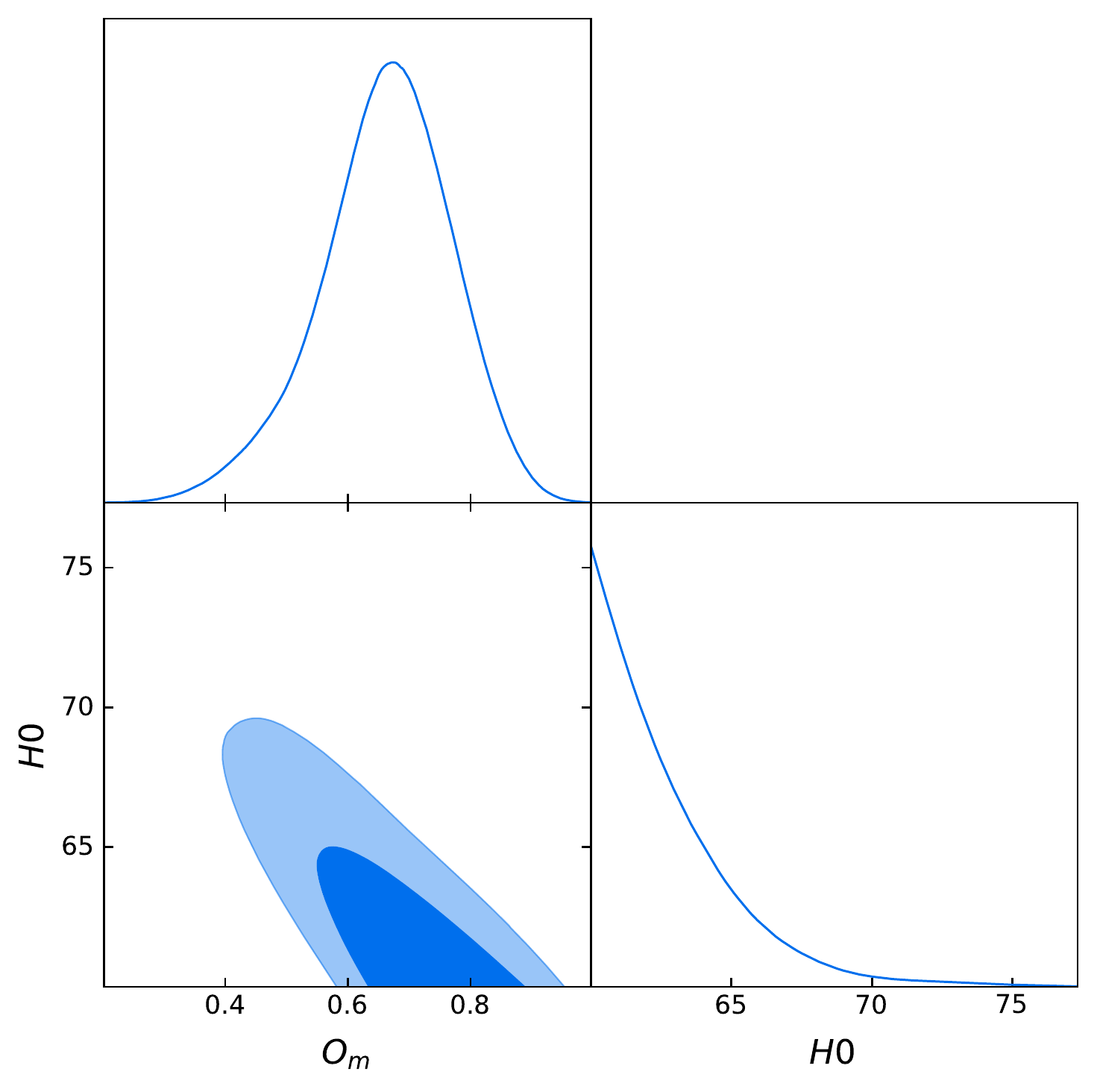}{0.2\textwidth}{(a) Only QSOs calibrated without evolution}
          \fig{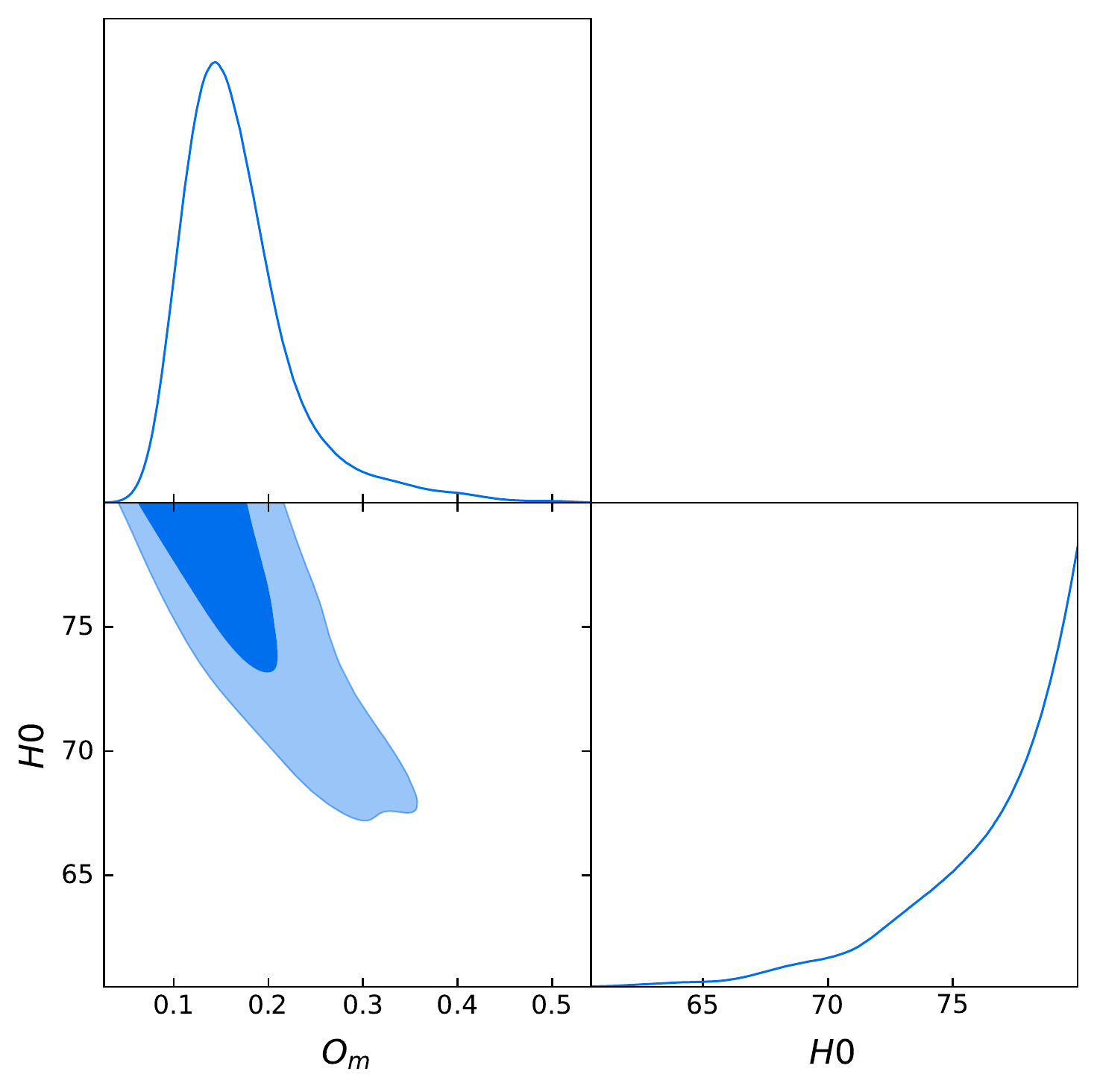}{0.2\textwidth}{(b) Only QSOs calibrated with fixed evolution}
          \fig{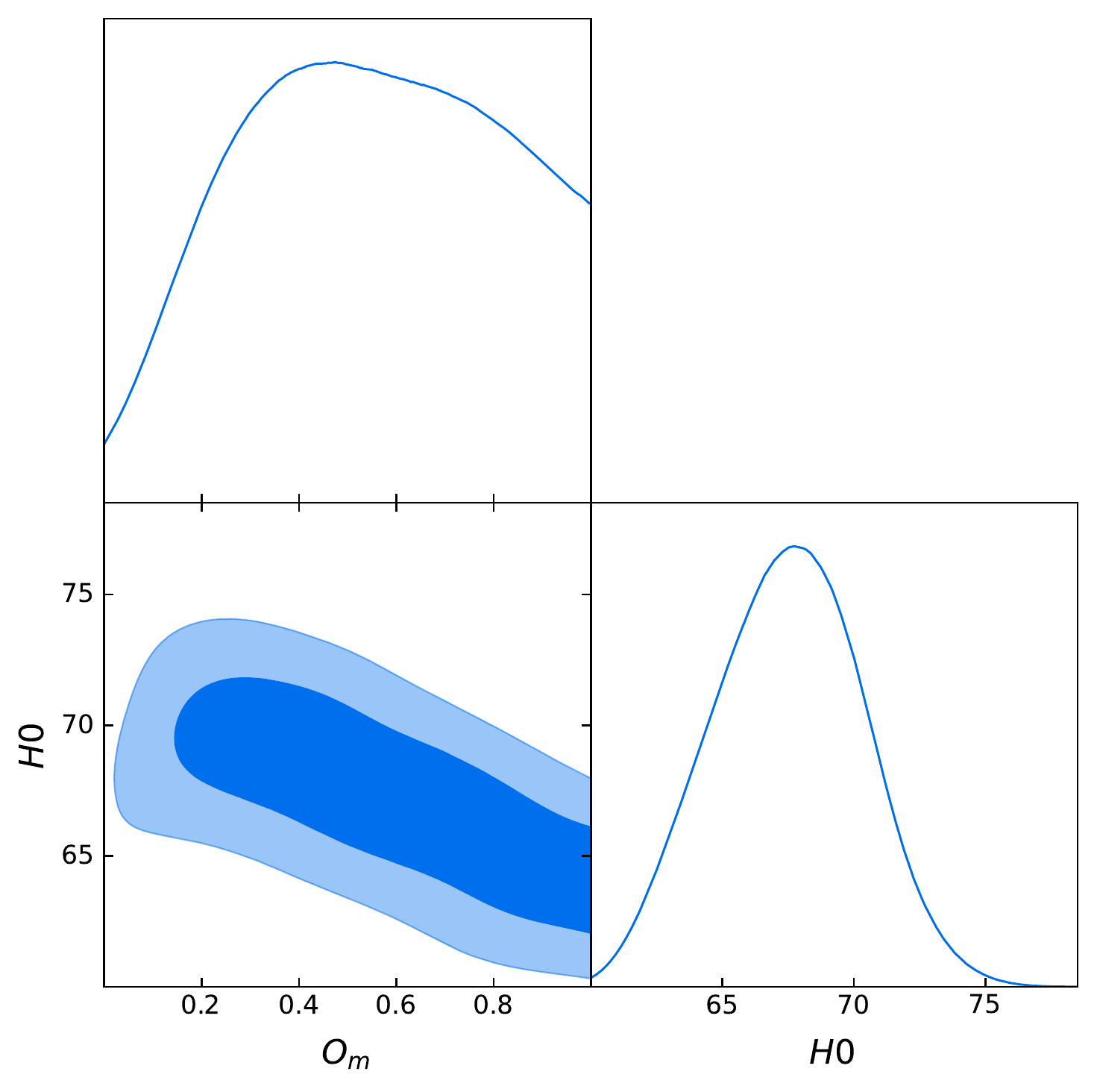}{0.2\textwidth}{(c) Only QSOs calibrated with varying evolution}}
\gridline{\fig{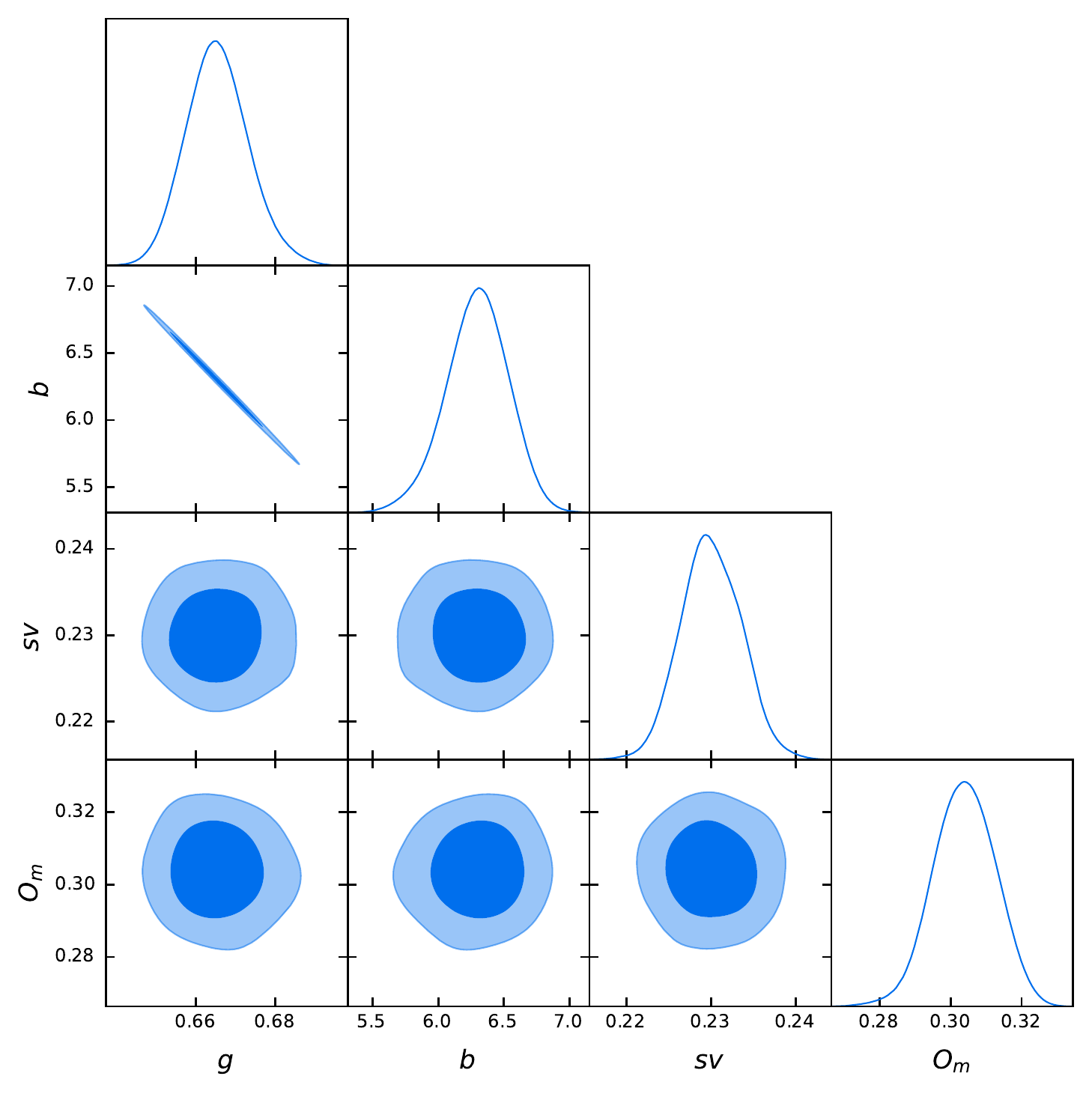}{0.28\textwidth}{(d) Non-calibrated QSOs + SNe Ia without evolution, only $\Omega_M$}
          \fig{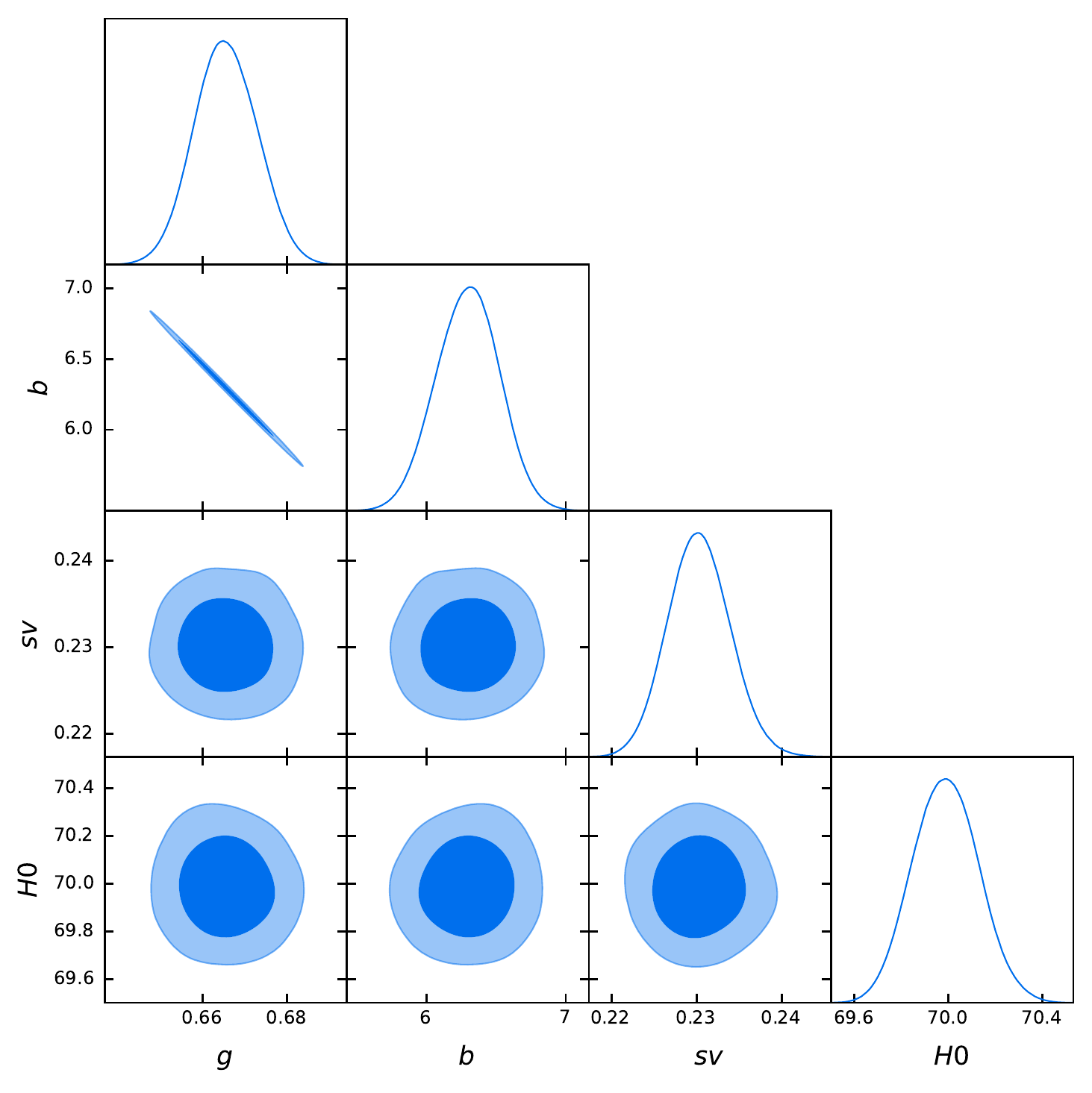}{0.28\textwidth}{(e) Non-calibrated QSOs + SNe Ia without evolution, only $H_0$}
          \fig{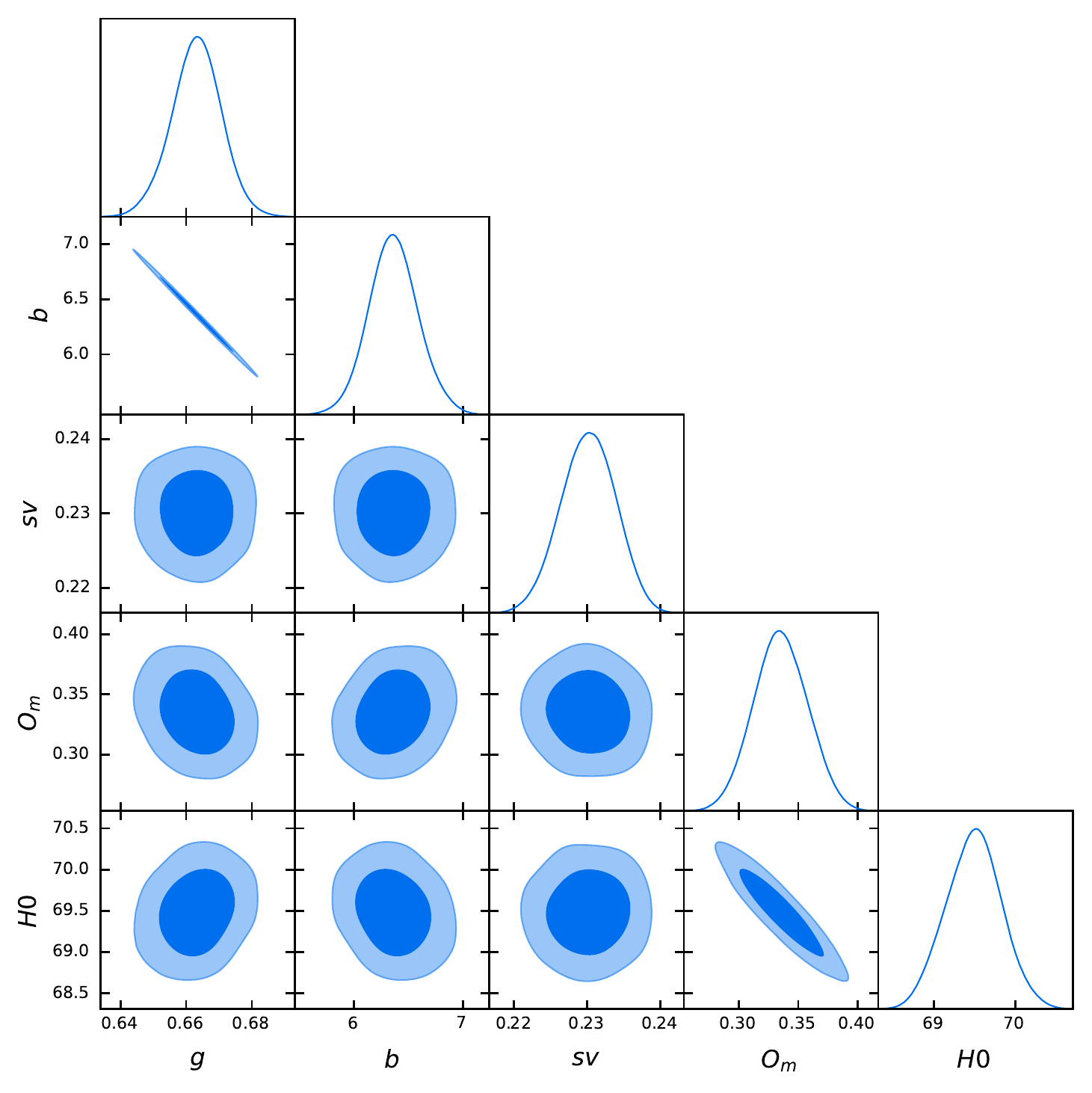}{0.28\textwidth}{(f) Non-calibrated QSOs + SNe Ia without evolution, both $\Omega_M$ and $H_0$}}
\gridline{\fig{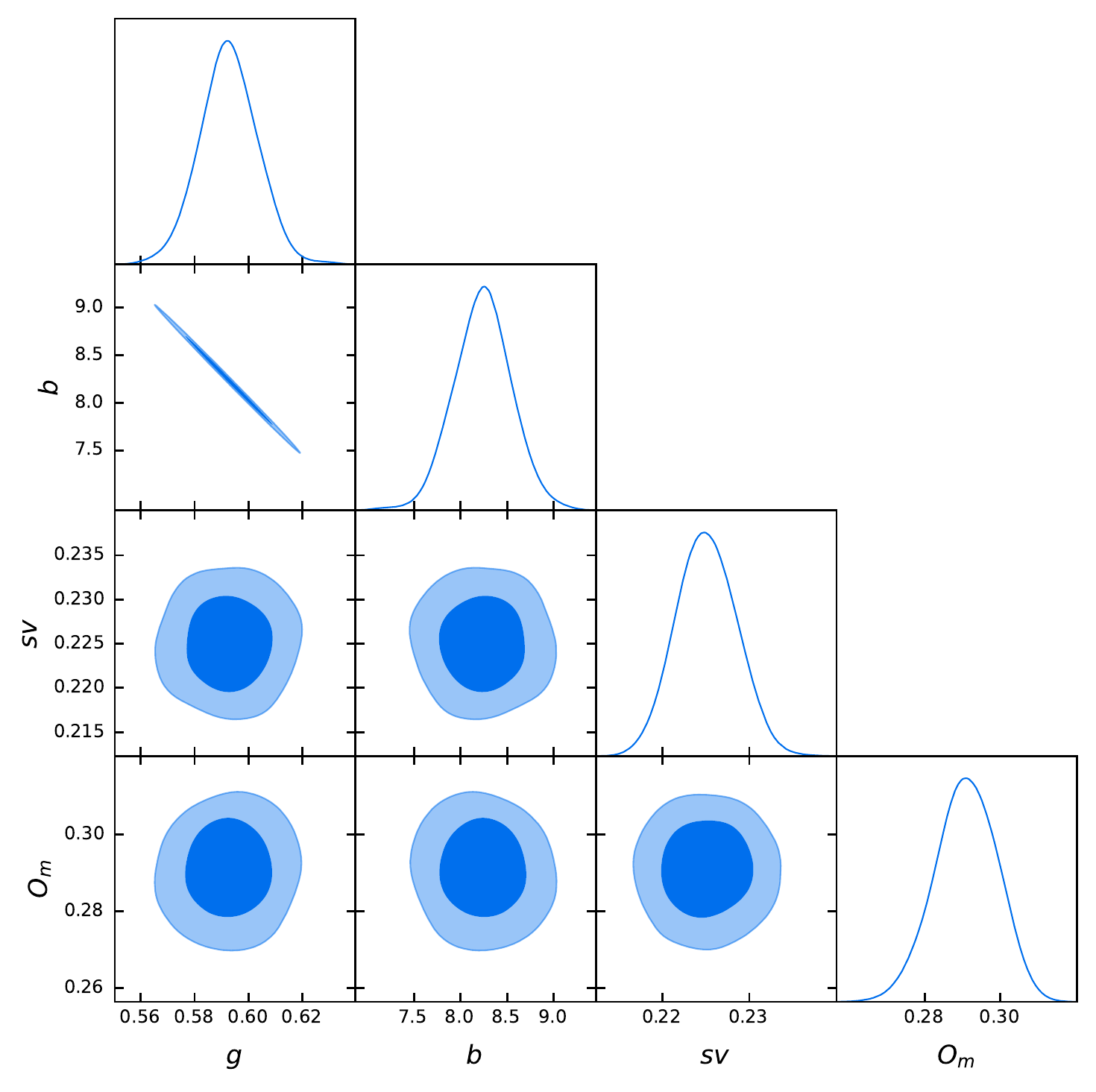}{0.28\textwidth}{(g) Non-calibrated QSOs + SNe Ia with fixed evolution, only $\Omega_M$}
        \fig{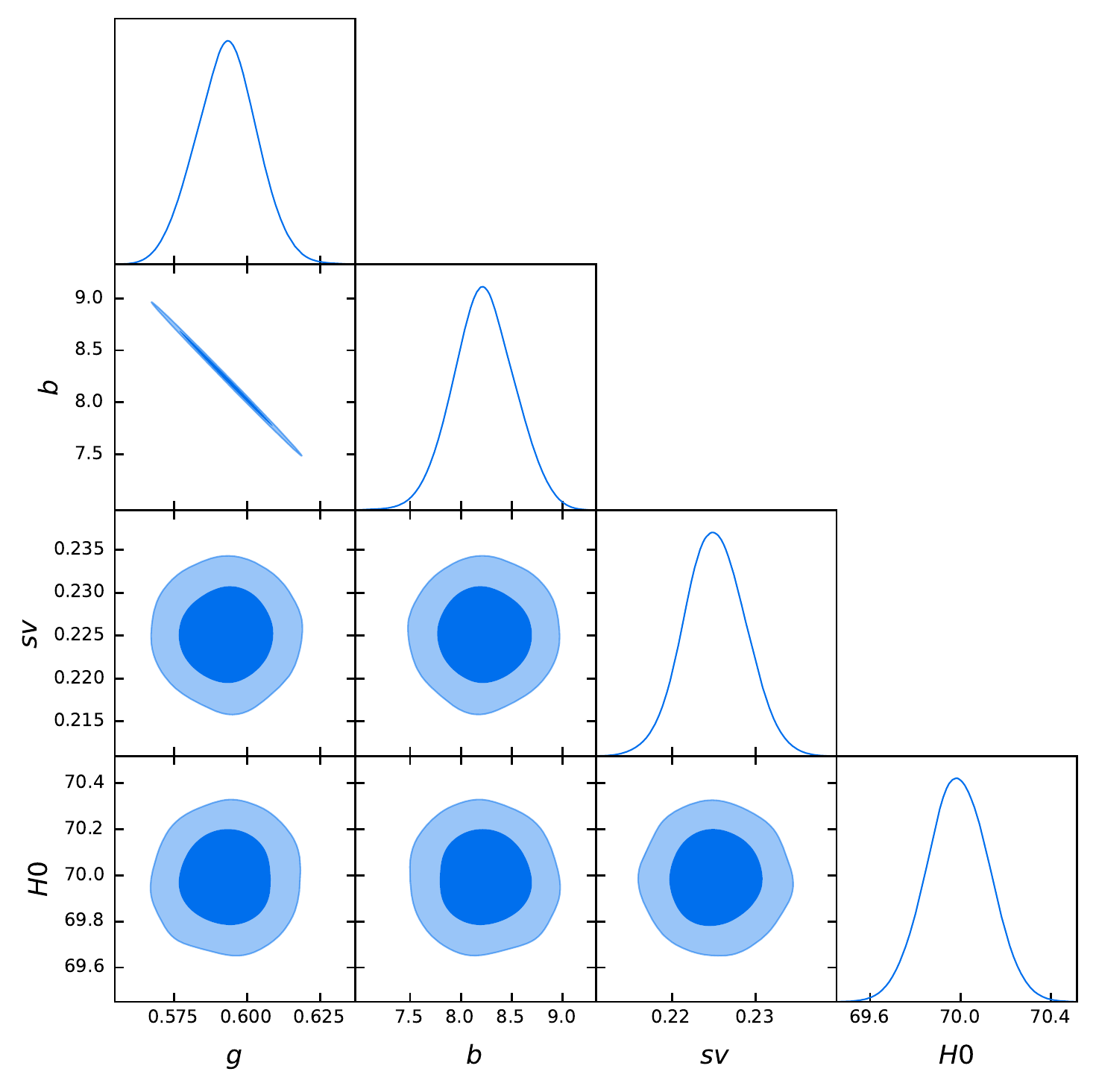}{0.28\textwidth}{(h) Non-calibrated QSOs + SNe Ia with fixed evolution, only $H_0$}
        \fig{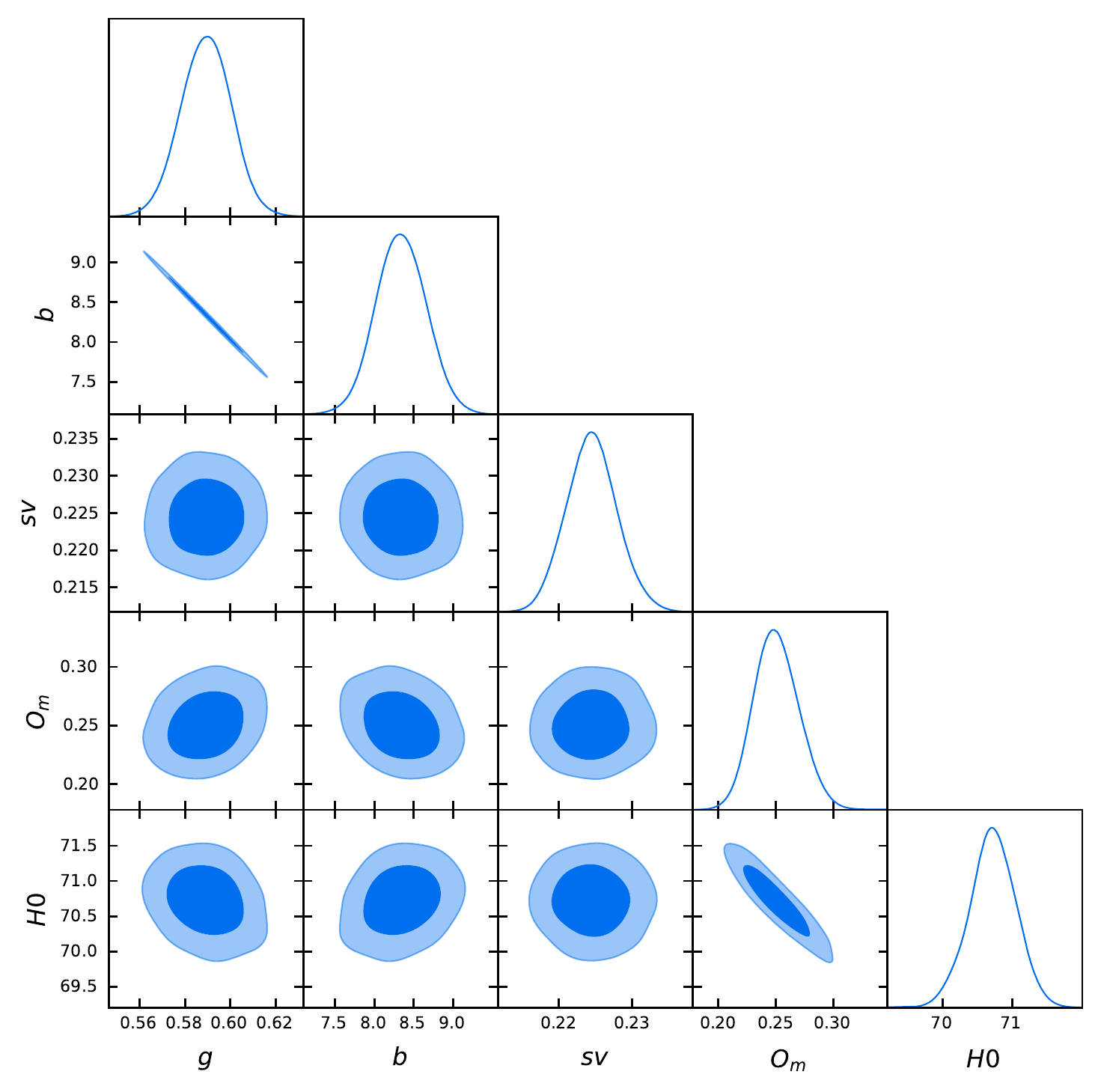}{0.28\textwidth}{(i) Non-calibrated QSOs + SNe Ia with fixed evolution evolution, both  $\Omega_M$ and $H_0$}}
\gridline{\fig{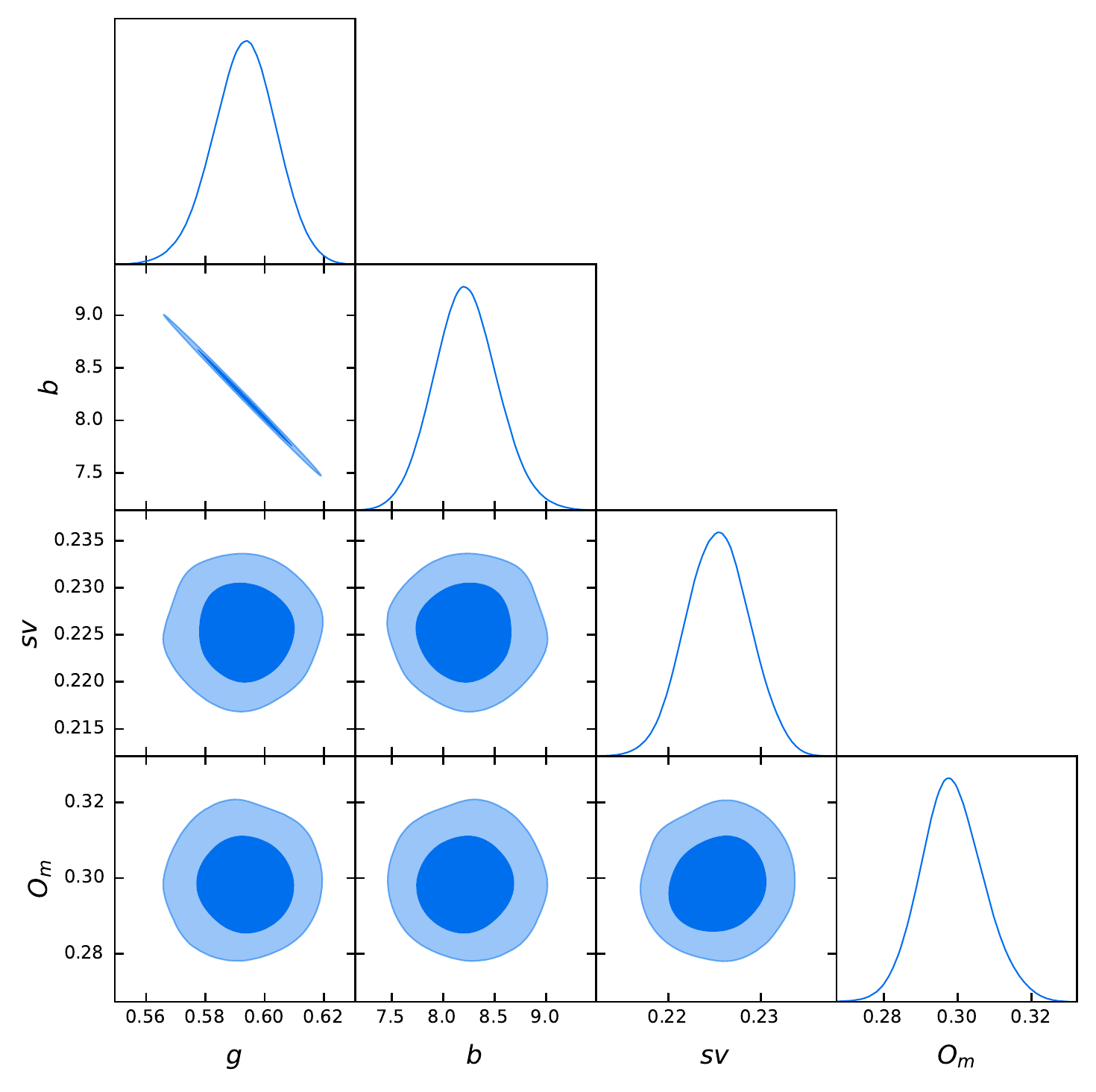}{0.28\textwidth}{(j) Non-calibrated QSOs + SNe Ia with varying evolution, only $\Omega_M$}
        \fig{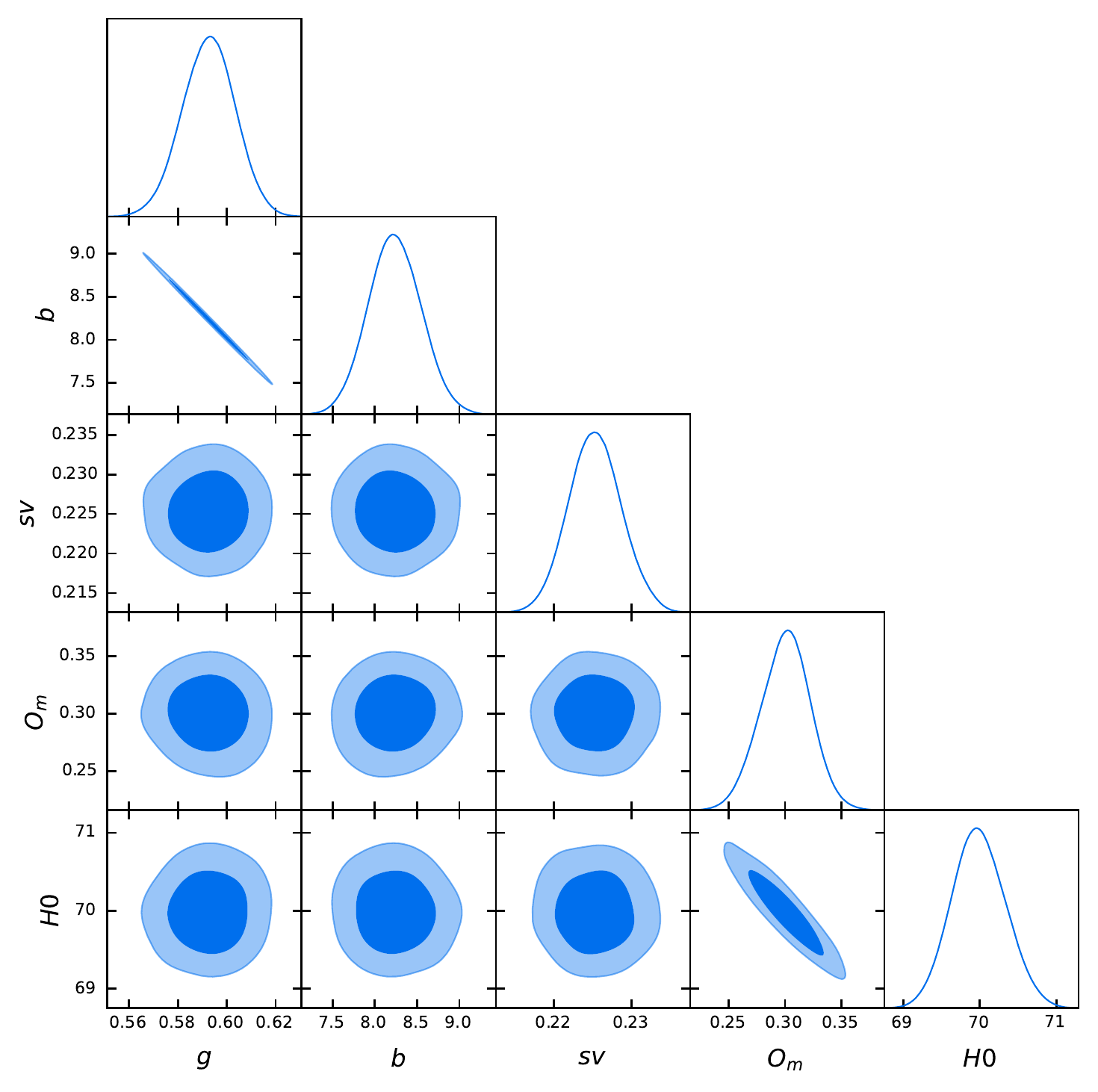}{0.28\textwidth}{(k)Non-calibrated QSOs + SNe Ia with varying evolution evolution, both $\Omega_M$ and $H_0$}
        }
\caption{Corner plots obtained under the assumption of the flat $\Lambda$CDM model.}
\label{cornerplot1}
\end{figure}

\begin{figure}
\gridline{\fig{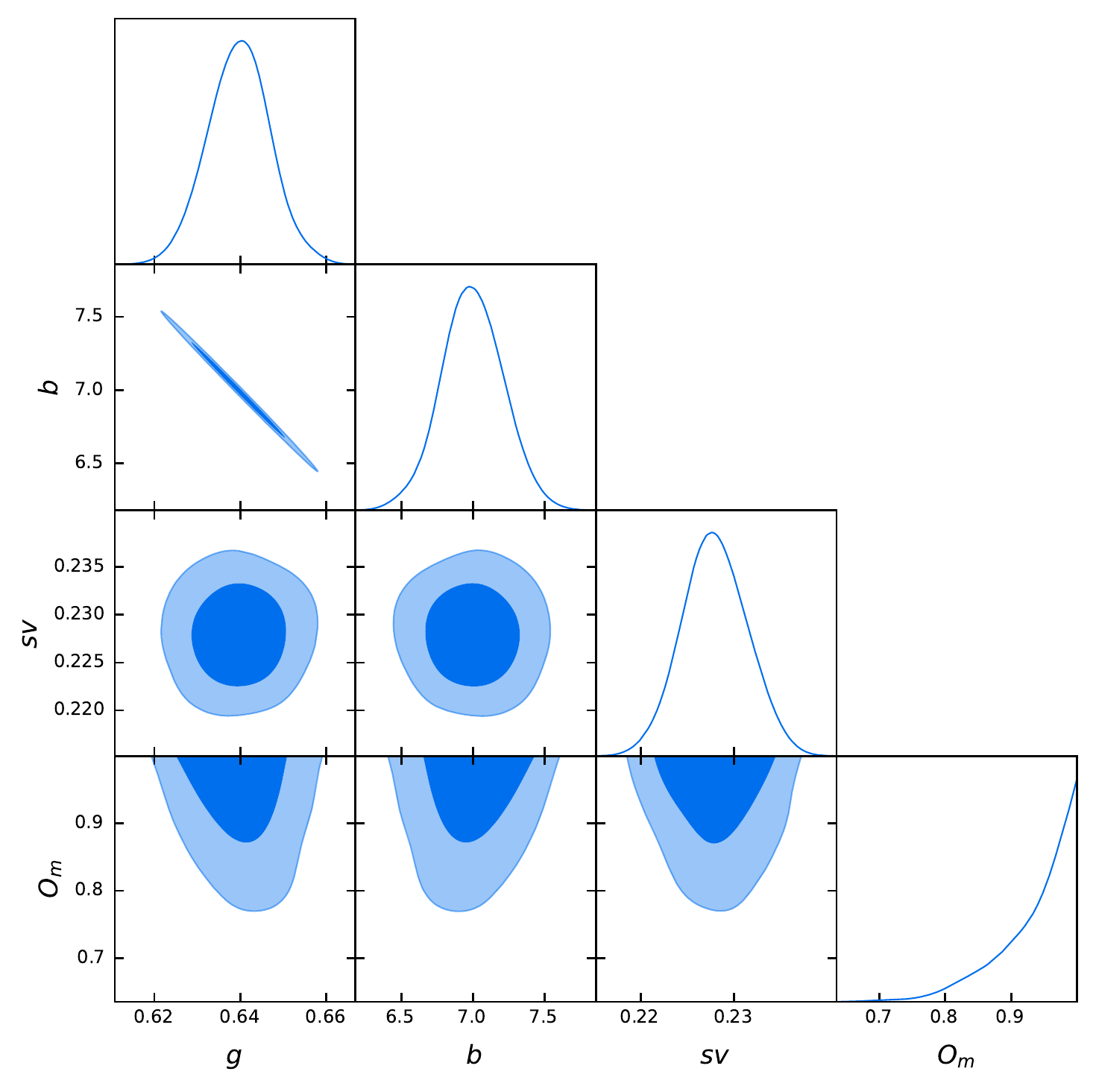}{0.33\textwidth}{(a) Only non-calibrated QSOs without evolution}
          \fig{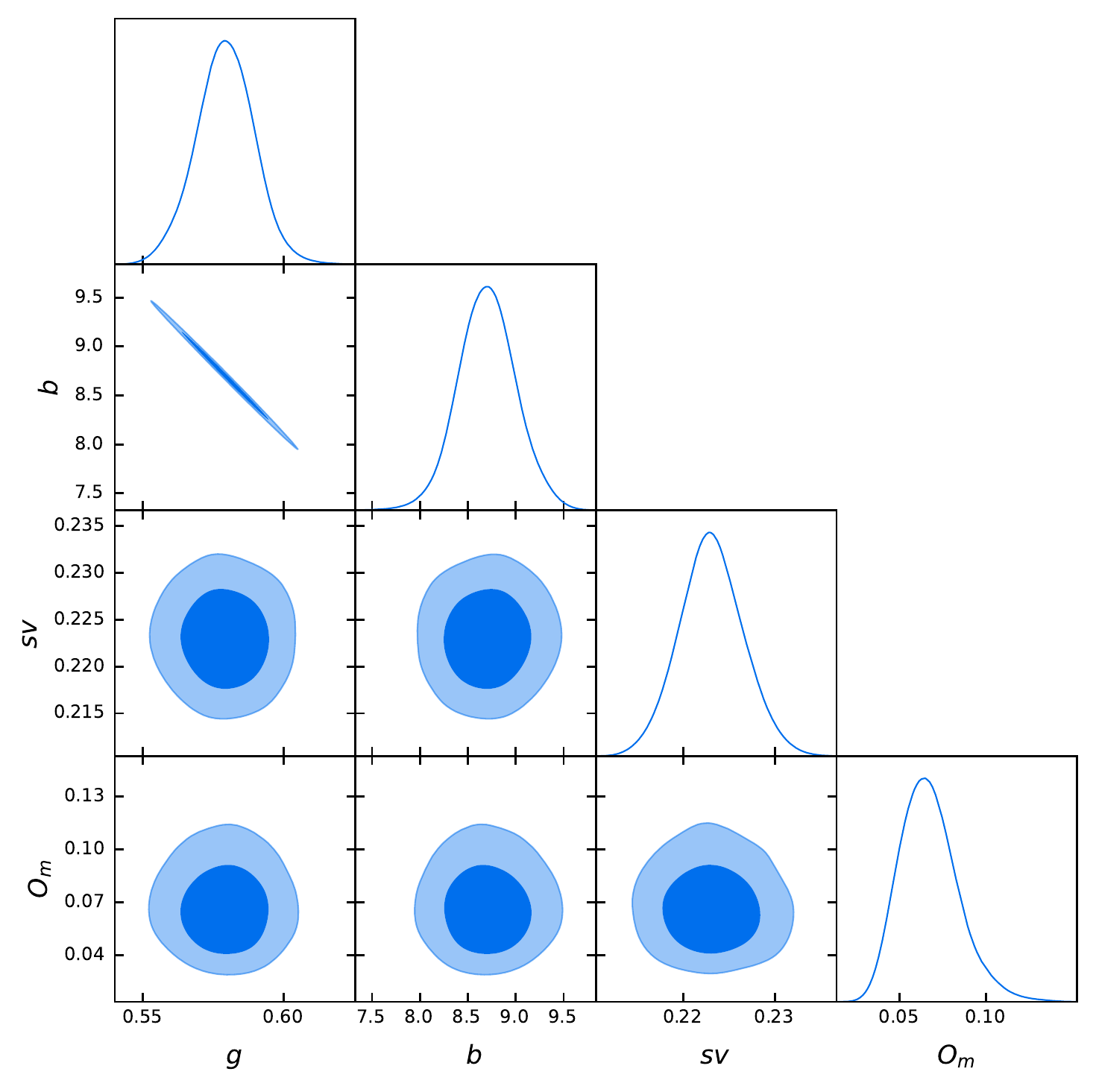}{0.33\textwidth}{(b) Only non-calibrated QSOs with fixed evolution}
          \fig{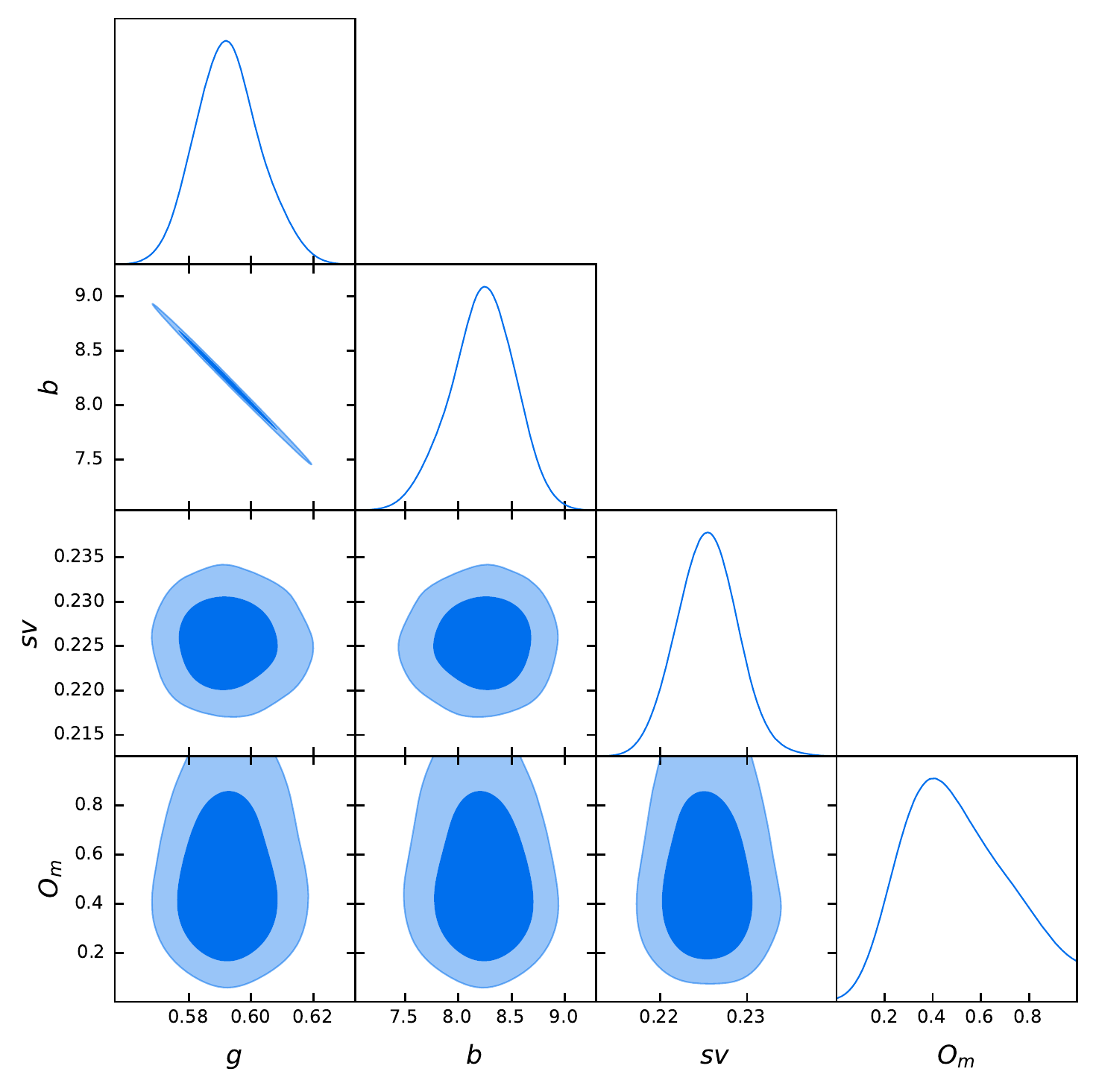}{0.33\textwidth}{(c) Only non-calibrated QSOs with varying evolution}}

\caption{Corner plots obtained under the assumption of the flat $\Lambda$CDM model using non-calibrated QSOs alone.}
\label{cornerplot1.1}
\end{figure}

\section{Summary \& Conclusions}
\label{conclusions}

In this work, we analyzed the flat and non-flat $\Lambda \mathrm{CDM}$ models and the flat $w$CDM model using SNe Ia and the most updated sample of QSOs, as cosmological probes in view of testing QSOs as distance indicators and alleviating the $H_0$ tension issue. This study is strongly motivated by the need for testing the predictions of the spatially flat $\Lambda \mathrm{CDM}$ model and searching for possible deviations to explain the theoretical and observational shortcomings of this model, both in the direction of a non-flat Universe and of a potential new physics. The inclusion of QSOs in the cosmological analysis is crucial to this aim as they extend the Hubble-Lema\^itre diagram of SNe Ia up to a higher-redshift range ($z=2.26-7.54$) in which predictions from different cosmological models can be better distinguished and more easily compared to observational data. We explored and compared different approaches to apply the RL relation for including QSOs in the cosmological analyses. Specifically, we used this relation both in its original form and with luminosities corrected for selection biases and evolution in redshift. In this latter approach, making use of the EP method, we investigated both a fixed evolution and an evolution that depends on the cosmological parameters of the model studied to overcome the circularity problem. The latter technique has recently been applied to GRBs \citep{grbcosmology}, but it is the first time in the literature that it is used for QSOs. This study also answers some previous criticisms concerning the application of the X-UV relation in cosmology \citep[e.g.][]{2022ApJ...935L..19P}. All these approaches have been also explored using both non-calibrated QSOs combined with SNe Ia and QSOs alone calibrated with SNe Ia. Below, we summarise our main results.

\begin{itemize}
    \item In the effort of overcoming the circularity problem, we apply the innovative method in which the correction of luminosities for the evolution in redshift varies together with the cosmological parameters of the assumed cosmological model. Our investigation proves that $k_{L_X}$ and $k_{L_{UV}}$ significantly depend on the cosmological parameters, as shown in Figures \ref{fig:EPvsOmOk} and \ref{fig:EPOk}. Assuming a non-flat $\Lambda$CDM model with $\Omega_M = 0.3$ the values of $k$ do not vary much for $\Omega_k$ close to $\Omega_k = 0$, being compatible within $1 \, \sigma$ with the value obtained for $\Omega_k = 0$ up to $\Omega_k \sim -0.4$ for both UV and X-ray wavelengths (upper row of Figure \ref{fig:EPOk}). Moreover, if we consider the whole parameter space (i.e. $k(\Omega_M, \Omega_k)$ in the upper row of Figure \ref{fig:EPvsOmOk}), we conclude that, in reasonable regions of this space, the values of $k$ are compatible with the ones obtained for $\Omega_{k}=0$, $\Omega_{M}=0.3$ for both UV and X-ray cases. Similar conclusions can be drawn for $k(w)$ and $k(\Omega_M,w)$ in the flat $w$CDM model. Indeed, if we fix $\Omega_M =0.3$, $k$'s values are compatible with the ones obtained with $w = -1$ in the whole range of $w$ explored, as shown in the bottom row of Figure \ref{fig:EPOk} for both the UV and X-ray cases. This explains why we do not observe any significant differences in the cosmological results between the computation of $w$ with fixed correction for evolution and the one with $k = k(w)$ (see Table \ref{tab:1}). As happens for the non-flat $\Lambda$CDM case, $k$'s values in the parameter space $\{\Omega_M, w \}$ result compatible within $2\,\sigma$ with the ones obtained for $\Omega_{M}=0.3$ and $w=-1$ for both UV and X-ray cases (see bottom row of Figure \ref{fig:EPvsOmOk}).
   Nevertheless, since we explore the behaviour of $k$ in very wide ranges of the cosmological parameters $\Omega_M$, $\Omega_k$, and $w$, we do observe significant deviation ($> 3 \, \sigma$) of $k$'s values from the one expected for $\Omega_M=0.3$, $\Omega_k=0$, and $w=-1$ in some extreme regions of the parameter spaces. Thus, in these particular regions, although one may argue the need of exploring such exotic parameter space, the dependence of $k$ on the cosmological parameters is significant and should be taken into account in the computations.

    \item Under the assumption of a flat $\Lambda$CDM model, results from QSOs alone calibrated with SNe Ia (the left side of Table \ref{tab:1} and panels (a)-(c) of Figure \ref{cornerplot1}) do not show a common trend, with values of $\Omega_M$ and $H_0$ that have large uncertainties, span a wide range of values, and are not well constrained.
    These issues are completely removed by joining non-calibrated QSOs and SNe Ia in the analyses. Indeed, with this sample, we do not encounter any convergence issue and, in addition, we always find a common trend toward $\Omega_M = 0.3$ and $H_0 = 70$ that becomes tighter (within $1 \, \sigma$), with the application of a correction for the evolution in the redshift of luminosities that varies together with the cosmological parameters (see the right side of Table \ref{tab:1} and panels (j) and (k) in Figure \ref{cornerplot1}). Within this model, considering the data set of QSOs alone not calibrated on SNe Ia and fixing $H_{0}=70\, \mathrm{km} \, \mathrm{s^{-1}} \,\mathrm{Mpc^{-1}}$, presents interesting results on the $\Omega_M$ parameter (see Table \ref{tab:3} and Figure \ref{cornerplot1.1}). More specifically, this investigation shows that also QSOs alone, without any calibration, well constrain $\Omega_M$ once we include a correction for the evolution with redshift that varies with the cosmological parameter (i.e. $k=k(\Omega_M)$). This case results in $\Omega_M$ compatible within $1 \, \sigma$ with $\Omega_M=0.3$, but with a preference toward higher values.
    
    \item When allowing for a possible curvature of the Universe, QSOs calibrated alone present, as in the previous case, some convergence issues. Nevertheless, when considering a fixed correction for evolution and both $\Omega_M$ and $\Omega_k$ as free cosmological parameters of the fit, we obtain $\Omega_M$ compatible with $\Omega_M = 0.3$ within $2 \, \sigma$. These results are shown in the left side of Table \ref{tab:1} and panels (a)-(c) of Figure \ref{cornerplot2}.
    When fitting together QSOs non- calibrated and SNe Ia (the right side of the Table \ref{tab:1} and panels (d)-(i) in Figure \ref{cornerplot2}), $\Omega_k$ remains compatible within $1 \, \sigma$ with $\Omega_k = 0$ if we fix $\Omega_M = 0.3$, both correcting and not correcting for the evolution, but $\Omega_k$ approaches very different values once $\Omega_M$ is free to vary.

    \item Assuming a flat $w$CDM model, calibrated QSOs alone do not clearly distinguish between the quintessence and the phantom DE scenario, as shown in the left side of Table \ref{tab:1} and in panels (a)-(c) in Figure \ref{cornerplot3}.
    Instead, when considering non-calibrated QSOs together with SNe Ia, in all cases, we obtain compatibility in $1 \, \sigma$ with $w=-1$ when $\Omega_M = 0.3$ (as expected in a flat $\Lambda$CDM scenario), while varying also $\Omega_M$ results in significant deviations from $w=-1$, except for the case with a correction that varies with the cosmological parameters, in which we recover compatibility in $2 \, \sigma$ with $\Omega_M = 0.3$ and $w=-1$. This is shown in the right side of Table \ref{tab:1} and panels (d)-(i) in Figure \ref{cornerplot3}.

\begin{figure}
\gridline{\fig{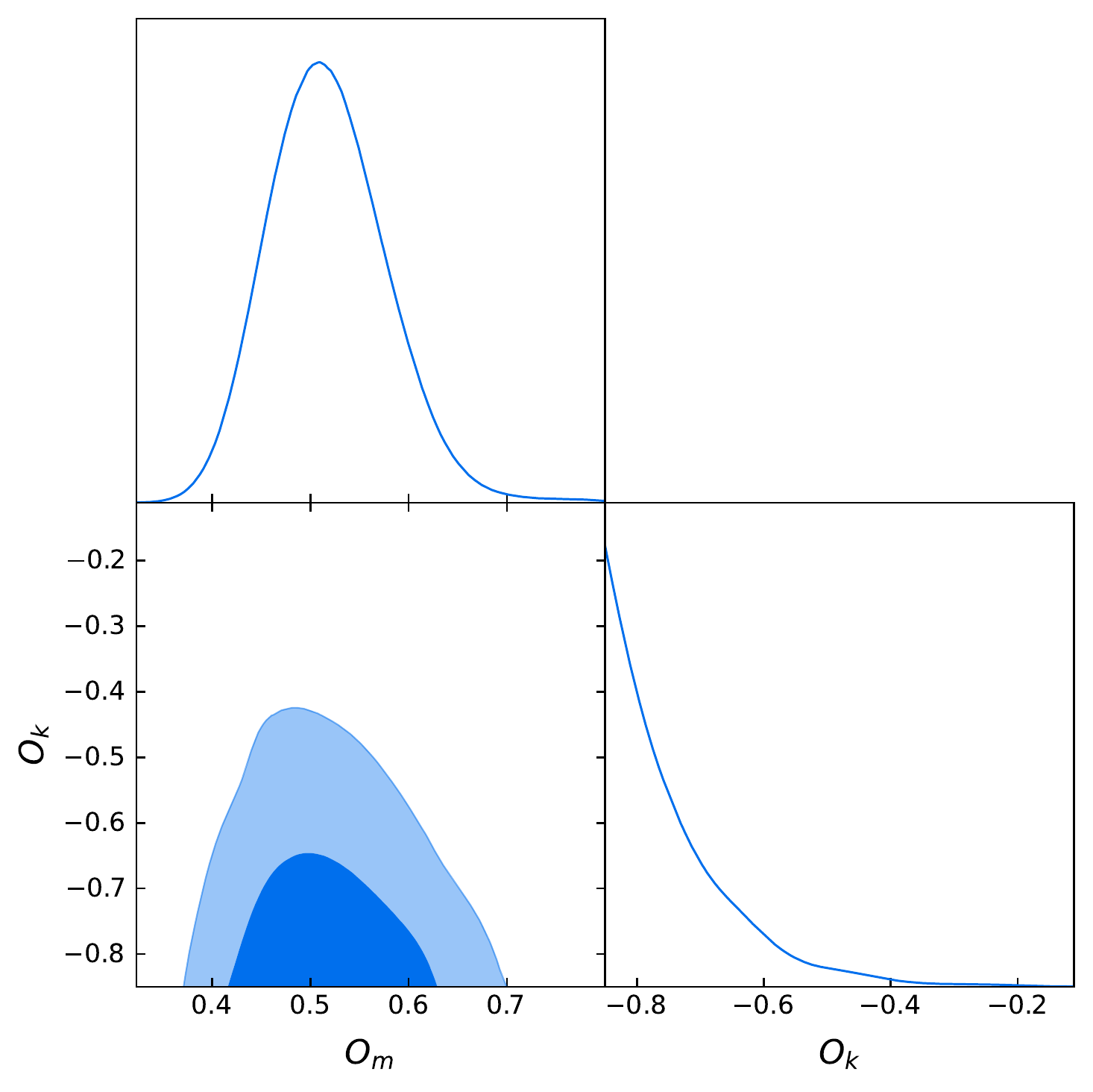}{0.3\textwidth}{(a) Only QSOs calibrated without evolution}
          \fig{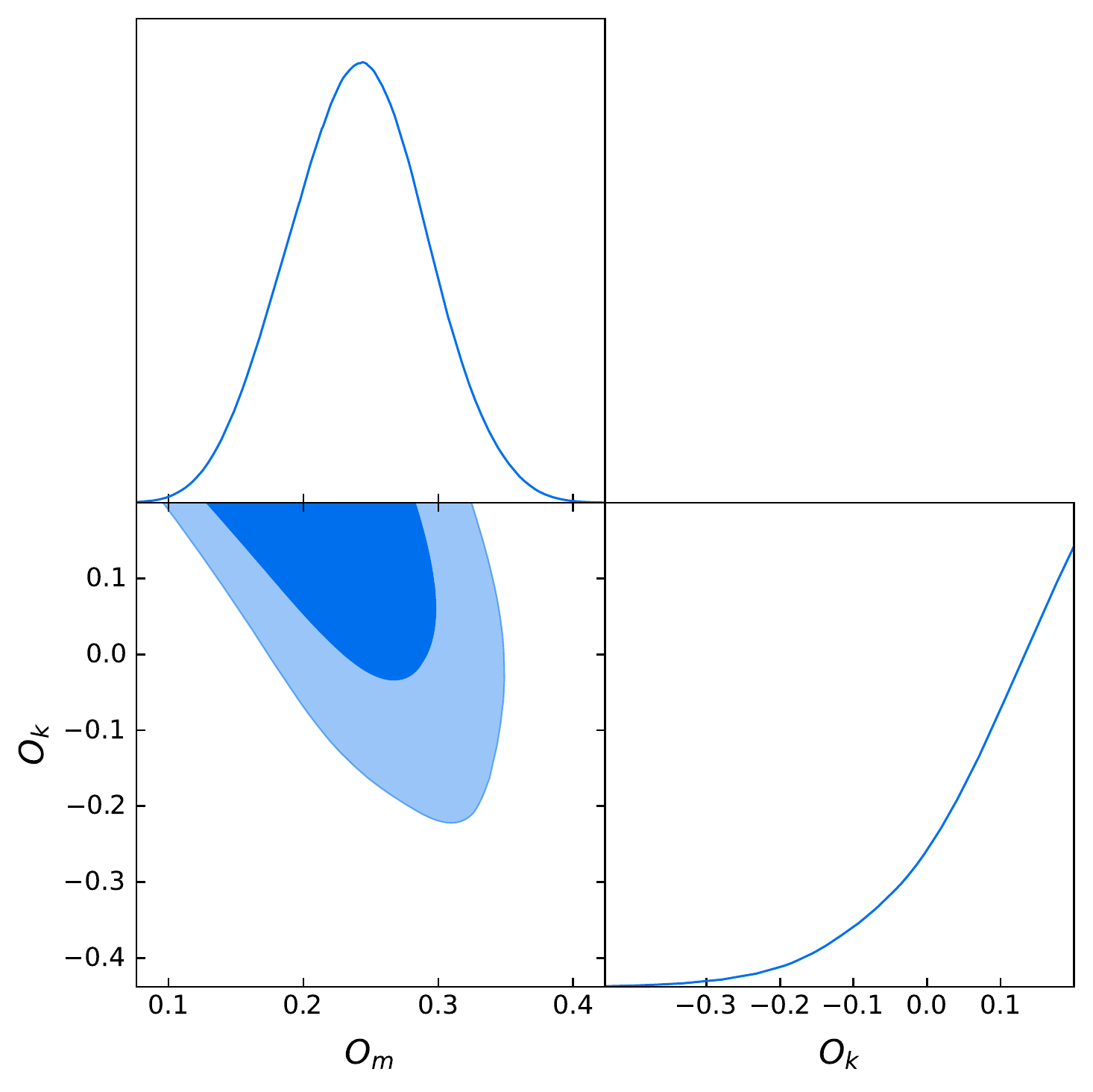}{0.3\textwidth}{(b) Only QSOs calibrated with fixed evolution}
          \fig{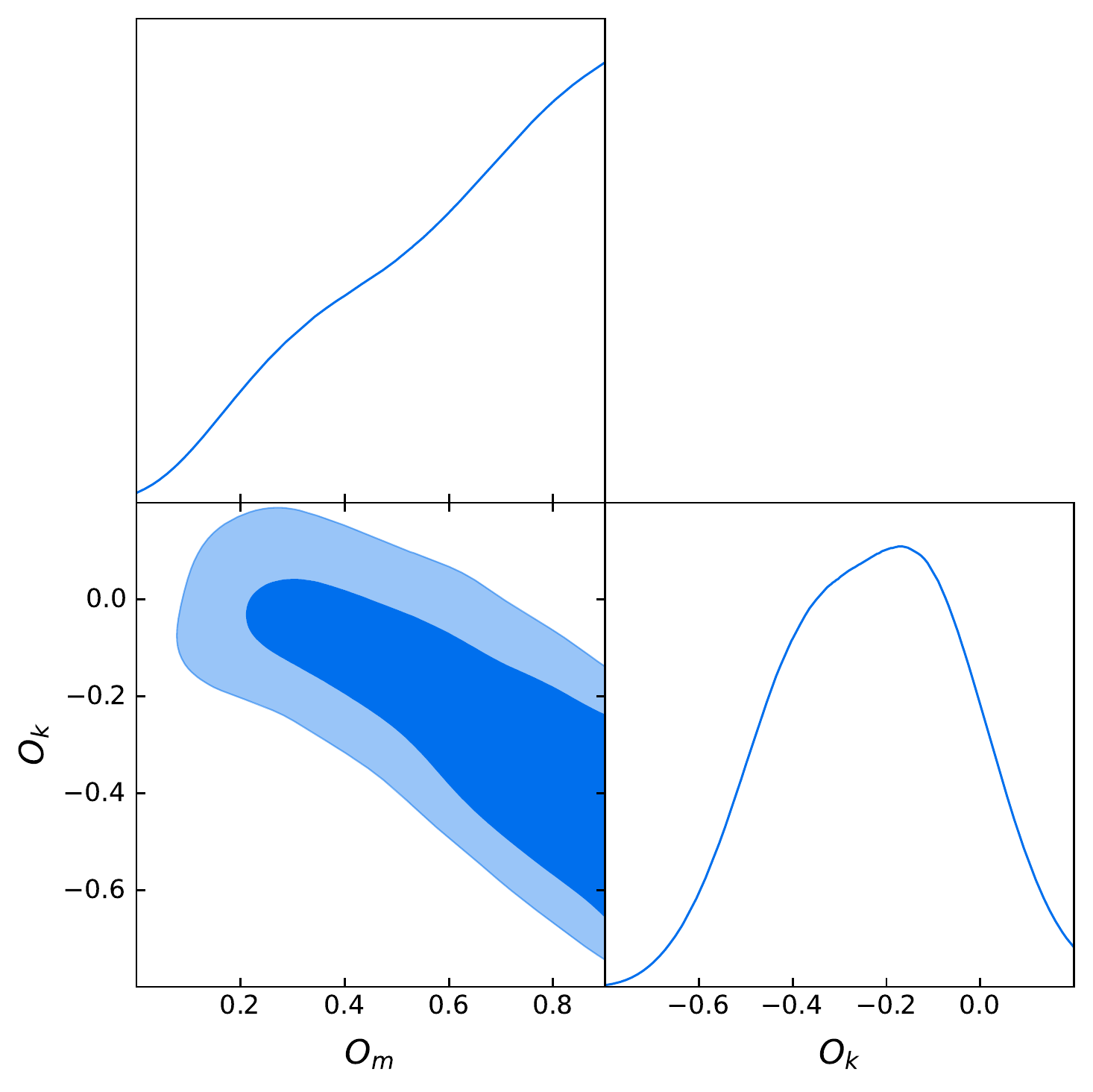}{0.3\textwidth}{(c) Only QSOs calibrated with varying evolution}}
\gridline{\fig{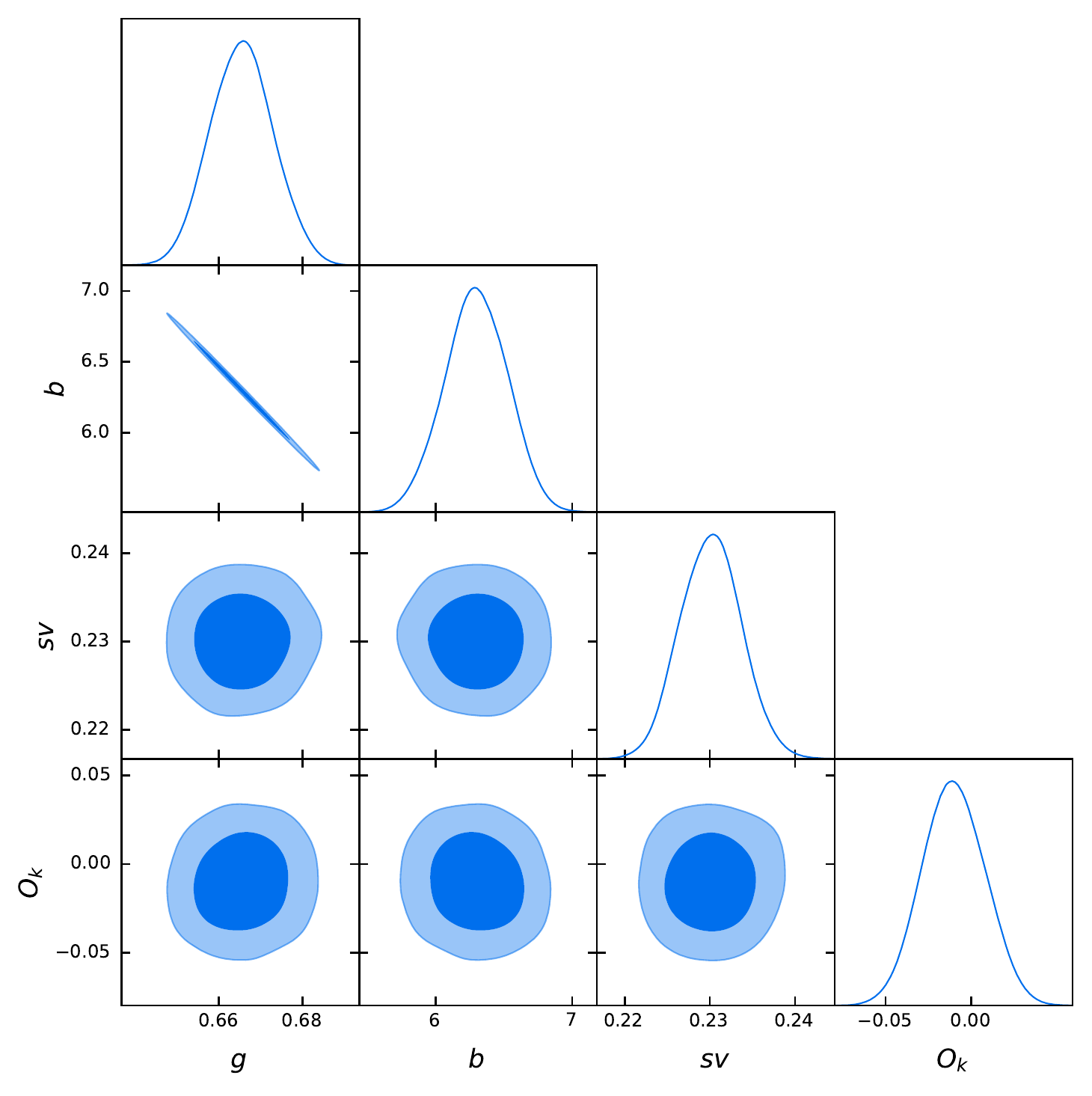}{0.3\textwidth}{(d) Non-calibrated QSOs + SNe Ia without evolution, only $\Omega_k$}
          \fig{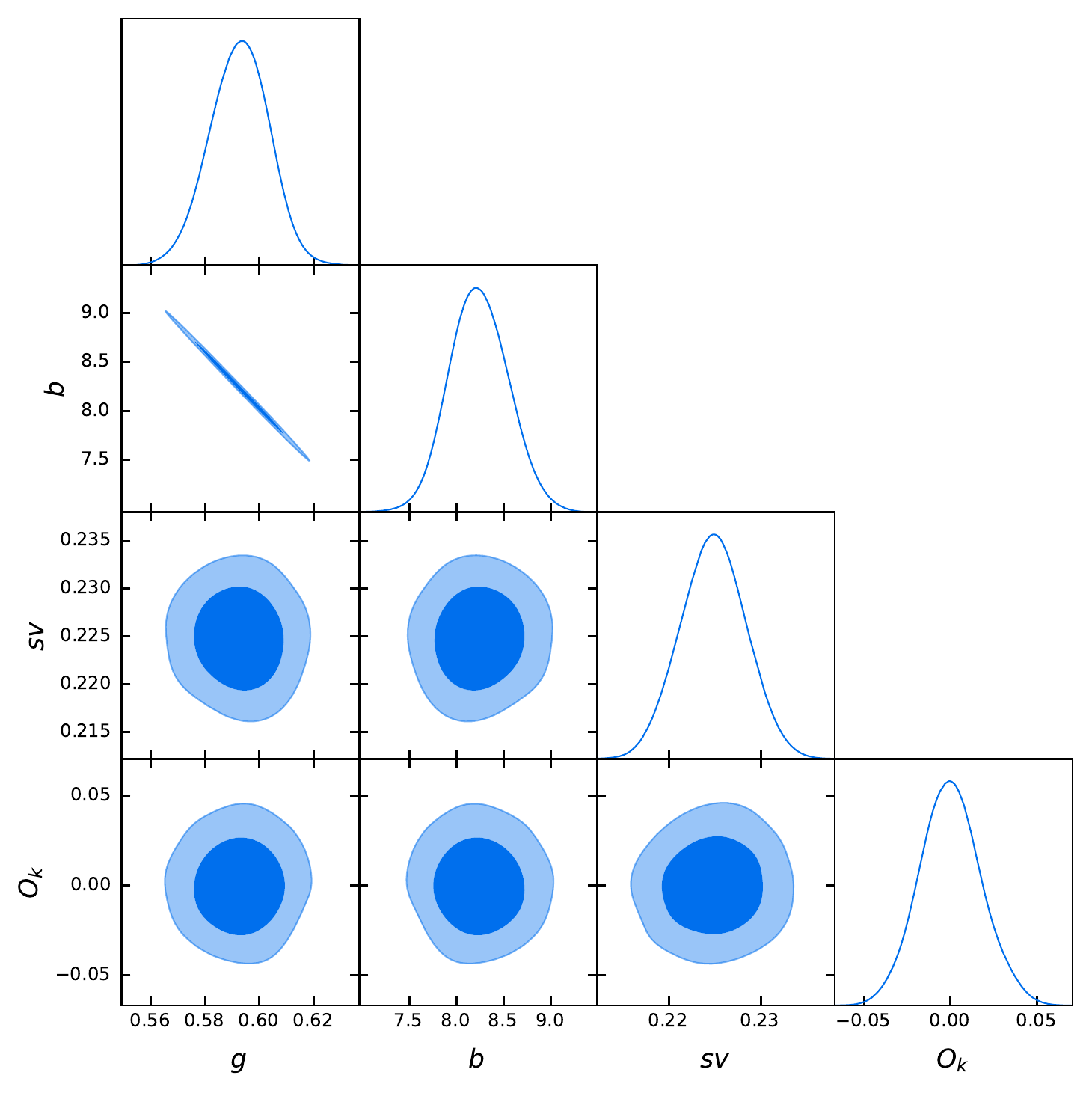}{0.3\textwidth}{(e) Non-calibrated QSOs + SNe Ia with fixed evolution, only $\Omega_k$}
          \fig{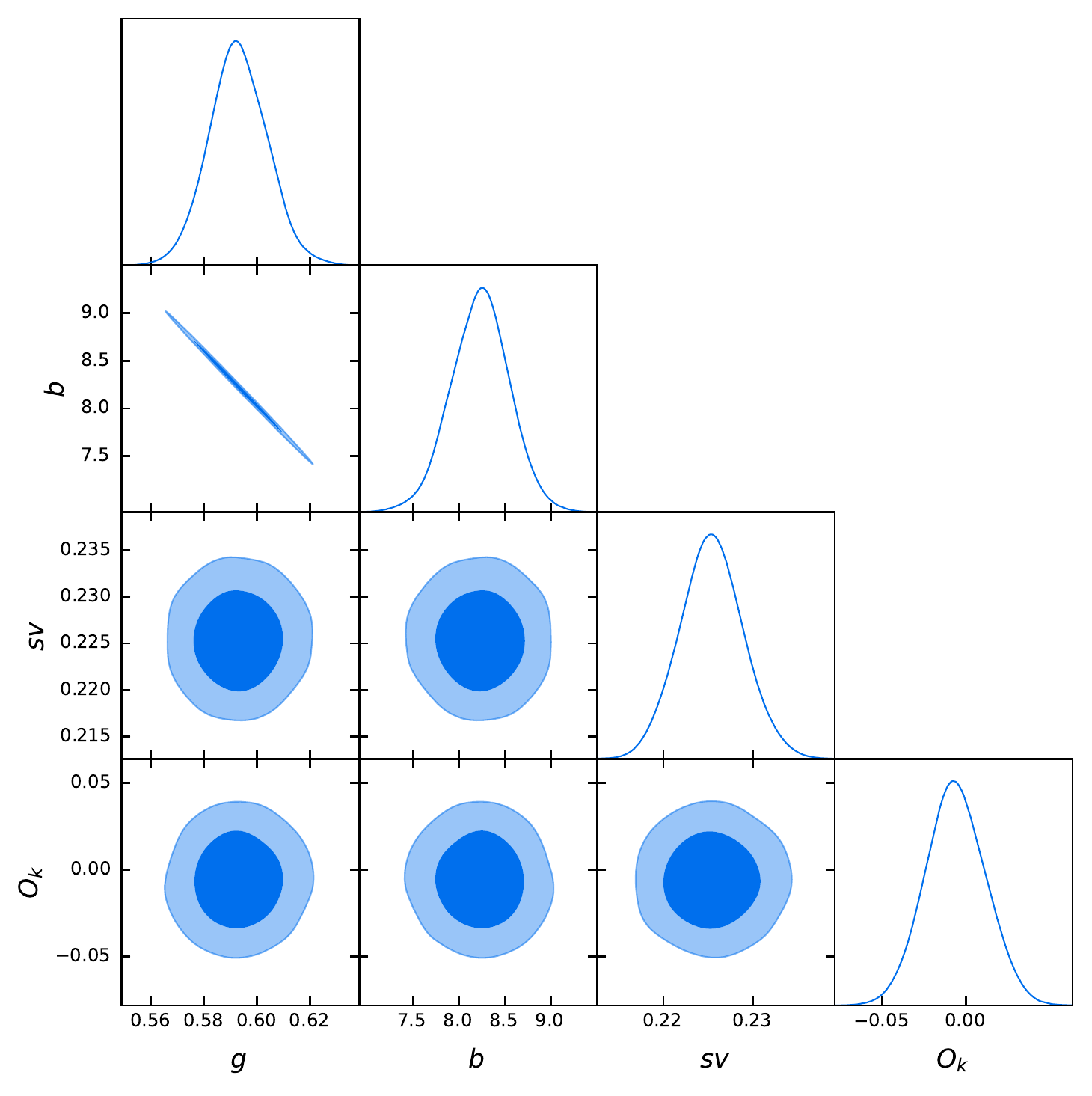}{0.3\textwidth}{(f) Non-calibrated QSOs + SNe Ia with varying evolution, only $\Omega_k$}}
\gridline{\fig{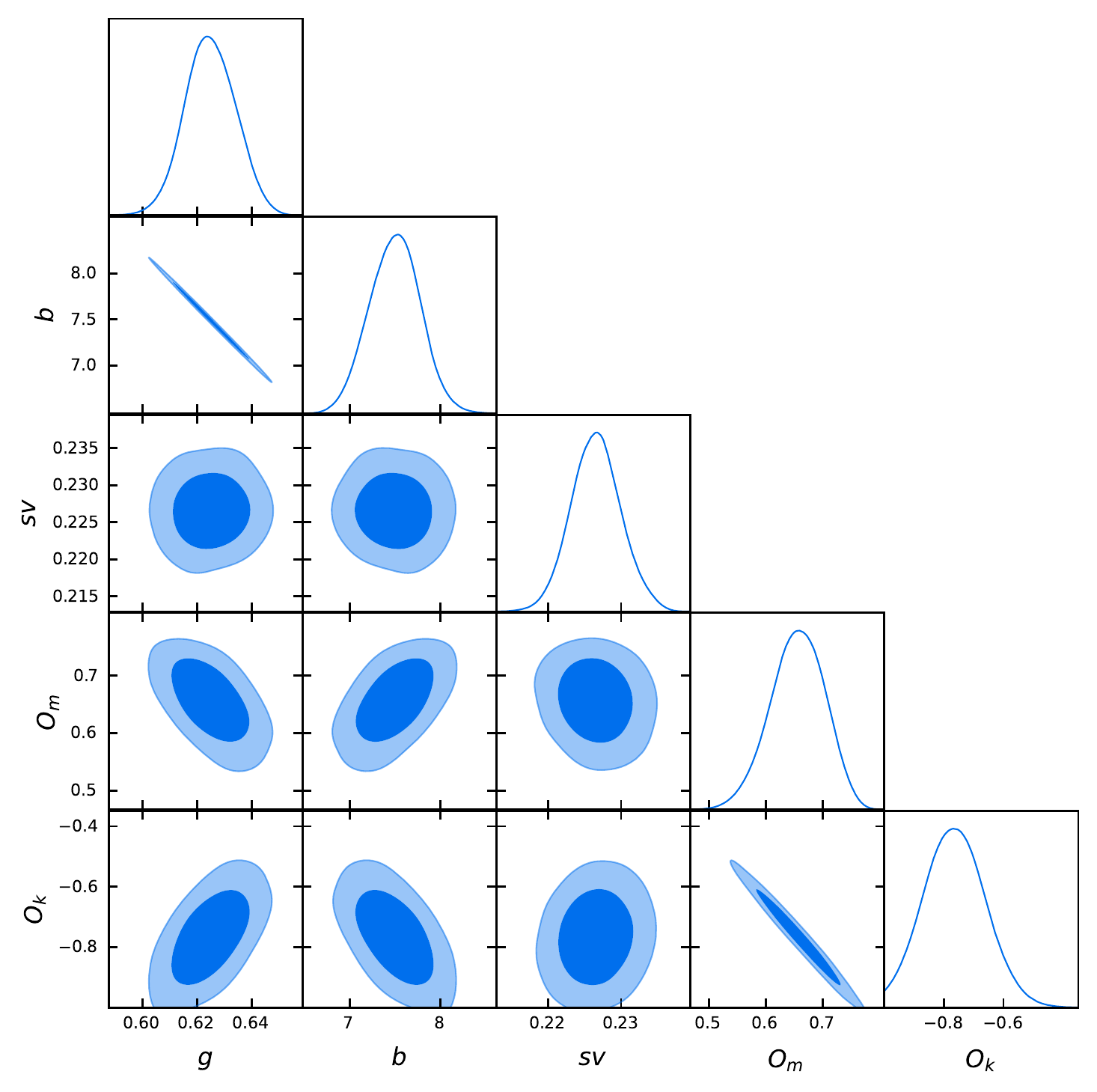}{0.3\textwidth}{(g) Non-calibrated QSOs + SNe Ia without evolution, both $\Omega_M$ and $\Omega_k$}
        \fig{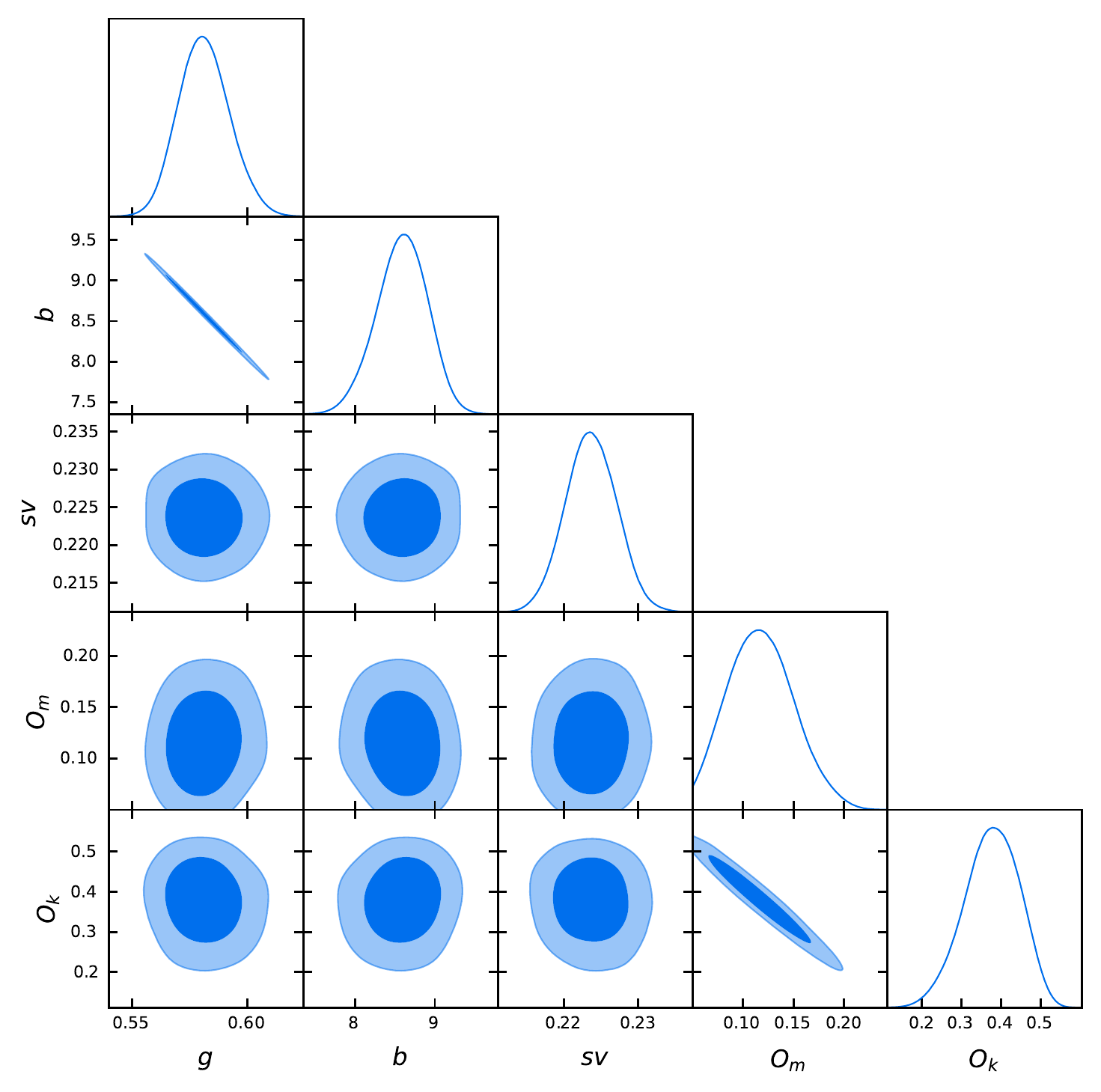}{0.3\textwidth}{(h) Non-calibrated QSOs + SNe Ia with fixed evolution, both $\Omega_M$ and $\Omega_k$}
        \fig{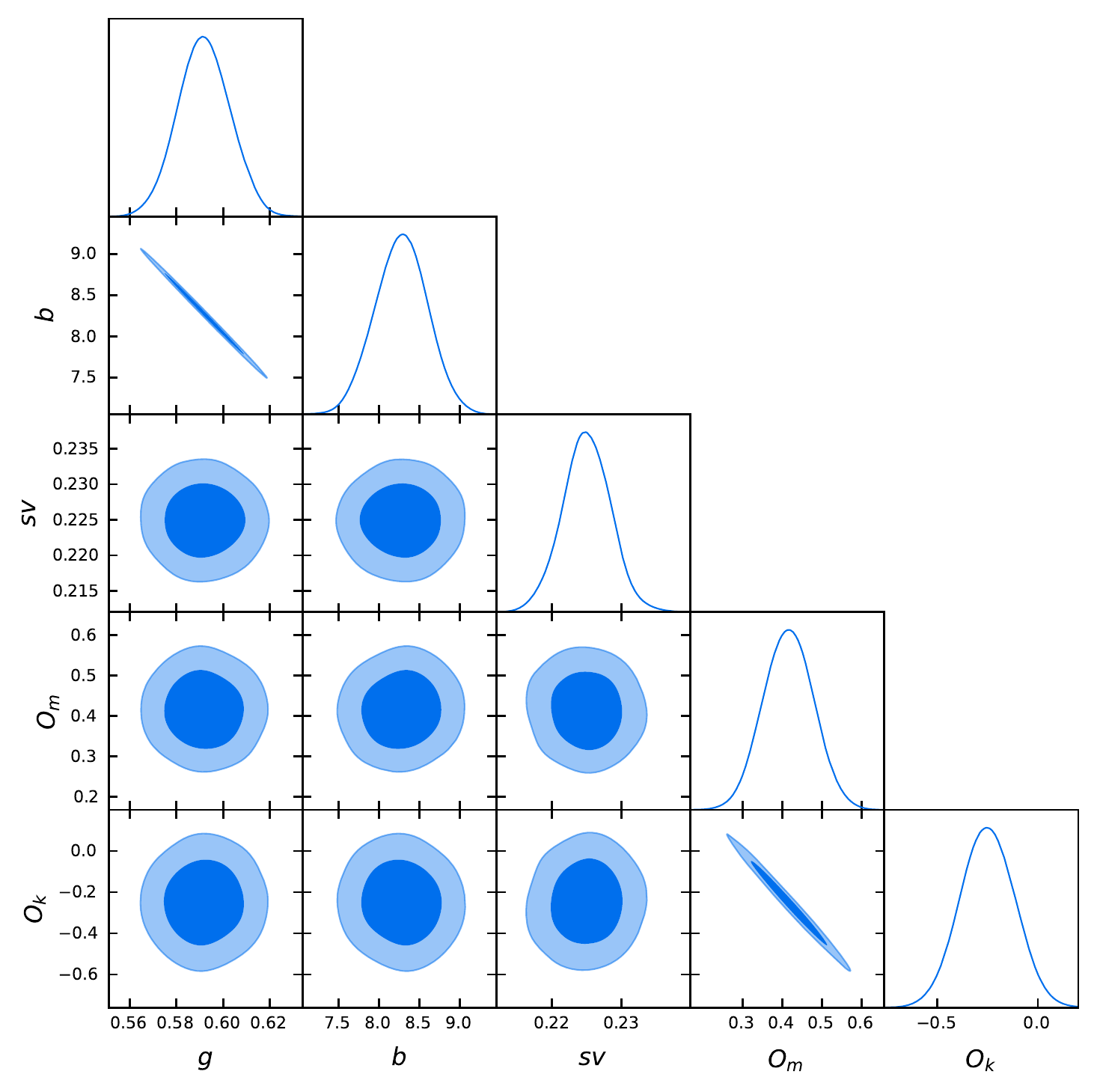}{0.3\textwidth}{(i) Non-calibrated QSOs + SNe Ia with varying evolution, both $\Omega_M$ and $\Omega_k$}}
\caption{Corner plots obtained under the assumption of the non-flat $\Lambda$CDM model.}
\label{cornerplot2}
\end{figure}

\begin{figure}
\gridline{\fig{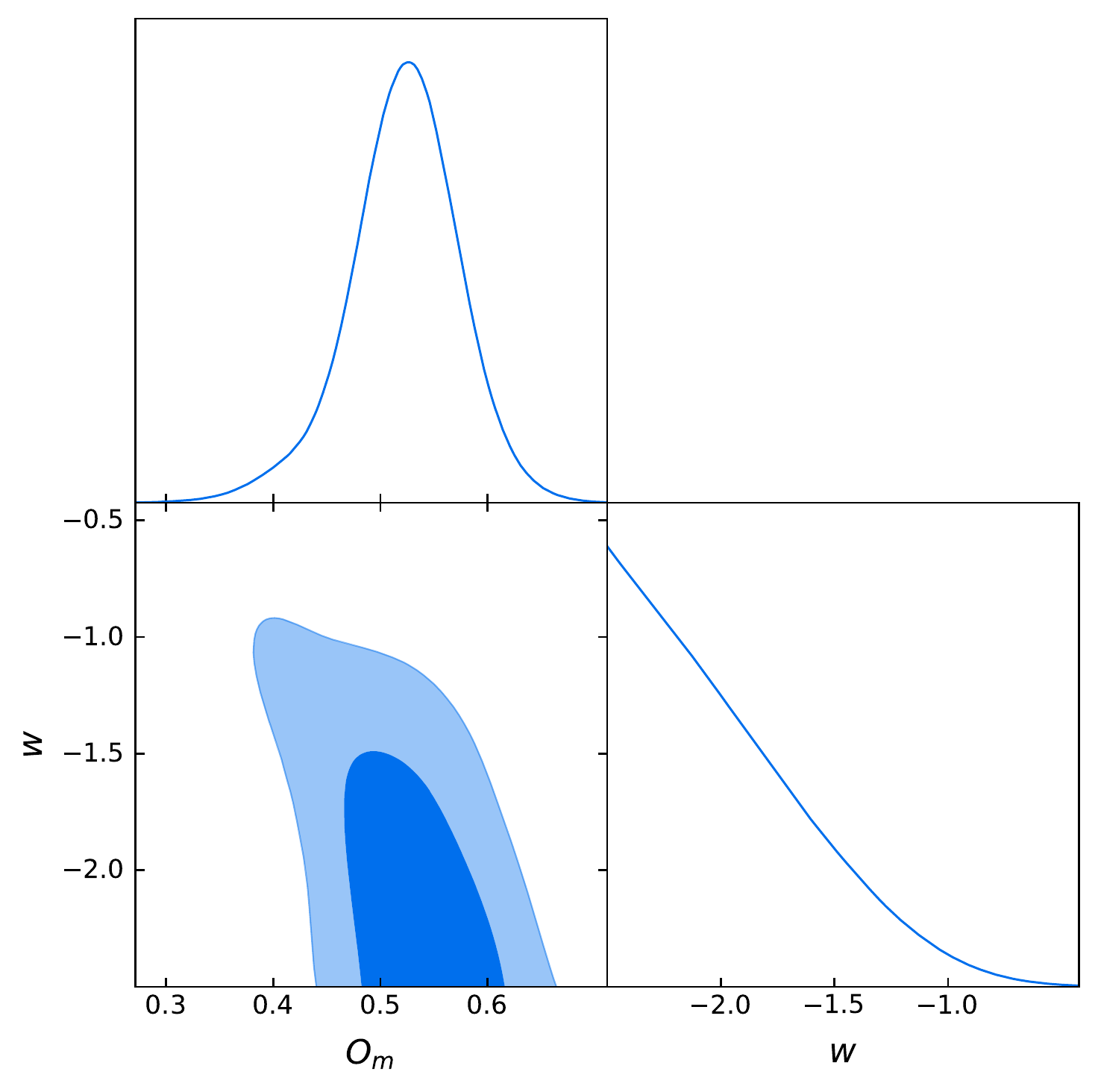}{0.3\textwidth}{(a) Only QSOs calibrated without evolution}
          \fig{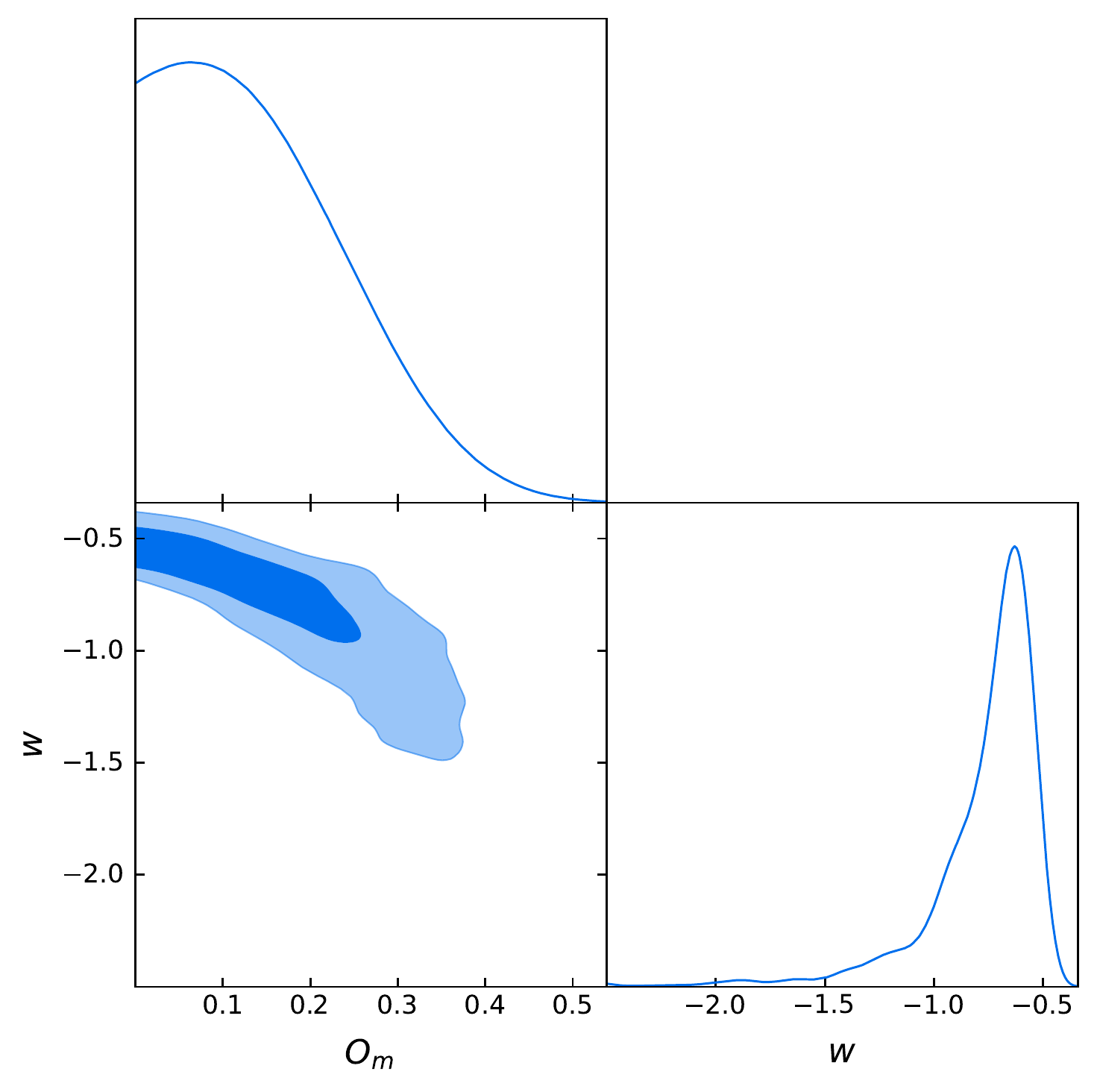}{0.3\textwidth}{(b) Only QSOs calibrated with fixed evolution}
          \fig{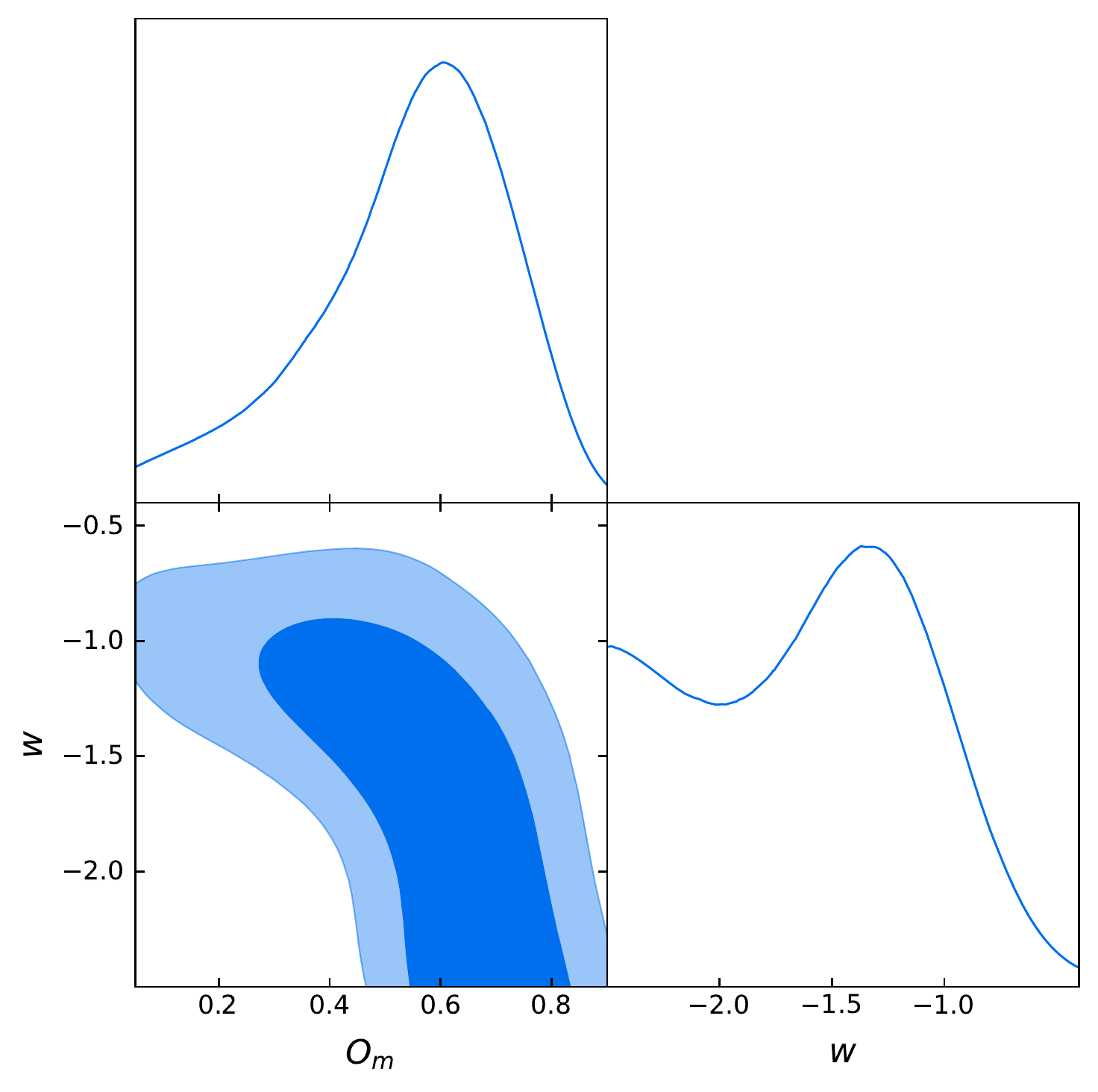}{0.3\textwidth}{(c) Only QSOs calibrated with varying evolution}}
\gridline{\fig{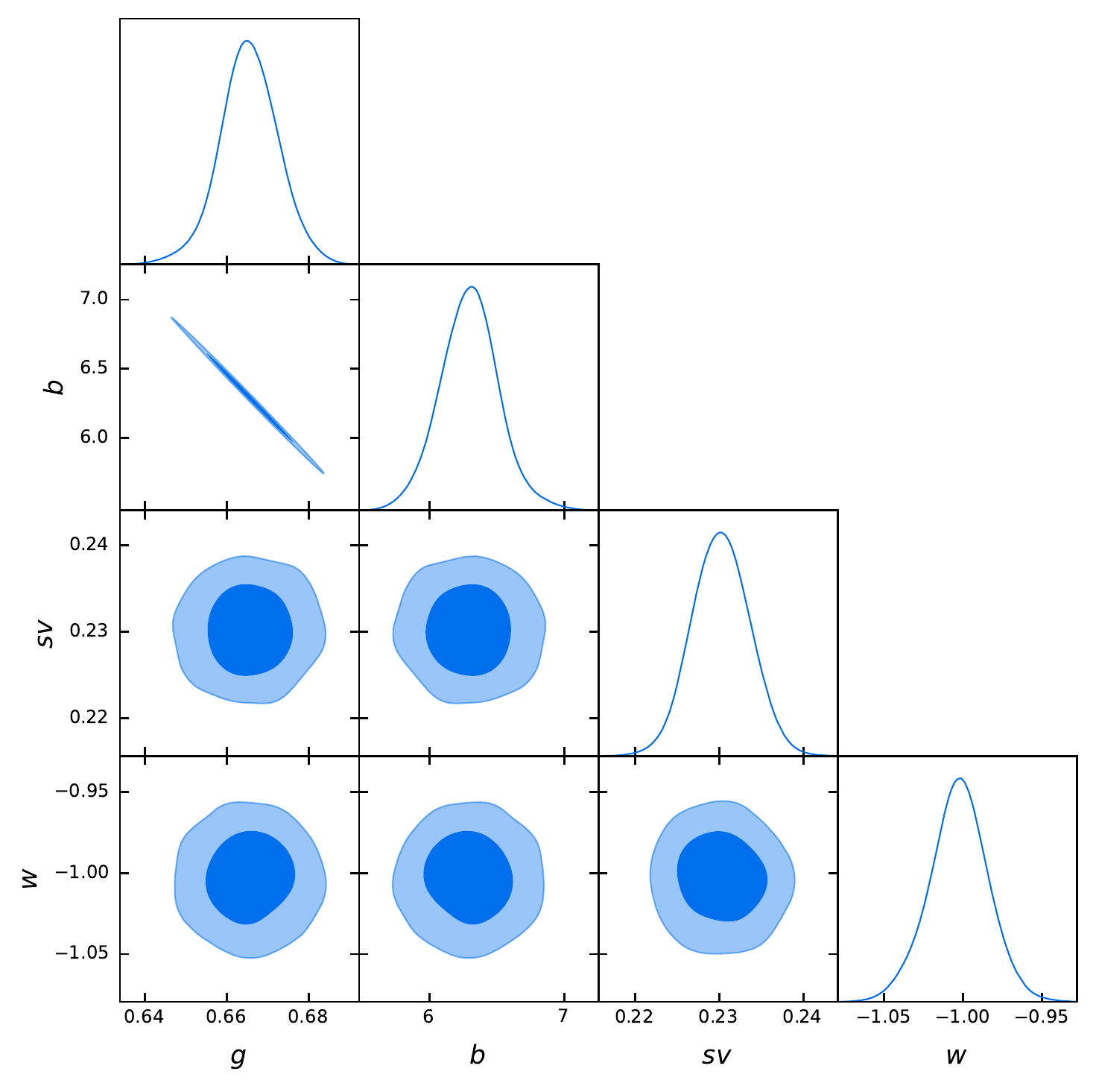}{0.3\textwidth}{(d) Non-calibrated QSOs + SNe Ia without evolution, only $w$}
          \fig{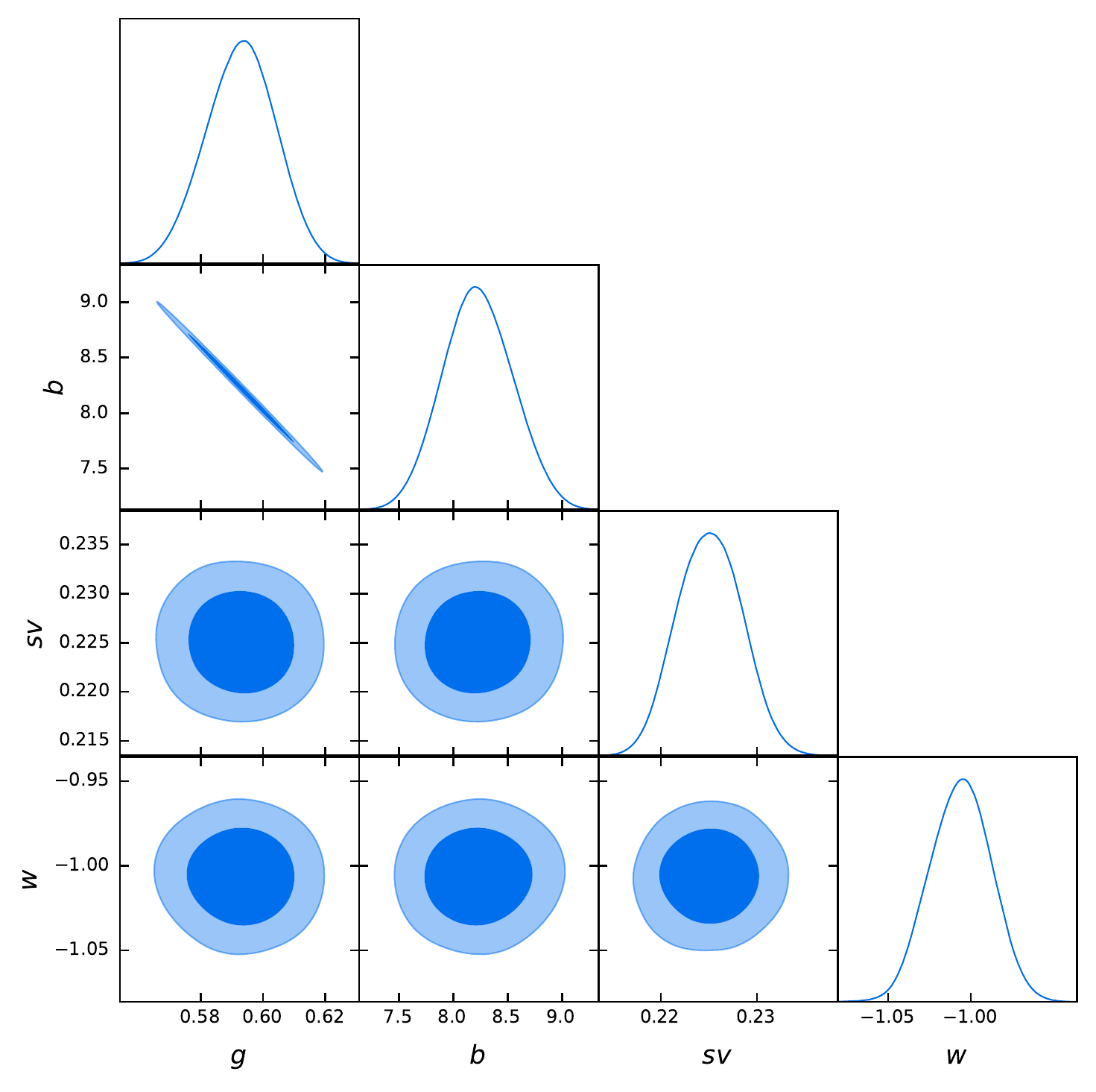}{0.3\textwidth}{(e) Non-calibrated QSOs + SNe Ia with fixed evolution, only $w$}
          \fig{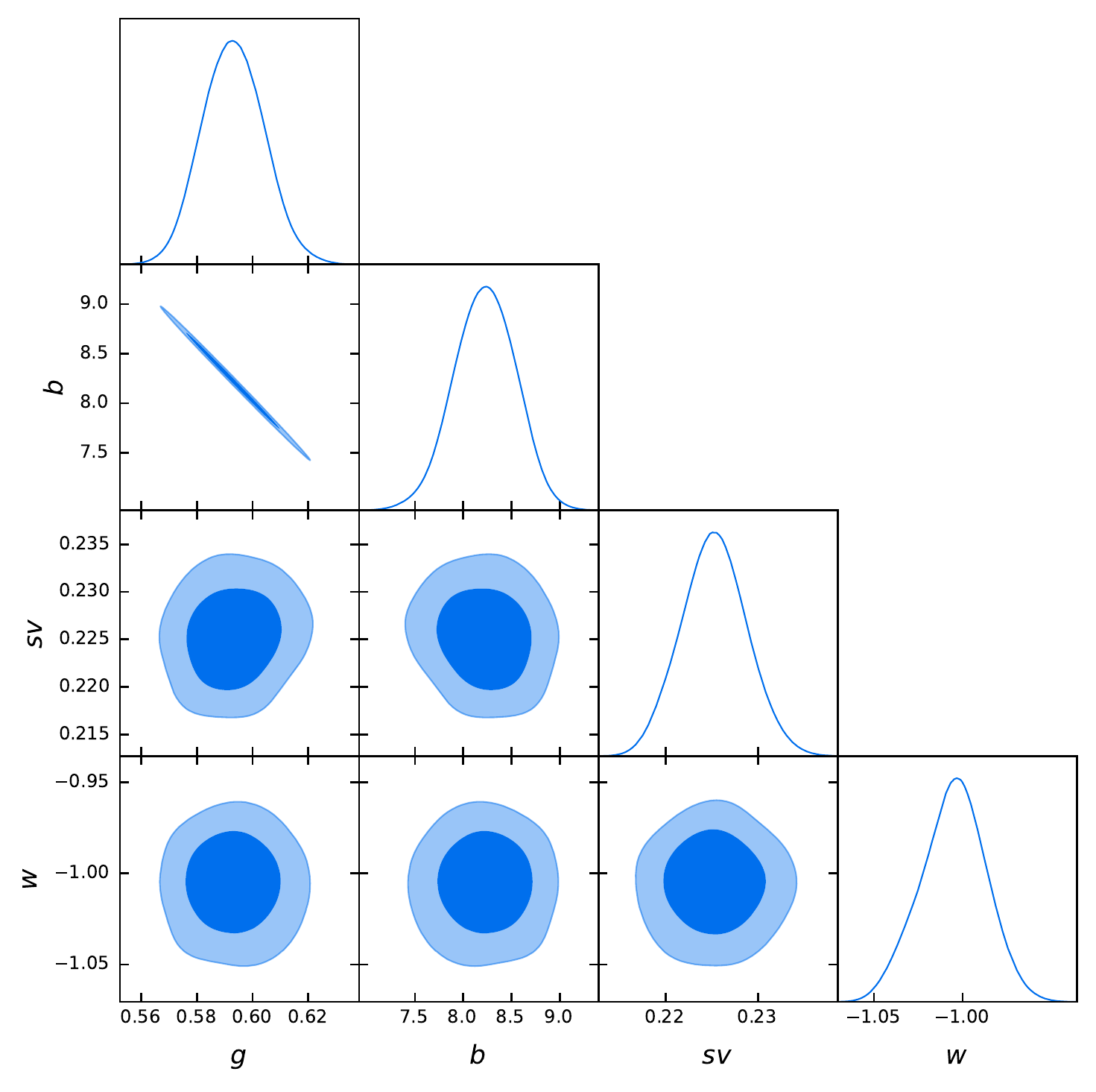}{0.3\textwidth}{(f)Non-calibrated QSOs + SNe Ia with varying evolution, only $w$}}
\gridline{\fig{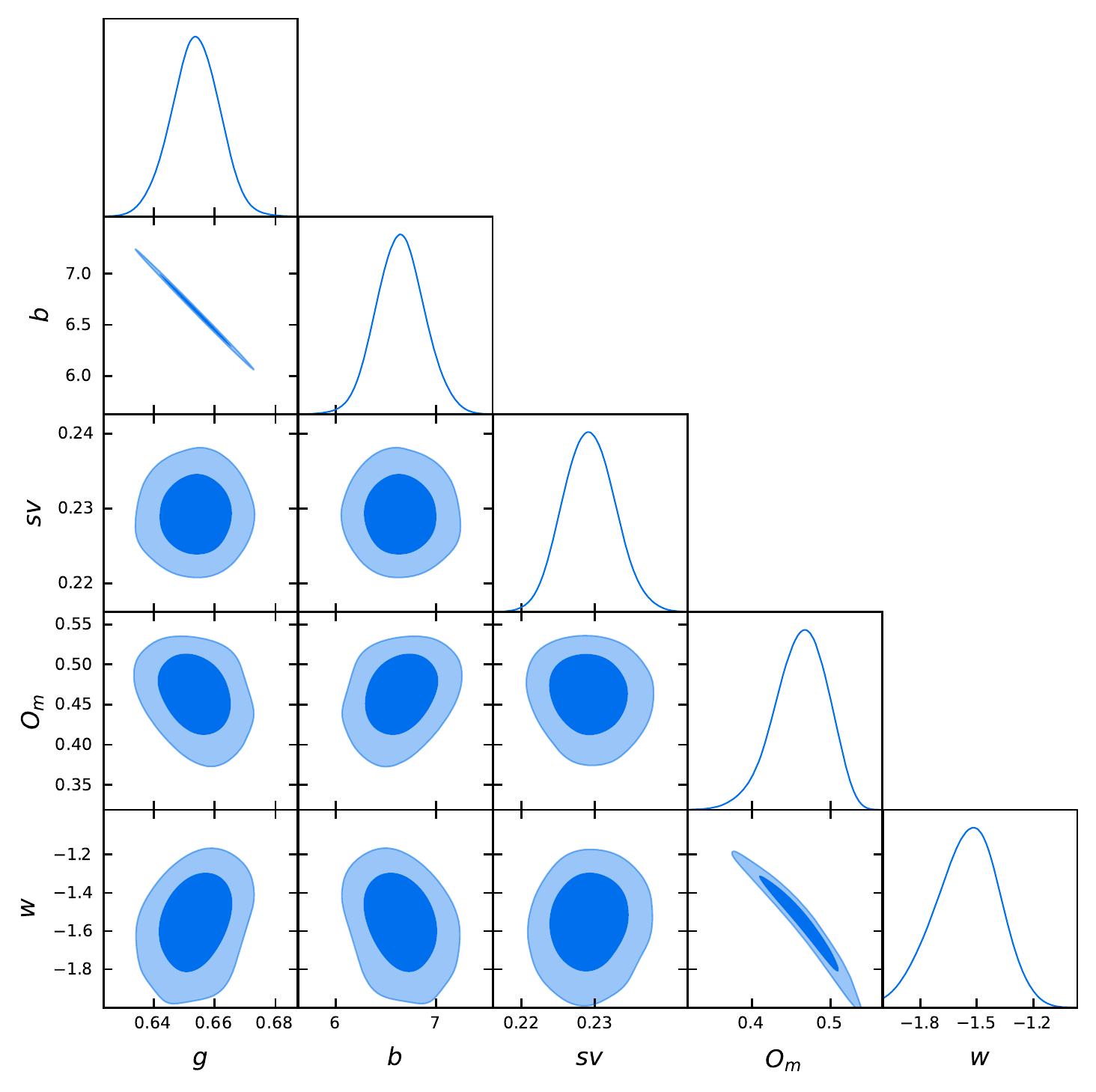}{0.3\textwidth}{(g) Non-calibrated QSOs + SNe Ia without evolution, both $\Omega_M$ and $w$}
        \fig{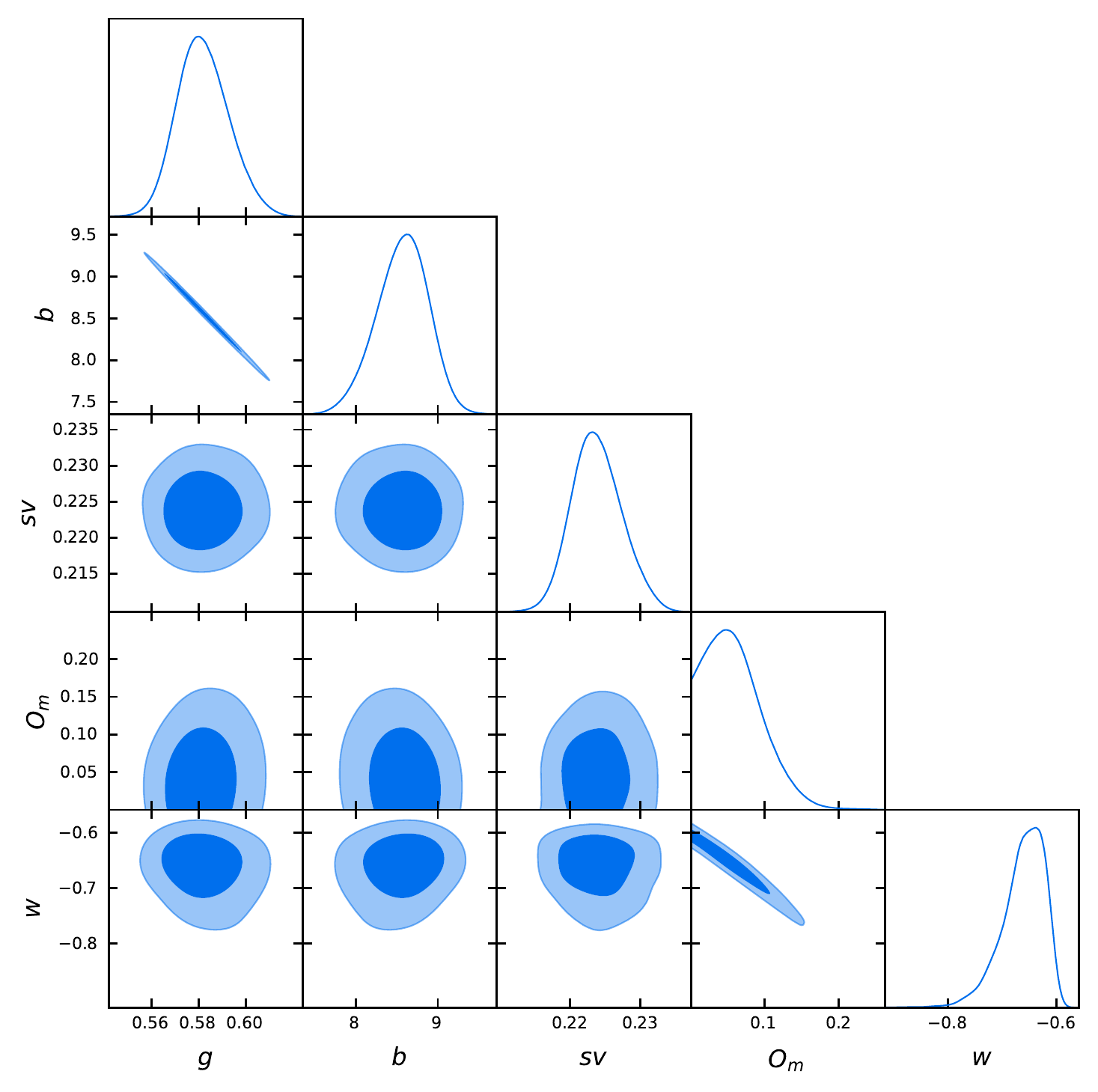}{0.3\textwidth}{(h) Non-calibrated QSOs + SNe Ia with fixed evolution, both $\Omega_M$ and $w$}
        \fig{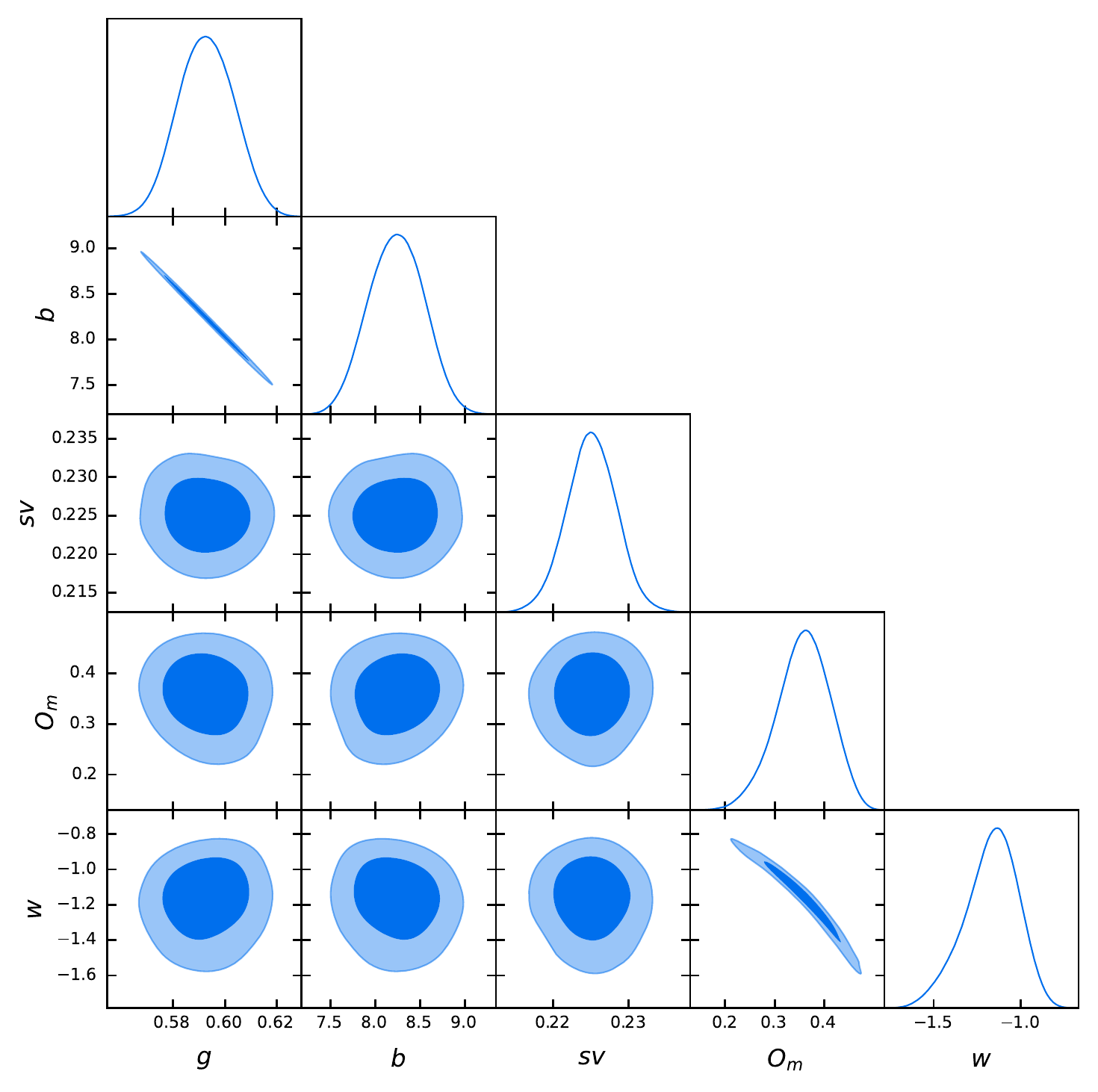}{0.3\textwidth}{(i)Non-calibrated QSOs + SNe Ia with varying evolution, both $\Omega_M$ and $w$}}
\caption{Corner plots obtained under the assumption of the $w$CDM model.}
\label{cornerplot3}
\end{figure}
    
    \item Investigating the results on the parameters of the RL relation when the QSO sample is not calibrated, we obtain that, by comparing cases with the same treatment of the correction for evolution, the fitted values of $g$, $b$, and $sv$ are always the same, independently on the cosmological model assumed and cosmological results obtained. Indeed, in our cosmological computations we do not obtain any significant correlation of the parameters of the RL relation with the cosmological ones, as visible in Figures \ref{cornerplot1}, \ref{cornerplot1.1}, \ref{cornerplot2}, and \ref{cornerplot3}. In addition, the application of the correction for the evolution, both fixed and free to vary with the cosmological parameters, reduces the intrinsic dispersion of the relation by a factor of $2.2 \%$ (from $sv= 0.230$ to $sv = 0.225$) making the RL relation even stronger. 
    
    \item Focusing on how the $H_0$ tension is impacted by our analyses, we show in Figure \ref{fig:H0tension} the $H_0$ values with their corresponding $1 \, \sigma$ uncertainties obtained in all the cases studied in this work. These values can be compared to the reference measurements from SNe Ia ($H_0 = 74.03 \, \pm 1.42 \,  \mathrm{km} \, \mathrm{s^{-1}} \,\mathrm{Mpc^{-1}}$) and CMB ($H_0 = 67.4 \, \pm 0.5 \, \mathrm{km} \, \mathrm{s^{-1}} \,\mathrm{Mpc^{-1}}$) shown with vertical dashed gray lines. Beside each point, we also report the computed z-score $z_s$ with respect to the CMB. This is defined as $z_s = \frac{|H_{0,CMB}-H_{0}|}{\sqrt{\sigma_{H_0,CMB}^2+\sigma_{H_0}^2}}$, where $H_{0,CMB} = 67.4$, $\sigma_{H_0,CMB} = 0.5$, and $H_0$ and $\sigma_{H_0}$ refer to the value and $1 \, \sigma$ error obtained from our fit, respectively. This computation allows us to compare each of the case results with the one obtained from the CMB. Referring to Figure \ref{fig:H0tension} and to Table \ref{tab:1}, we can note that the $1 \, \sigma$ errors on $H_0$ strongly decrease when we consider non-calibrated QSOs combined with SNe Ia, compared to the cases with calibrated QSOs alone, due to the significant contribution of SNe Ia. Indeed, the inclusion of SNe Ia in the cosmological computations guarantees tighter constraints on the free cosmological parameters, removing also convergence issues that are instead present when we do not consider the SNe Ia sample. In addition, calibrated QSOs alone do not show any precise hint on the $H_0$ favoured value, as the best-fit values span a very wide range of values, even from values lower than the one of CMB to values greater than the one of SNe Ia (blue, orange, green, red, and black points). Conversely, all the $H_0$ values obtained with non-calibrated QSOs combined with SNe Ia (purple, brown, pink, dark blue, and olive points) are compatible within $2 \, \sigma$ with each other pointing clearly to the region intermediate between the one of CMB and SNe Ia, with a z-score that increases significantly. Notably, the $1 \, \sigma$ uncertainties on $H_0$ obtained with non-calibrated QSOs together with SNe Ia are even smaller than $\sigma_{H_0,CMB} = 0.5$, reaching a value of $\sigma_{H_0}=0.14$ in the case with a fixed correction for evolution with $H_0$ the only free parameter of the fit. These results, concerning both the $H_0$ value and its uncertainty, prove the possible strong impact of our work on the $H_0$ tension. In the light of the newest results of binned analysis of the SNe Ia sample \citep{2021ApJ...912..150D} and SNe Ia together with BAOs \citep{2022arXiv220109848D} showing the existence of an evolution of $H_{0}$ parameter with redshift, our results make it even more significant to study such a model, since the high redshift QSOs could show more clearly if such an evolution exists and whether it could explain the current $H_{0}$ tension problem.
\end{itemize}

\begin{figure}[b!]
    \centering
    \includegraphics[width=.9\columnwidth]{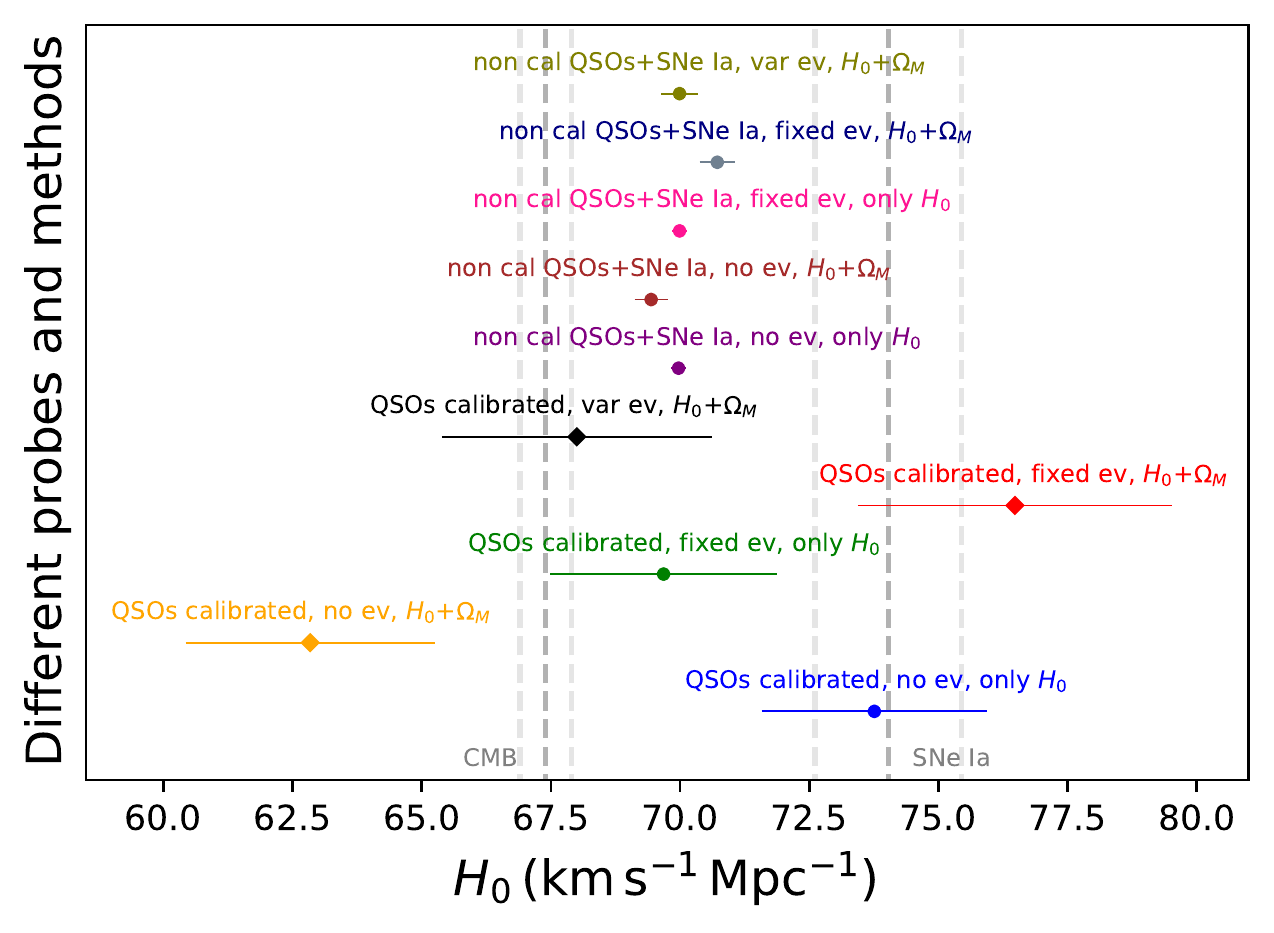}
    \caption{Comparison between the values of $H_0$ obtained in all our analyses. Beside each point, we report both the corresponding data set and methodology used for the computation and the z-score $z_s$ with respect to the $H_0$ value obtained from the CMB, defined as $z_s = \frac{|H_{0,CMB}-H_{0}|}{\sqrt{\sigma_{H_0,CMB}^2+\sigma_{H_0}^2}}$, where $H_0$ refers to the value obtained from our fit. The two vertical dashed gray lines with the corresponding dashed light gray lines are the reference points of CMB and SNe Ia measurements with their $1 \, \sigma$ uncertainties. The points marked with '$\diamondsuit$' correspond to the cases in which the analyses do not converge for one or more parameters.}
    \label{fig:H0tension}
\end{figure}

In conclusion, among all the data sets and methodologies investigated in this work, the case with non-calibrated QSOs combined with SNe Ia and with the correction for the evolution in redshift of luminosities varying together with the cosmological parameters is the one that turned out to be the most reliable and complete, as expected. Indeed, including the SNe Ia sample and taking into account this type of redshift evolution removes convergence issues and results in tighter constraints on the fitted parameters in all cosmological models studied, also allowing us to overcome the circularity problem. 
Considering this case, our results are always compatible with a flat $\Lambda$CDM model with $\Omega_M = 0.3$ and $H_0 = 70  \,  \mathrm{km} \, \mathrm{s^{-1}} \,\mathrm{Mpc^{-1}}$, without any hint of a DE with $w \neq -1$ nor a non-flat Universe. More precisely, we have always compatibility in $1 \, \sigma$, except for two cases: the flat $w$CDM and the non-flat $\Lambda$CDM models with both cosmological parameters free to vary, that shows a $2 \, \sigma$ discrepancy from $\Omega_M = 0.3$ and $w=-1$, and from $\Omega_M = 0.3$ and $\Omega_k=0$, respectively. We can notice that our approach has lead to values of cosmological parameters closer to the ones obtained with alone SNe Ia assuming a flat $\Lambda$CDM model than the ones presented in the QSO literature.
In addition, this case also allowed us to investigate how the values of the parameters of the RL relation are impacted by the assumption of a specific model. About this, we have strongly proved that these parameters do not show any dependence on the choice of the cosmological model.  

This manuscript shows that, when it comes to cosmological measurements, one should be extremely careful in assuming the absence of selection biases and redshift evolution. Indeed, it is necessary to study their possible presence and behaviour, as corrections for such effects can push further the constraining of cosmological parameters. Future analyses should investigate those biases also for other probes such as SNe Ia. We showed that also QSOs alone can be used as standalone probes without any cut in redshift nor calibration once the correction for the evolution is accounted for. Indeed, for the first time in the literature for such a sample we obtained closed contours in $2\,\sigma$ for the computation of $\Omega_{M}$. 
In the end, this work could represent a leap forward to shed a light on the $H_0$ tension and to investigate if the current tension could be ascribed to an evolution of $H_0$ with redshift or to a constant $H_0$ value that stands between the one of SNe Ia and CMB.
\section*{Acknowledgements}

A.L. would like to acknowledge the financial support from the Programme Council of Studies in Mathematics and the Natural Sciences (SMP) at the Jagiellonian University. S.N. is supported financially by JSPS KAKENHI (A) Grant Number JP19H00693, Interdisciplinary Theoretical and Mathematical Sciences Program (iTHEMS), and the Pioneering Program of RIKEN for Evolution of Matter in the Universe (r-EMU). GB acknowledges the Istituto Nazionale di Fisica Nucleare (INFN), sezione di Napoli, for supporting her visit at NAOJ. GB is grateful to be hosted by Division of Science. MGD acknowledges the Division of Science and NAOJ.


\bibliography{bibliografia}{}
\bibliographystyle{aasjournal}



\end{document}